\documentclass[preprint,amsmath,amssymb,nofootinbib]{revtex4}
\usepackage[utf8]{inputenc}
\usepackage{rotating}
\usepackage{hyperref}
\usepackage{array}
\usepackage{subfig}
\usepackage{float}
\usepackage{fancyhdr}
\usepackage{slashed}
\usepackage{scalerel}
\usepackage[inline]{enumitem}
\usepackage{dcolumn}
\usepackage{caption}
\usepackage{booktabs}
\usepackage{makecell}
\usepackage{placeins}
\usepackage[export]{adjustbox}
\usepackage{bm}
\usepackage{xcolor}
\usepackage{graphicx}
\usepackage{diagbox}
\usepackage{multirow}
\usepackage{ctable}
\usepackage[normalem]{ulem}
\usepackage[T1]{fontenc}

\definecolor{orange}{rgb}{1,0.5,0}

\begin{document}

%\title{Majorana neutrino effective interactions in ep colliders.}
\title{Sensitivity prospects for lepton-trijet signals in the $\nu$SMEFT at the LHeC}

\author{Gabriel Zapata}
\affiliation{Instituto de F\'{\i}sica de Mar del Plata (IFIMAR)\\ CONICET, UNMDP\\
Funes 3350, (7600) Mar del Plata, Argentina} 
\affiliation{Instituto de F\'{\i}sica La Plata (IFLP) CONICET, UNLP\\
Diagonal 113 e/ 63 y 64, (1900) La Plata, Argentina.
}

\author{Tom\'as Urruzola}
\affiliation{Instituto de F\'{\i}sica, Facultad Ingenier\'ia,
 Universidad de la Rep\'ublica \\ Julio Herrera y Reissig 565,(11300) 
Montevideo, Uruguay.}

\author{Oscar A. Sampayo}
%\email{sampayo@mdp.edu.ar}
\affiliation{Instituto de F\'{\i}sica de Mar del Plata (IFIMAR)\\ CONICET, UNMDP\\ Departamento de F\'{\i}sica,
Universidad Nacional de Mar del Plata \\
Funes 3350, (7600) Mar del Plata, Argentina}

\author{Luc\'{\i}a Duarte}
\email{lucia@fisica.edu.uy}
\affiliation{Instituto de F\'{\i}sica, Facultad de Ciencias,
 Universidad de la Rep\'ublica \\ Igu\'a 4225,(11400) 
Montevideo, Uruguay.}

%%%%%%%%%%%%%%%%%%%%%%%%%%%%%%%%%%%%%%%%%%%%%%%%%%%%%%%%%%%%%%%%%%%%%%%%%%%%%%%%%%%%%%%%
\begin{abstract}

The observation of neutrino oscillations and masses motivates the extension of the standard model with right handed neutrinos, leading to heavy neutrino states possibly in the electroweak scale, which could be impacted by new high-scale weakly coupled physics. A systematic tool for studying these interactions is the neutrino-extended standard model effective field theory $\nu$SMEFT. In this work we study the prospects of the future LHeC electron-proton collider to discover or constrain the $\nu$SMEFT interactions, performing the first dedicated and realistic analysis of the well known lepton-trijet signals, both for the lepton flavor violating $p ~ e^{-} \rightarrow \mu^{-} +  3 \mathrm{j}$ (LFV) and the lepton number violating $p ~ e^{-} \rightarrow \mu^{+} + 3 \mathrm{j}$ (LNV) channels, for HNLs masses in the electroweak scale range: $100 ~\rm GeV \leq m_N \leq 500 ~\rm GeV$. The obtained sensitivity prospects show that the LHeC with $100 ~\rm fb^{-1}$ luminosity could be able to probe the scenario of a heavy $N$ and constrain the effective couplings to a region of the parameter space as tight as the bounds that are currently considered for the $\mathcal{O}(10)$ GeV scale masses, with effective couplings of $\mathcal{O}(10^{-1})$ for NP scale $\Lambda=1 \rm TeV$.

\end{abstract}
%%%%%%%%%%%%%%%%%%%%%%%%%%%%%%%%%%%%%%%%%%%%%%%%%%%%%%%%%%%%%%%%%%%%%%%%%%%%%%%%%%%%%%%%

\maketitle

\section{Introduction}{\label{intro}}

The existence of light neutrino masses and the oscillation phenomena can be accounted for in a minimal extension of the SM Lagrangian with sterile right-handed neutrinos $N_R$, which allow for a lepton number violating Majorana mass term, as in the Type-I seesaw \cite{Minkowski:1977sc, Mohapatra:1979ia, Yanagida:1980xy, GellMann:1980vs, Schechter:1980gr}. This Majoarana mass scale is a parameter, not related to electroweak symmetry breaking, and in the naive (high-scale) seesaw it is taken to be large, so that it suppresses the induced mass for the mostly active light neutrinos, thus leading to very heavy massive states in addition to the light ones. However, the lightness of the known neutrinos could also be explained by symmetry principles, which is the argument of the linear and inverse seesaw variants \cite{Malinsky:2005bi, Mohapatra:1986bd} to lower the mass scale of the heavy neutrinos, and allow to probe their phenomenology at laboratory energies. Indeed, even if the heavy neutral leptons (HNLs) or heavy neutrinos are accesible, their interactions with the SM particles in this scenarios are also suppressed by their small mixing $U_{\ell N}$ with the active neutrinos $\nu_{\ell L}$, strongly constrained by experiments \cite{Abdullahi:2022jlv}, and thus would have undetectably weak interactions.

Yet, a variety of new physics may be hidden at energies well above the EW scale: its possible effects on the SM degrees of freedom are systematically studied with the use of the SM effective field theory (SMEFT). This new physics can also impact the behavior of the heavy neutrinos if they are light enough to be included in the low energy spectrum, and if present, would probably dominate over the interactions due to their mixing with the active neutrino states. Thus the HNLs interactions with the SM particles can be seen as the remnant of new UV physics and described by an effective field theory including them also as part of its building blocks. This is the Standard Model Effective Field Theory framework extended with right-handed Neutrinos $\nu$SMEFT, with operators known up to mass dimension $d=9$ \cite{Anisimov:2006hv,Graesser:2007yj,Graesser:2007pc,delAguila:2008ir,Aparici:2009fh,Liao:2016qyd,Bhattacharya:2015vja,Li:2021tsq}.\footnote{Also called SMNEFT, $N_R$SMEFT and $\nu_R$SMEFT in the literature.}

This EFT includying the $N_R$ as accesible states has received increasing attention since it offers an efficient tool to parameterize the effects of UV physics and learn from prospective studies how to discover the HNLs by their new interactions, or to constrain the Wilson coefficients of the distinct operators consistent with the SM symmetry at a given mass dimension $d$ and energy scale $\Lambda$. Since its first introduction by the authors of \cite{delAguila:2008ir} as an explicit EFT to study neutrino interactions, much progress has been made in the $\nu$SMEFT framework both from the phenomenology and from the theory side \cite{Peressutti:2014lka,Duarte:2014zea,Duarte:2015iba,Duarte:2016miz,Duarte:2016smd,Duarte:2016caz,Caputo:2017pit,Duarte:2018xst,Yue:2018hci,Duarte:2018kiv,Duarte:2019rzs,Bischer:2019ttk,Alcaide:2019pnf,Butterworth:2019iff,Jones-Perez:2019plk,Chala:2020vqp,Dekens:2020ttz,Barducci:2020ncz,Duarte:2020vgj,Biekotter:2020tbd,DeVries:2020jbs,Barducci:2020icf,Dekens:2021qch,Cirigliano:2021peb,Cottin:2021lzz,Beltran:2021hpq,Zhou:2021ylt,Zhou:2021lnl,Beltran:2022ast,Delgado:2022fea,Barducci:2022gdv,Zapata:2022qwo,Barducci:2022hll,Talbert:2022unj,Mitra:2022nri,Beltran:2023nli,Fernandez-Martinez:2023phj,Beltran:2023ymm,Beltran:2023ksw,Duarte:2023tdw}. 

Many of these recent efforts have pointed to the phenomenology of HNLs within the $\nu$SMEFT framework at the LHC and future lepton colliders. 
Here we focus in the study of the sensitivity projections for the HNL $N$ with effective interactions at the LHeC, an $e^{-}p$ collider to be built at the LHC tunnel \cite{LHeC:2020van, AbelleiraFernandez:2012cc, Bruening:2013bga}. For this study we consider a simplified scenario with only one heavy neutrino, neglecting its mixing with the active states and tackling an encompassable parameter space involving the HNL mass and its effective couplings. We consider the $N$ decay to muons and jets, usually called the lepton-trijet final state: $p ~ e^{-} \rightarrow \mu^{\pm} +  3 \mathrm{j}$. The process with final muons ($p ~ e^{-} \rightarrow \mu^{-} +  3 \mathrm{j}$) conserves lepton number but violates flavor (LFV), while the process with final anti-muons ($p ~ e^{-} \rightarrow \mu^{+} +  3 \mathrm{j}$) violates also lepton number by two units (LNV). 

Previous studies of the lepton-trijet signal given by the Type-I seesaw mixing interactions at the LHeC can be found in \cite{Li:2018wut, Antusch:2019eiz, Gu:2022muc}, originally tackled in \cite{Antusch:2016ejd, Blaksley:2011ey, Liang:2010gm, Ingelman:1993ve, Buchmuller:1991tu}. In ref. \cite{Duarte:2014zea} our group studied for the first time the potential of the LHeC to discover Majorana neutrinos in the $\nu$SMEFT context, and in ref. \cite{Duarte:2018xst} we explored the possibility to disentangle the contributions of effective operators with different Dirac-Lorentz structure to the LNV lepton-trijet process with the aid of angular distributions and polarization effects. Here we improve those results with up-to-date and realistic simulation and analysis at the reconstructed level. A recent prospective study for the Electron-Ion Collider (EIC) at Brookhaven can be found in \cite{Batell:2022ogj}. Also a diversity of related LHeC sensitivity studies can be found in the recent literature: a search for HNLs with additional Leptoquark interactions giving displaced fat-jets can be found in \cite{Cottin:2021tfo}, prospects for charged leptons flavor violating signals are given in \cite{Antusch:2020vul}, as well as fat-jet searches of very heavy massive neutrinos in \cite{Das:2018usr}.

The paper is organized as follows: in section \ref{sec:eff_form} we describe the $\nu$SMEFT formalism and overview the phenomenology associated with the different effective interactions. In section \ref{sec:constraints} we inspect the current existing bounds on the effective couplings to be considered for the benchmark scenarios probed in this study. In section \ref{sec:Collider analysis} we discuss our search strategy for the LHeC, characterizing the $N$ production mechanism kinematics in $ep$ colliders (\ref{sec:signals}). We also revisit the $N$ decay channels branching ratios and total width (\ref{sec:NWidth}), discuss the SM processes considered as background at the reconstructed level, and discuss the multivariate analysis performed to separate the signals. The sensitivity prospects for the heavy $N$ in $\nu$SMEFT at the LHeC are shown in figure \ref{fig:contmuamu}. We close with a summary in section \ref{sec:summary}.

%%%%%%%%%%%%%%%%%%%%%%%%%%%%%%%%%%%%%%%%%%%%%%%%%%%%%%%%%%%%%%%%%%%%%%%%%%%%%%%%%%%%%%%%
\section{Effective interactions formalism\label{sec:eff_form}}

We consider the SM Lagrangian to be extended with only \emph{one} right-handed neutrino $N_R$ with a Majorana mass term ($\sim M_N$).\footnote{At least two heavy $N$ states are required to reproduce the measured masses and mixings with light neutrinos, but this simplifying assumption retains the main phenomenology and corresponds to scenarios where the additional $N$ are too heavy to impact in low-energy observables.} The renormalizable $d=4$ Lagrangian extension reads 
\begin{eqnarray}\label{eq:LagSeesaw}
\mathcal{L}_{\rm d=4}= \overline{N_R} i\slashed{\partial} N_R - \left( \frac{M_N}{2} \overline{N^{c}_R} N_R + \sum_{\ell} Y_\ell ~ \overline{L_\ell}\tilde{\phi} N_R + \text{ h.c.}\right).
\end{eqnarray}
After diagonalization, one gets a massive state $N$ as an observable degree of freedom, together with the three known light neutrino states (with masses $m_{\nu}\sim 0.1$ eV), which are all of Majorana nature. The flavor neutrino eigenstates contain some part of the heavy $N$ due to the mixing $U_{\ell N} = Y_\ell ~v/\sqrt{2} M_N$: 
\begin{equation*}\label{eq:mixing}
    \nu_{\ell L}=\sum_{i=1}^{3}U_{\ell i}\nu_i+U_{\ell N}N.
\end{equation*}
In turn, the heavy state $N$ is mostly composed of the right-handed state $N \simeq N_R$ with negligible mixing with the active $\ell$ flavor states $\nu_{\ell L}$, constrained by the naive seesaw relation $U_{\ell N}\lesssim \sqrt{ \frac{m_{\nu}}{M_{N}}}$, and thus with negligible interaction through the SM electroweak currents when its mass is above the GeV scale.

\begin{table}[tbp]
\begin{adjustbox}{width=\textwidth,center} % You. Shall. Not. Pass!
\begin{tabular}{|>{\raggedright\arraybackslash}p{3cm}|l l|>{\centering\arraybackslash}p{5.8cm}|l|}
\hline
\textbf{Type} & \textbf{Operator} & 
\phantom{X} & \textbf{Interactions} & \textbf{Coupling} \\ \hline\hline

\phantom{X} $N$ mass $d=5$ 
& $\mathcal{O}^{d=5}_{N\phi}$ ($\mathcal{O}^{d=5}_{\rm Higgs}$)
& $(\bar{N}N^{c})(\phi^{\dagger} \phi)$  
& $h N N$ and Majorana mass term  
& $\alpha^{d=5}_{N\phi}$ \\ \hline 

\phantom{X} Dipole $d=5$ 
& $\mathcal{O}^{(5)}_{NB}$  
& $(\bar{N}_a \sigma_{\mu \nu} N^{c}_b) B^{\mu \nu}$, $a \neq b$ 
& Dipoles $d_{\gamma}, d_Z$ 
& $\alpha^{d=5}_{NB}$ \\ \hline \hline

$h$-dressed mixing   
& $\mathcal{O}^{(i)}_{LN\phi}$ ($\mathcal{O}_{\rm LNH}^\beta$)  
& $(\phi^{\dagger}\phi)(\bar L_i N \tilde{\phi})$  
& Yukawa + doublet ($U_{\ell N}$ and $m_{\nu}$) 
& $\alpha^{(i)}_{LN\phi}$ \\ \hline 

\phantom{X} Bosonic 
& $\mathcal{O}_{NN\phi}$ ($\mathcal{O}_{\rm HN}$) 
& $i(\phi^{\dagger}\overleftrightarrow{D_{\mu}}\phi)(\bar N \gamma^{\mu} N)$                                
& Neutral current ($NNZ$) 
& $\alpha_{NN\phi}=\alpha_{Z}$ \\ \cline{2-3} \cline{5-5}

\phantom{X} Currents   
& $\mathcal{O}^{(i)}_{Nl\phi}$ ($\mathcal{O}_{\rm HN\ell}^{\beta}$) 
& $i(\phi^T \epsilon D_{\mu}\phi)(\bar N \gamma^{\mu} l_i)$  
& Charged current ($N l W$) 
& $\alpha^{(i)}_{Nl\phi}=\alpha^{(i)}_{W}$ \\ \hline

\phantom{X} Dipoles 
& $\mathcal{O}^{(i)}_{NB}$ ($\mathcal{O}_{\rm NB}$)  
& $(\bar L_i \sigma^{\mu\nu} N) \tilde \phi B_{\mu\nu}$ 
& One-loop level generated 
& $\alpha^{(i)}_{NB}/(16\pi^2)$ \\ \cline{2-3} \cline{5-5}

\phantom{X}  
& $\mathcal{O}^{(i)}_{NW}$ ($\mathcal{O}_{\rm NW}^\beta$)    
& $(\bar L_i \sigma^{\mu\nu} \sigma^I N) \tilde \phi W_{\mu\nu}^I$ 
& $d_{\gamma}, d_Z, d_W$ 
& $\alpha^{(i)}_{NW}/(16\pi^2)$ \\ \hline 

\phantom{X}  
& $\mathcal{O}^{(i)}_{QNN}$ ($\mathcal{O}_{\rm QN}$)  
& $(\bar{Q_i} \gamma^\mu Q_i) (\bar{N} \gamma_\mu N)$  
& 4-fermion 
& $\alpha^{(i)}_{QNN}$ \\ \cline{2-3} \cline{5-5}

\phantom{X} 4-fermion NC 
& $\mathcal{O}^{(i)}_{LNN}$ ($\mathcal{O}_{\rm LN}^{\beta}$) 
& $(\bar{L_i} \gamma^\mu L_i) (\bar{N} \gamma_\mu N)$  
& vector-mediated  
& $\alpha^{(i)}_{LNN}$ \\ \cline{2-3} \cline{5-5}

\phantom{X} 
& $\mathcal{O}^{(i)}_{fNN}$ ($\mathcal{O}_{\rm ff}$) 
& $(\bar f_i \gamma^{\mu}f_i) (\bar N \gamma_{\mu}N)$  
& $f=u, d, l$  
& $\alpha^{(i)}_{fNN}$ \\ \hline

\phantom{X} 4-fermion CC 
& $\mathcal{O}^{(i, j)}_{duNl}$ ($\mathcal{O}_{\rm duN\ell}^{\beta}$) 
& $(\bar d _j \gamma^{\mu} u _j) (\bar N \gamma_{\mu} l_i)$ 
& 4-fermion vector-mediated  
& $\alpha^{(i, j)}_{duNl}= \alpha^{(i, j)}_{V_0}$ \\ \hline 

\phantom{X} 
& $\mathcal{O}^{(i, j)}_{QuNL}$ ($\mathcal{O}_{\rm QuNL}^\alpha$) 
& $(\bar Q _i u _i)(\bar N L_j)$ 
& 4-fermion 
& $\alpha^{(i,j)}_{QuNL}=\alpha^{(i,j)}_{S_1}$ \\ \cline{2-3} \cline{5-5} 

\phantom{X} 4-fermion 
& $\mathcal{O}^{(i, j)}_{LNQd}$ ($\mathcal{O}_{\rm LNQd}^\alpha$)  
& $(\bar L_i N) \epsilon (\bar Q _j d _j)$  
& scalar-mediated  
& $\alpha^{(i,j)}_{LNQd}=\alpha^{(i,j)}_{S_2}$ \\ \cline{2-3}  \cline{5-5}

\phantom{X} CC/NC 
& $\mathcal{O}^{(i, j)}_{QNLd}$ 
& $(\bar Q _i N)\epsilon (\bar L_j d_j)$  
& \phantom{X}
& $\alpha^{(i,j)}_{QNLd}= \alpha^{(i,j)}_{S_3}$ \\  \cline{2-3}  \cline{5-5}

\phantom{X} 
& $\mathcal{O}^{(i, j)}_{LNLl}$ ($\mathcal{O}_{\rm LNL\ell}^{\delta\beta}$) 
& $(\bar L_i N)\epsilon (\bar L_j l_j)$  
&  \phantom{X}
& $\alpha^{(i, j)}_{LNLl}=\alpha^{(i, j)}_{S_0}$ \\ \hline
\end{tabular} 
\end{adjustbox}
\caption{{Basis of $d=5$ and $d=6$ operators with a right-handed neutrino $N$ \cite{delAguila:2008ir, Liao:2016qyd}. Here $l_i$, $u _i$, $d _i$ and $L_i$, $Q _i$ denote the right handed singlets and the left-handed $SU(2)$ doublets, respectively. The field $\phi$ is the scalar doublet, $B_{\mu\nu}$ and $W_{\mu\nu}^I$ are the $U(1)_{Y}$ and $SU(2)_{L}$ field strengths. Also $\sigma^{\mu \nu}=\frac{i}{2}[\gamma^{\mu}, \gamma^{\nu}]$ and $\epsilon=i\sigma^{2}$ is the anti symmetric symbol in two dimensions. We follow the notation in \cite{delAguila:2008ir} and quote the names in (\cite{Fernandez-Martinez:2023phj}). See Appendix \ref{app:explicitLagOpe} for the explicit Lagrangian terms.}}
\label{tab:Operators}
\end{table}

In our simplified setup, we will not include the renormalizable Lagrangian in \eqref{eq:LagSeesaw} and thus neglect the mixings $U_{\ell N}$. In this way, the new physics effects on the heavy state $N$ (possibly due to the presence of new mediators in the UV scale $\Lambda$) are parameterized by a set of effective operators $\mathcal{O}_\mathcal{J}$ constructed with the SM and the $N_R$ fields and satisfying the $SU(2)_L \otimes U(1)_Y$ gauge symmetry \cite{delAguila:2008ir, Liao:2016qyd, Wudka:1999ax}. The total Lagrangian we consider is organized as follows:
\begin{eqnarray}\label{eq:Lagrangian}
\mathcal{L}=\mathcal{L}_{SM}+\sum_{d=5}^{\infty}\frac1{\Lambda^{d-4}}\sum_{\mathcal{J}} \alpha_{\mathcal{J}} \mathcal{O}_{\mathcal{J}}^{d}
\end{eqnarray}
where $d$ is the mass dimension of the operator $\mathcal{O}_{\mathcal{J}}^{d}$, $\alpha_{\mathcal{J}}$ are the effective (Wilson) couplings and the sum in $\mathcal{J}$ goes over all independent interactions at a given dimension $d$.

There are only three dimension 5 effective operators in \eqref{eq:Lagrangian}. The well known Weinberg operator $\mathcal{O}_{W}=(\bar{L}\tilde{\phi})(\phi^{\dagger}L^{c})$ \cite{Weinberg:1979sa} involving only left-handed neutrino fields, which violates lepton number and contributes to the light neutrino masses. The Anisimov-Graesser operator $\mathcal{O}_{N\phi}=(\bar{N_R}N^{c}_{R})(\phi^{\dagger} \phi)$ \cite{Anisimov:2006hv, Graesser:2007yj} which also contributes to the Majorana mass term for $N_R$ (when it is considered, in the renormalizable $d=4$ Lagrangian in \eqref{eq:LagSeesaw}) and gives Higgs-$N$-$N$ interactions. And finally the dipole operator $\mathcal{O}^{(5)}_{NB}= (\bar{N}_{R}\sigma_{\mu \nu}N^{c}_{R}) B^{\mu \nu}$ inducing magnetic moments for the heavy neutrinos, which is identically zero if we include just one sterile neutrino $N_{R}$ in the theory, and was first studied in ref. \cite{Aparici:2009fh}. Their phenomenology both regarding Higgs and heavy $N$ physics in colliders together and in comparison to the simple seesaw mixing interactions has been studied in \cite{Delgado:2022fea, Barducci:2020icf, Butterworth:2019iff, Jones-Perez:2019plk, Caputo:2017pit, Graesser:2007yj}. We refer the reader to \cite{Delgado:2022fea} for a detailed description of the interplay between the Type-I seesaw Lagrangian and the dimension 5 operators in the generation of light and heavy neutrino masses and the heavy neutrinos decays, in the case where three right-handed neutrinos are added to the SM field content. Here we will not consider the $d=5$ operators, as in the simplified scenario with only one right-handed neutrino state added we will focus on, they make no contributions to the studied processes at electron-proton colliders when discarding the heavy-active neutrino mixings $U_{\ell N}$, neither to the $N$ decay. Thus we will only consider the contributions of the $d=6$ operators, following the treatment presented in \cite{delAguila:2008ir, Liao:2016qyd}, and shown in table \ref{tab:Operators}. Our implementation of the effective Lagrangian in \texttt{FeynRules 2.3} has been discussed in \cite{Zapata:2022qwo}. Full expressions for each explicit Lagrangian term can be found in the Appendix \ref{app:explicitLagOpe}, and we briefly discuss them in the following. 

The Higgs dressed mixing operator $\mathcal{O}_{LN\phi}$ is a Yukawa interaction for the $N$ dressed with a Higgs doublet pair, and thus contributes with extra terms at tree-level to light neutrino masses $m_{\nu}$ and mixings \cite{Mitra:2022nri}. It also leads to new $N \nu$-Higgs interactions, as read from eq. \eqref{eq: app LNHiggs}. The neutral (NC) and charged (CC) bosonic currents $\mathcal{O}_{NN\phi}, \mathcal{O}_{N l \phi}$ are written in eqs. \eqref{eq: app BosonNC} and \eqref{eq: app BosonCC}. The charged current couples the $W$ boson with right-handed chiral leptons: a $(V+A)$ structure as opposed to the $(V-A)$ SM charged weak interaction. 

The explicit expressions of the $d=6$ dipole momenta operators $\mathcal{O}_{NB}, \mathcal{O}_{NW}$ are written in eqs. \eqref{eq: app ONB} and  \eqref{eq: app ONW}. They induce dipole interactions of the $N$ with the $Z, \gamma$ and $W$ bosons ($d_Z, d_{\gamma}$ and $d_W$) \cite{Fernandez-Martinez:2023phj, Barducci:2022gdv, Aparici:2009fh}. These operators are generated at one-loop level in the complete UV theory \cite{delAguila:2008ir, Arzt:1994gp}, and thus their contributions are suppressed by a loop factor $1/ (16 \pi^2)$ which is taken into account in our numerical calculations.  

We classify the four-fermion interactions in terms of possible UV mediators connecting the fermion lines in the Lagrangian terms given by each operator.\footnote{A detailed recent work on UV completions for the $\nu$SMEFT operators can be found in \cite{Beltran:2023ymm}.} 
There are three types of vector-mediated neutral currents involving two $N$ fields, one vector-mediated charged current, and as shown in eqs. \eqref{eq: app QuNL} to \eqref{eq: app LNLl} the last four-fermion operators induce Lagrangian terms that can be obtained from an UV completion where the fermion lines are mediated by neutral or charged scalars.

\subsection{Constraints on effective couplings}\label{sec:constraints}

In the last years many works on right-handed effective neutrino interactions ($\nu$SMEFT) including $d=6$ operators have derived bounds for the different effective couplings values $\alpha_{\mathcal{J}}$, or alternatively on the new physics scale $\Lambda$ given by existing experimental direct or indirect searches of BSM phenomena. Most of these constraints are applicable for $m_N$ masses below the range we consider in this work: $m_N =  100- 500$ GeV, but it is worth to make some comments for a review and comparison of the different approaches.

A very recent and illustrative review of constraints on the dim=6 $\nu$SMEFT operators coefficients can be found in \cite{Fernandez-Martinez:2023phj}. The authors neglect the seesaw mixings between the active and heavy $N$ neutrinos, in the same fashion we do here, and obtain bounds considering each effective operator acting separately at a time, thus avoiding cancellations. This approach generally leads to conservative constraints, as it does not allow for $N$ production and detection through different operators.
They show plots in the $m_N- C/\Lambda^2$ plane for each operator coupling and masses $m_N< 100$ GeV.
{\footnote{The new physics scale value used to obtain the couplings $\alpha$ can be changed to any other scale $\tilde{\Lambda}$ considering the relation $\tilde{\alpha}= \left(\frac{\tilde{\Lambda}}{\Lambda}\right)^2 \alpha$. In order to compare to the values discussed here for $\Lambda=1$ TeV, one can get an estimation at a glance considering that $\alpha/\Lambda^2 = 10^{-6}$ GeV$ ^{-2}$ corresponds to couplings of order one: $\alpha \sim 1$.}}
 
When more than one operator acting at a time is considered, bounds can be obtained from a variety of processes involving combinations of couplings (typically with different production and decay channels for the $N$). We refer the reader to ref. \cite{Mitra:2022nri} for some estimates obtained from collider searches: we must also be careful when considering their bounds values, as the authors do not neglect the mixings between heavy and active neutrinos in their calculations, which leads to a somewhat different phenomenology.

We discuss in the following the bounds that could be applicable for $N$ masses in the electroweak scale ($\simeq v=246$ GeV): $10^2 < m_N < 10^3$ GeV and give the values we take for our numerical calculations as estimates compatible with our simplified scenario. A detailed calculation of the bounds that could be considered for every effective operator from the plethora of existing and future experiments is beyond the scope of the present work.

\paragraph{Standard vector bosons operators: Higgs, neutral and charged currents.}

The operator $\mathcal{O}_{LN\phi}$ \eqref{eq: app LNHiggs} induces extra contributions at tree-level to light neutrino masses and their mixing with the $N$, and also an invisible Higgs decay channel into a light and a heavy neutrino. The bounds from  $h \to \nu N$, valid for $m_N<m_h$ translate to $\alpha^{(i)}_{LN\phi} \lesssim 0.3$ for $\Lambda= 1$ TeV, for every flavor $i$ when the operator is the only one active \cite{Fernandez-Martinez:2023phj}. Bounds on the coefficient of the neutral bosonic current operator $\mathcal{O}_{NN\phi}$ \eqref{eq: app BosonNC} can be obtained from invisible $Z$ decays and mono-photon searches at LEP, but these apply only for $m_N< m_Z$ \cite{Mitra:2022nri}. Also, the tensorial current (one-loop level generated) operators $\mathcal{O}_{NB}, \mathcal{O}_{NW}$ \eqref{eq: app ONB}  \eqref{eq: app ONW} can be constrained exploiting the known bounds on the $N$ dipole couplings to the bosons $W, Z$ and $ \gamma$ \cite{Magill:2018jla} obtained at LEP and the LHC: these give bounds above the values we consider (and obtain) here. In the case of the electron flavor, the coupling $\alpha^{(i=1)}_{NW}$ is bounded by its contribution to the unobserved neutrinoless double beta decay ($0\nu\beta\beta$-decay), as will be discussed below.

We now turn to discuss the bounds that can constrain the coefficient of the operator $\mathcal{O}_{N l \phi}$ which contributes to the charged bosonic current in eq. \eqref{eq: app BosonCC}
\begin{equation}
 \frac{\alpha_{N l \phi}}{\Lambda^2} \frac{g v^2}{2 \sqrt{2}} 
  \; (\overline N_R \gamma_{\mu} l_{R})\; W^{+ \mu}
\end{equation} 
with a similar structure to the CC of the SM, for each charged lepton flavor $l_i= e, \mu, \tau$ \cite{delAguila:2008ir}.

The most restrictive bound on the $\mathcal{O}_{N l \phi}$ operator coupling $\alpha^{(1)}_{N l \phi}$  for the electron flavor comes from the non-observation of neutrinoless double beta decay ($0\nu \beta \beta$-decay). The stringent limit on the lifetime $\tau_{{0\nu}_{\beta\beta}} \geq 1.1 \times 10^{26}$ years obtained by the KamLAND-Zen Collaboration \cite{KamLAND-Zen:2016pfg} gives us a mass-dependent bound $\alpha_{0\nu\beta\beta}(m_N) \leq 3.2\times 10^{-2}\left(\frac{m_N}{100 ~ ~\rm{GeV}} \right)^{1/2}$ for $\Lambda=1$ TeV on the coupling of every effective operator contributing to the vertex $u d N e$ (the details of the derivation can be followed form refs. \cite{Duarte:2016miz, Duarte:2014zea}, see also \cite{Beltran:2023ymm}).\footnote{The couplings of the operators contributing to this vertex are: $\alpha^{(1)}_{N l \phi}\; ,\alpha^{(1)}_{NW} $ which contribute through the interchange of a $W$ boson, and the four-fermion $\alpha^{(1, 1)}_{duNl}, ~\alpha^{(1, 1)}_{QuNL}, ~\alpha^{(1, 1)}_{LNQd} , ~\alpha^{(1, 1)}_{QNLd}$.} We will thus fix the coupling $\alpha^{(1)}_{N l \phi}(m_N) = \alpha_{0\nu\beta\beta}(m_N)$ throughout all the numerical calculations in this work.\footnote{This is in agreement with the $0\nu \beta \beta$-decay constraints on the mixing $U_{eN}$ given by the most recent literature \cite{Dekens:2023iyc, Antel:2023hkf, Bolton:2019pcu}.}

In the case of the muon and tau families, limits on this coupling can be obtained from the well known existing bounds on the seesaw mixings $U_{\ell N}$. One can consider the relation 
\begin{equation}\label{eq:miXalf}
U_{\ell N}\simeq  \frac{ v^2}{2}\frac{\alpha^{(i=\ell)}_{N l \phi}}{\Lambda^2}
\end{equation}
to derive bounds for the bosonic charged current effective coupling from the mixings with flavor $\ell \equiv l_i$.  

The most restrictive bounds for $m_N$ above the electroweak scale for the muon flavor come from the charged Lepton Flavor Violating (cLFV) processes induced by the quantum effects of the heavy neutrinos. The most stringent one can be derived from the MEG limit on the muon radiative decay branching fraction $Br(\mu^{+}\rightarrow e^{+}\gamma)\leq 4.2 \; 10^{-13}$ \cite{MEG:2016leq}. Following the treatment in ref. \cite{Blennow:2023mqx}, we can translate their bounds on the product of mixings as \footnote{We extract the values from Fig. 1 in ref. \cite{Blennow:2023mqx}.}
\begin{equation*}
|\eta_{e \mu}|= \frac{1}{2} U_{e N} U_{\mu N}= \frac{1}{2} \left( \frac{v^2}{2} \frac{\alpha^{(1)}_{N l \phi}}{\Lambda^2}\right) \left( \frac{v^2}{2} \frac{\alpha^{(2)}_{N l \phi}}{\Lambda^2}\right).
\end{equation*}
Given that we already consider $\alpha^{(1)}_{N l \phi}= \alpha_{0\nu\beta\beta}(m_N)$, this tiny value for the electron family coupling almost saturates the bound, leaving us with a possible limit on the muon flavor coupling which ranges from $\alpha^{(2)}_{N l \phi}(m_N=100 ~ \text{GeV})\leq 2.39$ to $\alpha^{(2)}_{N l \phi}(m_N=500 ~ \text{GeV})\leq 0.36$ for $\Lambda=1  ~  \text{TeV}$, which are above the considered values for the effective couplings in the current numerical analysis. 

For the tau flavor, the possible bounds that can be considered from $\tau \to e \gamma$ decays are even looser: taking the limit from the most stringent scenario in \cite{Blennow:2023mqx} for $|\eta_{e \tau}|\lesssim 10^{-5}$ and taking again $\alpha^{(1)}_{N l \phi}= \alpha_{0\nu\beta\beta}(m_N)$, we obtain a bound ranging from $\alpha^{(3)}_{N l \phi}(m_N=100 ~\text{GeV}) \lesssim 0.68$ to $\alpha^{(3)}_{N l \phi}(m_N=500 ~\text{GeV}) \lesssim 0.30$ for $\Lambda=1  ~ \text{TeV}$. 

The exercise of translating the bounds on the Type-I seesaw mixings to the effective coupling $\alpha^{(i=\ell)}_{N l \phi}$ using the relation in \eqref{eq:miXalf} can also be done with the recent results from the LHC experiments. For the masses considered in this work, the bounds on $|U_{\mu N}|^2$ from ref. \cite{CMS:2018jxx} on the same-sign dimuons signal are the strongest and give us couplings ranging roughly from $\alpha^{(2)}_{N l \phi}(m_N=100 ~ \text{GeV})\lesssim 2.1$ to $\alpha^{(2)}_{N l \phi}(m_N=1 ~ \text{TeV}) \lesssim 14.8$ for $\Lambda=1  ~ \text{TeV}$, which again are above the values considered in this work.

\paragraph{Four-fermion operators: charged and neutral currents.}

The existing constraints on the mixings $U_{\ell N}$ obtained from $N$ decay-in-flight and peak searches in meson decays can be used as well \cite{Fernandez-Martinez:2023phj} to constrain the four-fermion charged current operators  $\mathcal{O}_{duNl},\mathcal{O}_{QuNL}, \mathcal{O}_{LNQd}, \mathcal{O}_{QNLd}$ in table \ref{tab:Operators}, with a method similar to the one used in \cite{Beltran:2023nli} to rescale the existing bounds on each $U_{\ell N}$, given that the new operators may only generate a subset of the processes (allowed by the SM and the seesaw mixings) considered for each experiment. 

The four-fermion charged currents involving quarks could also induce mono-lepton processes in colliders ($pp \to N \ell$) in which the final state consists of a single observed lepton and missing energy. The bounds from LHC mono-lepton searches recast in ref. \cite{Alcaide:2019pnf} lead to flat bounds $\alpha^{(1)} < 0.1$ for the electron flavor operators. These bounds rely on the assumption that the $N$ are long-lived or decay invisibly in the detectors, which is not the case if one considers every effective interaction for the $N$ decay width (see section \ref{sec:NWidth}), so we do not take them directly into account for this study, since we consider the harder bounds coming from $0\nu \beta \beta-$decay. As explained above, the bound  $\alpha_{0\nu\beta\beta}(m_N)=3.2\times 10^{-2}\left(\frac{m_N}{100 ~ ~\rm{GeV}} \right)^{1/2}$ for $\Lambda=1$ TeV also applies to the couplings $\alpha^{(1, 1)}_{duNl}, ~\alpha^{(1, 1)}_{QuNL}, ~\alpha^{(1, 1)}_{LNQd} , ~\alpha^{(1, 1)}_{QNLd}$.

The four-leptons interactions from $\mathcal{O}_{eNN}, \mathcal{O}_{LNLl}$ are bounded considering mono-photon searches at LEP, with a single photon recoiling against invisible particles ($e^+ e^- \to N N \gamma$). For lower $m_N$, these operators can be bounded using tau and muon decays \cite{Alcaide:2019pnf, Fernandez-Martinez:2023phj}. 

The neutral current vectorial four-fermion operators involving two $N$ fields and light quarks $\mathcal{O}_{fNN}, f = u, d$ and $\mathcal{O}_{QNN}$ can induce quark scatterings in which the only visible signal is a single jet, produced mainly by a gluon emitted by any of the quarks, when the heavy $N$ escape undetected and give a missing energy signal. In ref. \cite{Alcaide:2019pnf} these bounds are obtained by recasting LHC mono-jet searches, and would give us flat bounds $\alpha< 0.4$ for $\Lambda=1$ TeV. Two comments are in order: as before, these bounds rely on the $N$ being long-lived or escaping undetected. Also, these operators are not involved in the $N$ decay or the processes considered for the LHeC phenomenology, so we do not take them into account in this study.

In our numerical setup, every $d=6$ operator in Tab. \ref{tab:Operators} for every flavor is turned on at the same time. We follow this democratic approach, since it leads to more realistic results, given that in most cases specific BSM UV models will generate not only one operator, but contribute to several operators when the correct matching for the model is calculated, due to operator mixing \cite{Chala:2020vqp, Beltran:2023ymm, Fernandez-Martinez:2023phj}.

We set the values of the effective couplings of the operators contributing to $0\nu \beta \beta-$decay for the first family to the value $\alpha_{0\nu\beta\beta}(m_N)=3.2\times 10^{-2}\left(\frac{m_N}{100 ~ ~\rm{GeV}} \right)^{1/2}$. These operators give contributions to the total decay width $\Gamma_N$, and to the $N$ production vertex in the considered processes. As we aim to obtain sensitivity estimates for heavy Majorana neutrinos $N$ in possible searches at the future LHeC, we do not impose constraints on the other effective operators. We set the other effective couplings to the same value $\alpha$ but only consider benchmark points with $\alpha \leq 0.3$ for masses in the range $m_N =  100- 500$ GeV to be conservative. The loop factor in the couplings of the $\mathcal{O}_{NB}$ and $\mathcal{O}_{NW}$ operators is considered in the numerical calculations: we fix $\alpha_{NB}=\alpha_{NW}=\alpha$, but include the loop factor in the interaction vertices, see \eqref{eq: app ONB} and \eqref{eq: app ONW}.

\section{Collider analysis}\label{sec:Collider analysis}

In this study we analyze the sensitivity projections for the HNL $N$ with effective interactions at the LHeC, considering its decay to muons and jets. The process with final muons ($p ~ e^{-} \rightarrow \mu^{-} +  3 \mathrm{j}$) conserves lepton number but violates flavor (LFV), while the process with final anti-muons ($p ~ e^{-} \rightarrow \mu^{+} +  3 \mathrm{j}$) violates also lepton number by two units (LNV). We will analyze both channels separately. 

The LHeC is proposed to be an $e^{-}p$ collider built at the LHC tunnel, using an electron beam with $60$ GeV energy in the the existing $7$ TeV proton beam, giving a center-of mass energy close to $1.3$ TeV. It is expected to achieve an integrated luminosity $\mathcal{L}=100 ~\rm fb^{-1}$ per operation year, and $1~ \rm ab^{-1}$ in total \cite{LHeC:2020van, AbelleiraFernandez:2012cc, Bruening:2013bga}. Here we consider an integrated luminosity of $100 ~ \rm fb^{-1}$ to calculate our physical numbers of events.

The numerical tools used for the calculations are the following. We use our implementation of the $d=6$ Lagrangian introduced in \cite{Zapata:2022qwo} in the \texttt{FeynRules 2.3} software \cite{Alloul:2013bka} and generate UFO files \cite{Degrande:2011ua} as output. The cross sections for the processes $p ~ e^{-} \rightarrow \mu^{\pm} + 3 \mathrm{j}$ are calculated  using  \texttt{MadGraph5\_aMC@NLO 3.4.1} \cite{Alwall:2014hca, Alwall:2011uj} generating LHE events at parton level, which are read by the embedded version of \texttt{PYTHIA 6} \cite{Sjostrand:2006za}, which is suited to handle proton-electron collisions, unlike the latest versions, as done in \cite{Antusch:2019eiz, Gu:2022muc}. Then the events are interphased to \texttt{Delphes 3.5.0} \cite{deFavereau:2013fsa} with the default \texttt{delphes-card-LHeC} card for a fast detector simulation. Jets are defined in \texttt{Delphes} using \texttt{FastJet} \cite{Cacciari:2011ma}, with the anti$-k_T$ algorithm. 

 The analysis of the generated events at the reconstructed level is made with the expert mode in \texttt{MadAnalysis5 1.8.58} \cite{Conte:2012fm}, and the multivariate statistical analysis with the \texttt{Root TMVA} package \cite{Hocker:2007ht}. 

For concreteness, we simplify the parameter space setting all the $d=6$ effective couplings $\alpha_{\mathcal{J}}$ in eq. \eqref{eq:Lagrangian} and table \ref{tab:Operators} to the same numerical value $\alpha$, and explore a grid of signal benchmark scenarios (see table \ref{tab:benchmarksAMU}). \footnote{Except for the operators with charged leptons of the first family constrained by the $0\nu\beta\beta$-decay bound, as explained in \ref{sec:constraints}. We also include the loop factor in \eqref{eq: app ONB} and \eqref{eq: app ONW}.} The total decay widths of the heavy Majorana neutrino $N$ are given to {\texttt{MadGraph}} from our analytical calculations updated from \cite{Duarte:2016miz} for each $m_N$ and $\alpha$ benchmark point (see section \ref{sec:NWidth}).\footnote{It is thus not easy to compare our results to the ones that would be obtained if we considered only one operator acting at a time for the $N$ production and decay, as the total width would change appreciably, besides the expected reduction of the $N$ production cross section. We leave the exploration of this scenario for future work.} We also fix the new physics scale $\Lambda=1$ TeV and show our results for different $N$ mass values $m_N = 100, 125, 150, 200, 300, 500$ GeV. In our numerical simulations we have considered hard scattering energies with $\sqrt{\hat{s}}< \Lambda$ in order to ensure the validity of the effective Lagrangian approach, by imposing the total transverse energy (TET) deposit to be below the new physics scale $\Lambda$ in every event.

\subsection{Signals characterization}\label{sec:signals}

\begin{figure}[tpb]
 \centering
  {\includegraphics[totalheight=5.8cm]{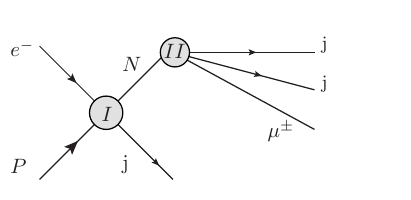}}
  \caption{ {\label{fig:eP_Nj_mu3j}}{Majorana neutrino $N$ production and semi-leptonic decay at an eP collider.}}
 \end{figure}

The LNV and LFV processes we want to study share the same production mechanism at electron-proton colliders, namely  $p ~ e^{-} \rightarrow N \mathrm{j}$ with the subsequent decays $N \to \mu^{\pm} \mathrm{j} \mathrm{j}$, pictured in figure \ref{fig:eP_Nj_mu3j}. 
The operators that explicitly contribute to the production mode considered $q e^{-} \rightarrow N \mathrm{j}$ are the same as the ones that contribute to the $N\to \mu^{\pm} \mathrm{j} \mathrm{j}$ decay (vertices $I$ and $II$ in figure \ref{fig:eP_Nj_mu3j} respectively). These are the vectorial $\mathcal{O}_{Nl\phi}$, which provides an $N l W$ vertex which combined with the SM $W q q'$ contributes to the production and decay of the heavy $N$, and the four-fermion  operators with a single $N$ and two quarks: the vectorial $\mathcal{O}_{duNl}$ and the scalar $\mathcal{O}_{QuNl}$, $\mathcal{O}_{LNQd}$ and $\mathcal{O}_{QNLd}$. The explicit analytical expressions for the cross-section $\sigma(p ~ e^{-} \to N \mathrm{j} \to \mu^{-} \mathrm{j} \mathrm{j})$ can be found in \cite{Duarte:2018xst}.

\begin{figure}[tpb]
\centering
{\includegraphics[width=0.48\textwidth]{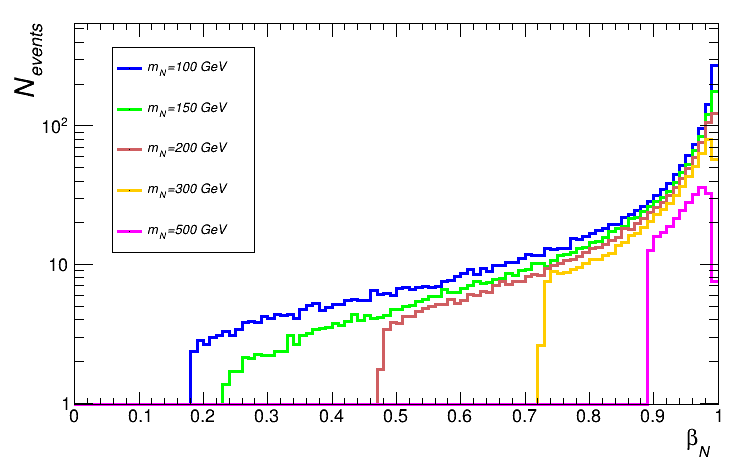}}~
{\includegraphics[width=0.48\textwidth]{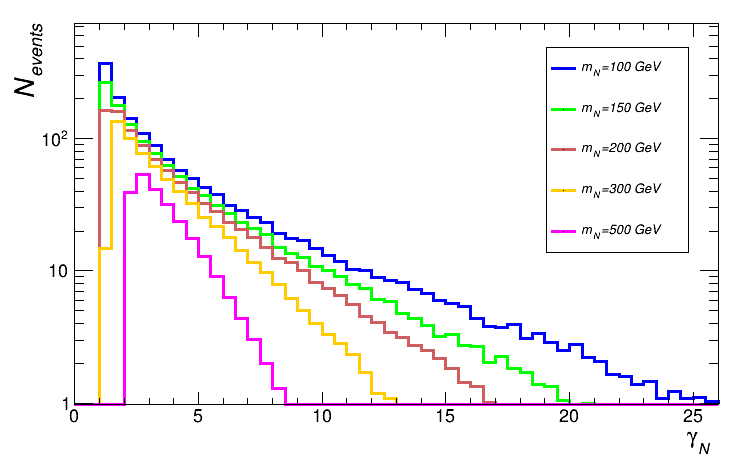}}

{\includegraphics[width=0.48\textwidth]{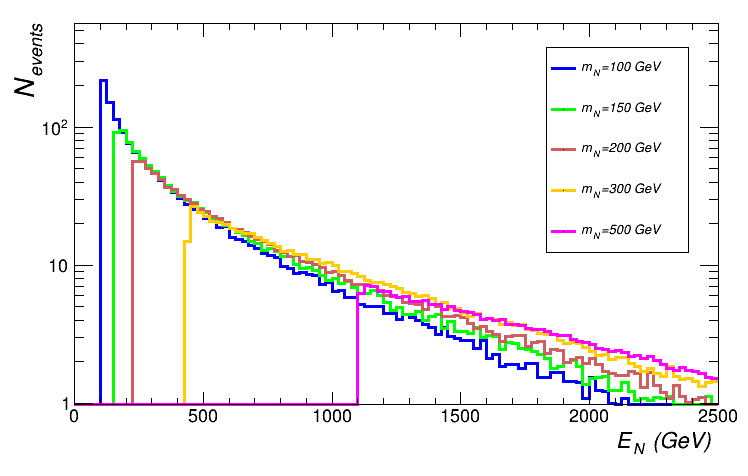}}~
{\includegraphics[width=0.48\textwidth]{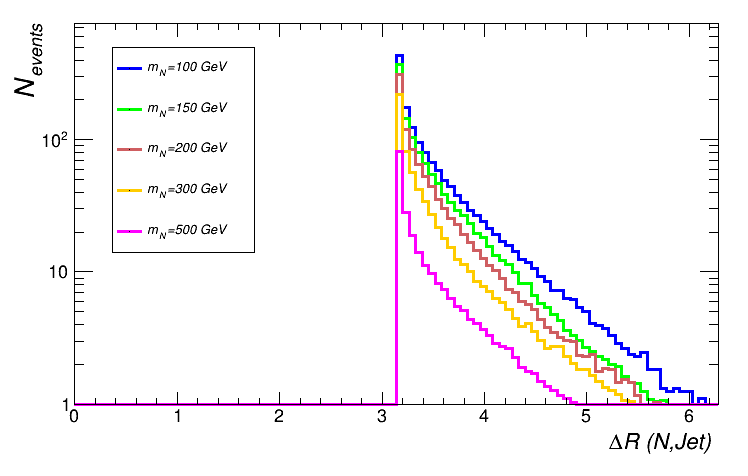}}

{\includegraphics[width=0.48\textwidth]{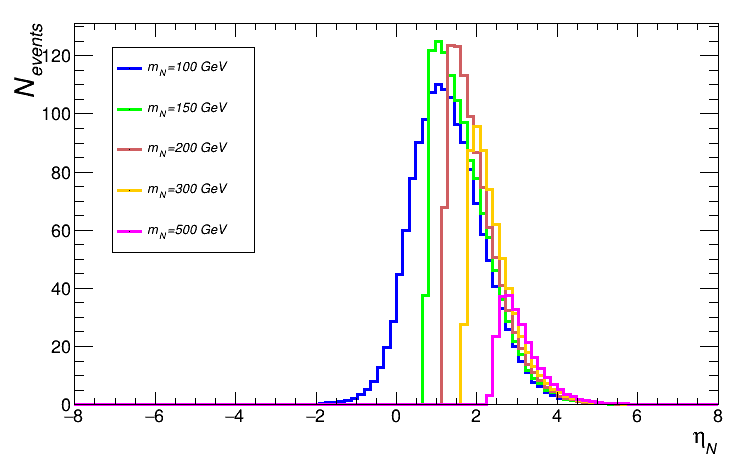}}~
{\includegraphics[width=0.48\textwidth]{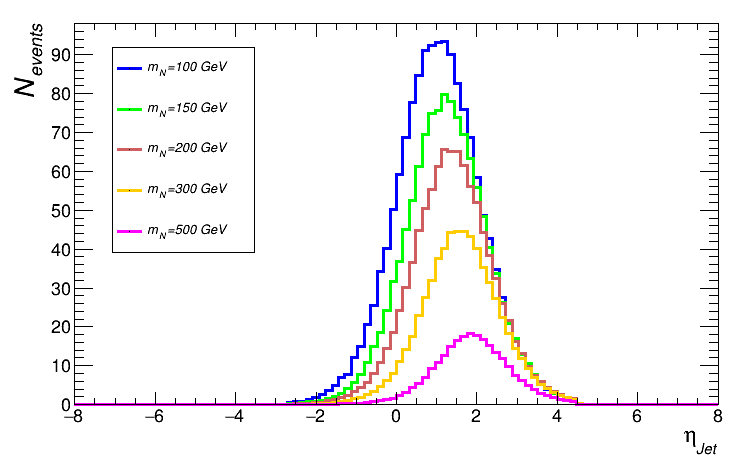}}
\caption{{\label{fig:ep-Nj_CHAR}}{ Kinematical characterization of $p ~ e^- \to N \mathrm{j}$ parton-level events at the LHeC.}}
 \end{figure}

To characterize the kinematics of the production of the heavy $N$ in the LHeC we generate $10^5$ $p ~ e^- \to N \mathrm{j}$ parton-level events in \texttt{MadGraph5\_aMC@NLO}, for the chosen $m_N$ benchmark points, with effective couplings fixed to $\alpha=0.2$. In figure \ref{fig:ep-Nj_CHAR} we show the Lab-frame physical events distributions of the produced $N$ boost velocity $\beta_N$, boost factor $\gamma_N$, energies $E_N$, and the distances between the $N$ and the beam-jet $\Delta R (N,\mathrm{j})$, as well as the pseudo-rapidities $\eta$ of the heavy $N$ and the jet (notice the positive $\hat{z}$ axis points in the incident proton direction, corresponding to $\theta=0$). 

%%%
%%%

The large asymmetry in the beam energies at electron-proton colliders tends to boost the final particles into the proton beam direction, identified with large positive pseudo-rapidity $\eta$ values, affecting the angular correlations in the Lab-frame. The center of mass energy at the LHeC collider is projected to be $\sqrt{S}= (4 E_p E_e)^{1/2} \sim 1.3$ TeV. This energy allows to produce boosted heavy neutrinos in the Lab-frame, with threshold energies corresponding to the production of the $N$ at rest in the CM-frame. The boost velocity of the threshold CM-frame in the Lab can be obtained as a function of the $N$ mass $m_N$ and the electron beam energy as
\begin{equation*}
\beta_{Th}= \frac{m_N^2-4 E_e^2}{m_N^2 +4 E_e^2},
\end{equation*}
corresponding to a minimal value of the Bjorken variable $x_{{Th}}= m_N^2/ 4 E_p E_e$. For higher $x$ values, with $m_N^2/ 4 E_p E_e < x < 1$, the $N$ and the jet are produced back-to-back in the CM-frame.

%% %  

The Lab-frame decay length of the $N$ is given by $d_N^{~Lab}= \beta \gamma c / \Gamma_N$, with $\gamma = E_N/m_N$. Given the values of the $N$ decay widths (see figure \ref{fig:Ndec}), we find all our benchmark signal scenarios produce prompt $N$ decays, with typical mean decay lengths ranging from $d \sim 10^{-14}$ m for $m_N=500$ GeV to $d \sim 10^{-11}$ m for $m_N=100$ GeV. 

The $m_N=100$ GeV sample has some events with negative $\eta_N$ pseudo-rapidity values, as the boost velocity of the threshold CM-frame is negative. Thus in some events the heavy $N$ is going backwards in the Lab, and also, in some events the beam jet is turned backwards. 

As the $N$ has opposite azimuthal angle to the beam jet, and can be very boosted in the Lab-frame, we expect to probe the LHeC ability to disentangle this signal with a strategy focusing on the separation of the $N$ decay products from the beam jet. When simulating the full production and decay processes, together with the detector simulation and considering possible backgrounds, we find a simple cut-and-count strategy does not lead to statistically significant signal to background separation, suggesting a deeper analysis. This can be understood by considering that while for higher $m_N$ values the kinematical distributions of the signal tend to show a better separation from the backgrounds, their cross-sections diminish, making difficult to separate both contributions (see figure \ref{fig:crossect}).

\subsection{N decay}\label{sec:NWidth}

\begin{figure}[t]
\centering
{\includegraphics[width=0.48\textwidth]{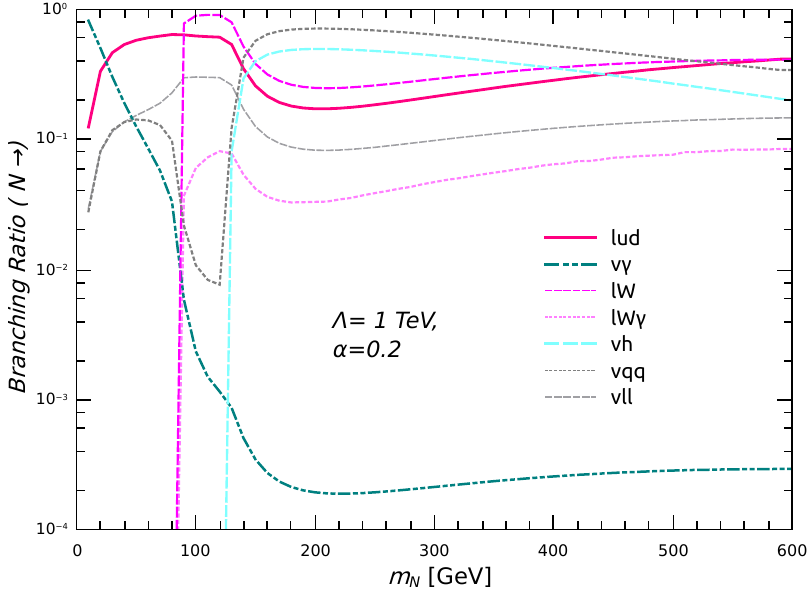}}~
{\includegraphics[width=0.48\textwidth]{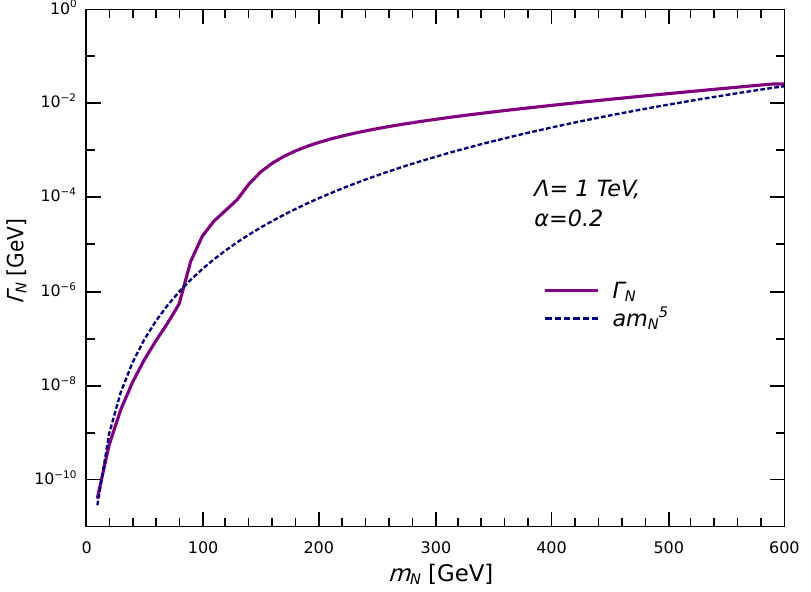}}
 \caption{ \label{fig:Ndec}{{Branching ratios of relevant channels (left) and total $N$ decay width (right) considering all the effective operators in table \ref{tab:Operators}, $\Lambda=1$ TeV and couplings $\alpha=0.2$ (see text).}} }
\end{figure}

We update here the calculation of the $N$ total decay width and branching ratios from our works in \cite{Duarte:2016miz,Duarte:2015iba}. In the $m_N$ mass range we are considering, the relevant decay channels of the heavy Majorana neutrino are those shown in figure \ref{fig:Ndec} (left), plotted here considering the contribution of every effective operator in table \ref{tab:Operators} contributing to the $N$ decay, with all couplings fixed to the same value $\alpha=0.2$ except for the ones contributing to the $0\nu\beta\beta$-decay as explained in section \ref{sec:constraints}. 

The most relevant channels for $m_N< m_W$ are those induced by four-fermion operators, which include the decays to charged (anti-)leptons and up and down type quarks  $N \to lud $, as well as the decays to light neutrinos plus quark or lepton pairs.\footnote{This channel also receives a contributon from the CC bosonic operator $\mathcal{O}_{Nl\phi}$ in \eqref{eq: app BosonCC}.} For even lower mass $m_N<30$ GeV, the decay $N \to \nu \gamma$ dominates the width. While the channels to $W$ bosons start to dominate at higher mass values, the decays to $Z$ bosons are always sub-dominant (below $10^{-5}$ for the $\alpha< 0.3$ $\Lambda=1$ TeV considered), and we do not show them here. The total $N$ decay width calculated with the full expressions updated from \cite{Duarte:2016miz,Duarte:2015iba} is compared in figure \ref{fig:Ndec} (right) with an $m_N^5$ dependence.

In our numerical simulations throughout this work we give the value of the total $N$ decay-width calculated considering all the effective operators couplings equal to the same value $\alpha$ for each mass $\Gamma_{N}(m_N, \alpha)$ as input for the Monte Carlo (MC) events generation in {\texttt{MadGraph}} for every signal benchmark point in table \ref{tab:benchmarksAMU}.\footnote{Except for the couplings bounded by $0\nu\beta\beta$-decay, see section \ref{sec:constraints}. Also the loop factors are included in the one-loop generated operators, see \eqref{eq: app ONB} and \eqref{eq: app ONW}.}

\subsection{Backgrounds}\label{sec:Backgrounds}

Although both studied signals are forbidden by flavor and lepton number conservation in the SM, and thus the signals have no SM irreducible backgrounds, one has to consider SM backgrounds which involve possible lepton charge misidentification, and final states with extra unobserved light neutrinos. We take into account the backgrounds in Table \ref{tab:Backgrounds}, for the analysis of both signal channels, following the discussions in refs. \cite{Antusch:2019eiz} and \cite{Gu:2022muc}.

Backgrounds ($B 1$), ($B 2$), and ($B 5$) in Tab \ref{tab:Backgrounds} arise from di-vector boson production together with a jet and and electron, when the vector bosons ($V=W$ or $V=Z$) decay into a pair of jets and a di-muon pair or (anti-)muon and a light neutrino. If the electron is soft, these processes can be confused with signals with extra radiated soft electrons, which cannot be rejected without decreasing the signal efficiency. 

The other important backgrounds come from di-vector boson production with a jet and a light neutrino, as in backgrounds labeled ($B 3$), ($B 4$) and ($B 6$) in table \ref{tab:Backgrounds}. Here, when one of the vectors is a $W$ decaying leptonically, the final state only differs from the signals in additional light neutrinos which escape undetected.

\renewcommand{\arraystretch}{0.8}
{\footnotesize{
\begin{table}[t]
 \centering
 \begin{tabular}{|l c c l c |}
 \firsthline 
Label   &   & Process &  & $\sigma_{(LHeC)}[Pb]$  \\
\hline
$B 1$ &   &  $p ~ e^{-}  \rightarrow  \mathrm{j} e^{-} (V V) \rightarrow \mathrm{j} e^{-} (\mathrm{j} \mathrm{j} \mu^{+} \mu^{-}) $  &  &  $1,054\times 10^{-4}$  \\
$B 2$ &  & $p ~ e^{-}  \rightarrow  \mathrm{j} e^{-} (V V) \rightarrow \mathrm{j} e^{-} (\mathrm{j} \mathrm{j} \mu^{-} \overline{\nu_{\mu}}) $  &   &  $1,801\times 10^{-3}$\\
$B 3$ &   & $p ~ e^{-}  \rightarrow  \mathrm{j} \nu_{e} (V V) \rightarrow \mathrm{j} \nu_{e} (\mathrm{j} \mathrm{j} \mu^{+} \mu^{-}) $  &  & $7,155\times 10^{-5}$ \\
$B 4$ &  & $p ~ e^{-}  \rightarrow  \mathrm{j} \nu_{e} (V V) \rightarrow \mathrm{j} \nu_{e} (\mathrm{j} \mathrm{j} \mu^{-} \overline{\nu_{\mu}}) $   &  & $5,716\times 10^{-4}$ \\
$B 5$ &  & $p ~ e^{-}  \rightarrow  \mathrm{j} e^{-} (V V) \rightarrow \mathrm{j} e^{-} (\mathrm{j} \mathrm{j} \mu^{+} {\nu_{\mu}}) $ &  & $1,879\times 10^{-3}$ \\
$B 6$ &  & $p ~ e^{-}  \rightarrow  \mathrm{j} \nu_{e} (V V) \rightarrow \mathrm{j} \nu_{e} (\mathrm{j} \mathrm{j} \mu^{+} {\nu_{\mu}}) $  &  & $2,776\times 10^{-4}$ \\
\lasthline 
 \end{tabular}
\caption{{Background processes considered for $p ~ e^{-} \rightarrow \mu^{\pm} + 3 \mathrm{j}$. }}\label{tab:Backgrounds}
\end{table}
}}

In fact, these last kind of processes could also be faked by the $N$ effective interactions, producing heavy neutrinos $N$ instead of light neutrinos, as in $p ~ e^{-}  \to  \mathrm{j} N (V V) \to \mathrm{j} N (\mathrm{j} \mathrm{j} \mu^{+} {\nu_{\mu}})$ that could, in principle, escape undetected or  also be virtual, contributing to the background amplitudes in diagrams as an internal line. We have checked that both contributions are negligible. In the first case, for the heavy $m_N$ mass values we are considering the $N$ must decay promptly in the detector volume and should be noticed, changing the final state. Second, the diagrams with virtual $N$ include two effective vertices (both with scalar and vector operators) which are suppressed by the neutrinoless double beta decay bound in the case of initial quarks of the first family. The second family initial quarks contribution is also suppressed by their PDF as sea-quarks inside the proton. Both effects force the effective contributions to the backgrounds to be negligible. 

One could in principle consider also backgrounds arising from single $W$ production with radiated jets, but we have checked they can be reduced, given that the events with radiated jets can be clearly distinguished from the signals and the backgrounds considered, as the radiated jets have very low $p_T$. The  background arising from single $Z$ production decaying to taus, and then muons, which can give a $\mathrm{j} \mathrm{j} \mu^{\pm}$ plus missing energy final state, has a much smaller cross section, and can be eliminated because these soft muons can be distinguished from the signal. Three vector boson production is not considered, due to its much smaller cross section.

\subsection{Pre-selection cuts and generated datasets}

We generate $10^{5}$ MC events for each LFV and LNV signal and the backgrounds in table \ref{tab:Backgrounds} in {\texttt{MadGraph}}. We adopt the following basic acceptance cuts on the generated final leptons $\ell$ and jets $\mathrm{j}$: $p_{T}(\ell)>2$ GeV, $p_{T}(\mathrm{j})> 5$ GeV, $|\eta_{(\ell, \mathrm{j})}| \leq 4.5$, and no cuts on the possible final photons. We keep the default isolation criteria between any jets and leptons ($\Delta R_{\mathrm{j} \mathrm{j}, \ell \ell, \ell \mathrm{j}} > 0.4$).

\begin{figure}[tpb]
\centering
\includegraphics[width=0.48\textwidth]{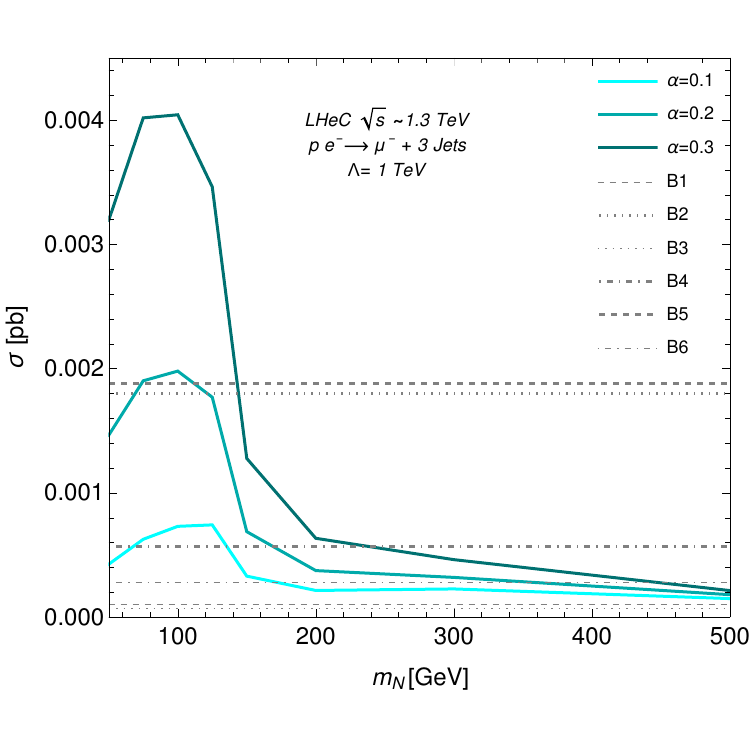}~~
\includegraphics[width=0.48\textwidth]{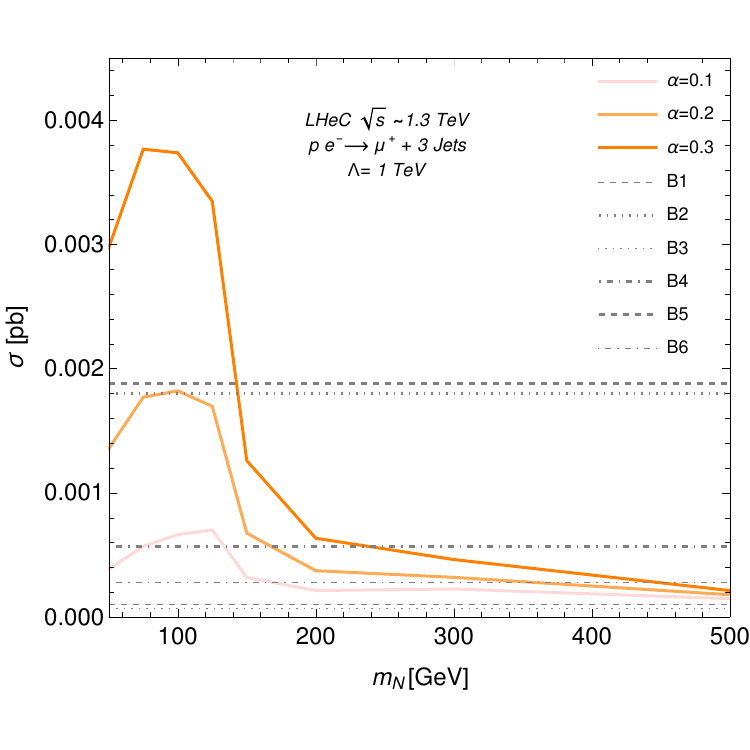}
 \caption{\label{fig:crossect}{{Parton-level cross sections for the LFV (left) and LNV (right) processes, and the six backgrounds considered (see text and table \ref{tab:Backgrounds}).}}}
 \end{figure}

In figure \ref{fig:crossect} we show the parton-level cross section values we obtain for both the LFV and LNV processes, generating signal events datasets with different values of the masses $m_N$ and effective couplings $\alpha$, together with the different backgrounds values. For $m_N$ near $120$ GeV, we find an enhancement of the signals cross section, partly due to the contribution of the $\mathcal{O}^{(i)}_{Nl\phi}$ operator to the $N$ decay when the $W$ is on-shell, which opens when $m_N> m_W$ and becomes a subleading decay channel when its decay to a Higgs boson and a light neutrino $N \to h \nu$ is open for $m_N> m_h$, as can be seen in figure \ref{fig:Ndec}.

After parton shower and hadronization in {\texttt{Pythia}} and fast detector simulation with {\texttt{Delphes}} we require (for signal reconstruction) a cut on the total transverse energy variable TET$<900$ GeV to insure the validity of the effective Lagrangian approach (with $\Lambda=1$ TeV). Also, we ask for only one final muon (or anti-muon), depending on the signal, and at least 3 jets, without cuts on their transverse momenta $p_T(\mathrm{j})$. The MC background sample after this pre-selection cuts has $3.7\times 10^5$ events for the LFV and $3.6\times 10^5$ for the LNV channels. The pre-selection cuts preserve more than 90\% of the generated events in every signal benchmark sample, for both channels. 

These are the pure-signals and pure-backgrounds datasets used to train the multivariate analysis (similarly generated but independent samples are mixed and used as input for the classification application).

\subsection{Multivariate analysis}\label{sec:TMVA_analysis}

We use many high level observables obtained from the information of the final reconstructed objects as input for the {\texttt{TMVA}} analysis to classify signal and background events, using a Boosted Decision Tree (BDT) algorithm.\footnote{We used 850 trees and the AdaBoost algorithm.} The BDT was trained with samples of $3\times 10^4$ MC signal and background events for each benchmark point, randomly chosen from the samples passing the pre-selection cuts described above. The rest of the events in each sample are used for testing. Then the method is applied to classify independently generated samples containing a mixture of $10^5$ MC events for every signal benchmark and each background process in table \ref{tab:Backgrounds}. 

\begin{figure}[tbp]
\centering
{\includegraphics[width=0.9\textwidth]{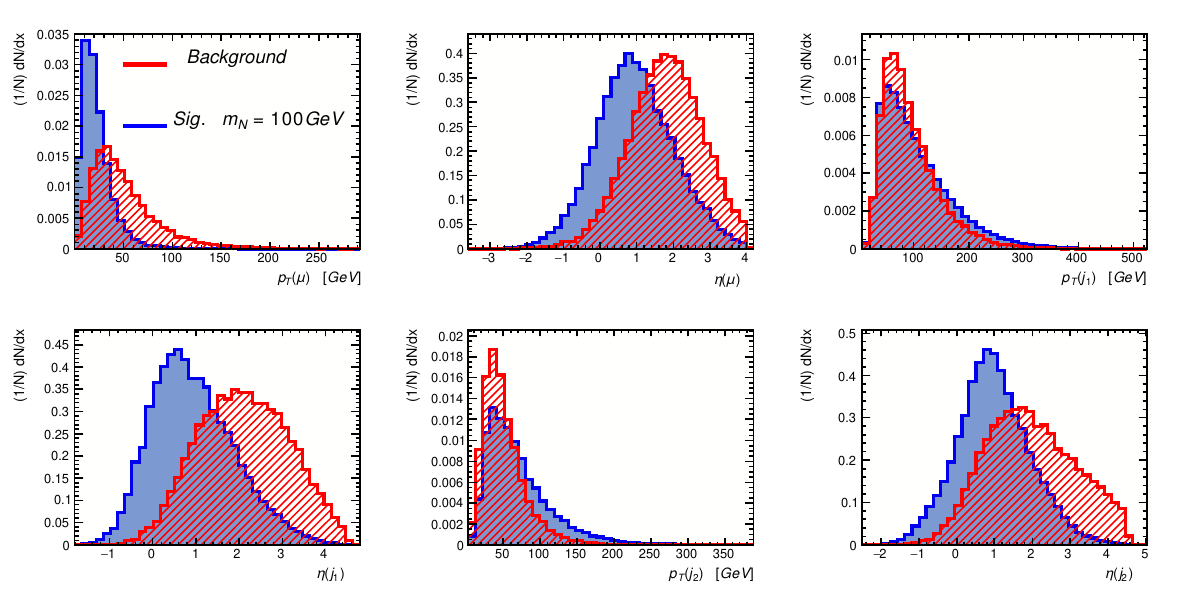}}

{\includegraphics[width=0.9\textwidth]{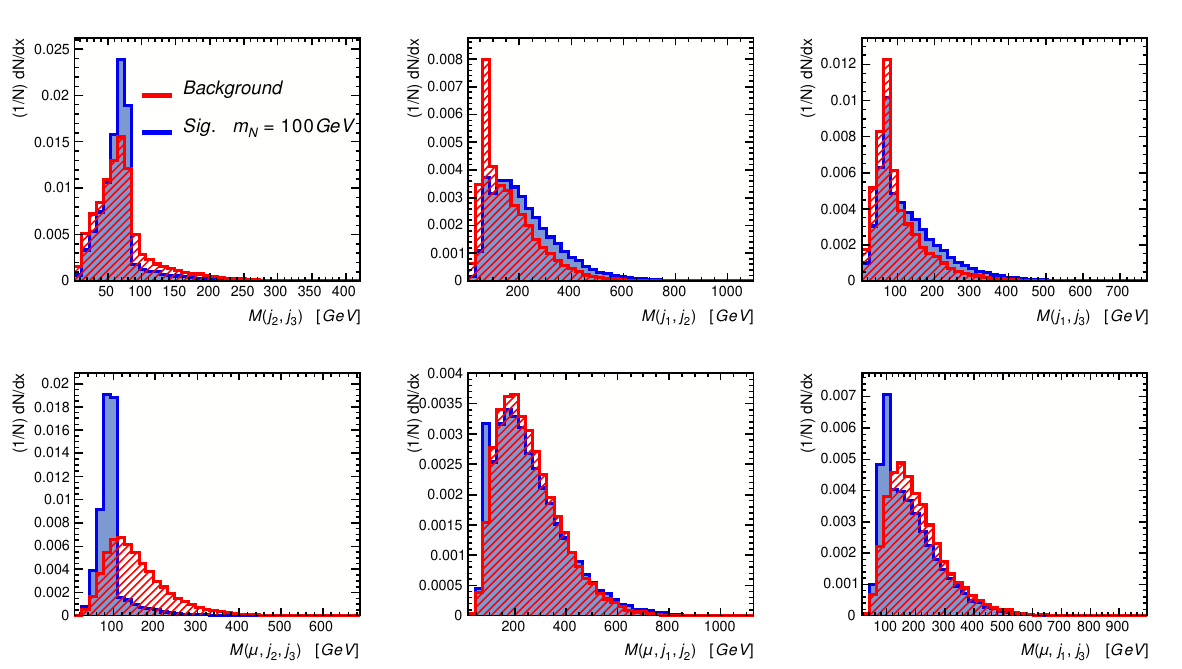}}
\caption{\label{fig:TMVAinput_100_02} {TMVA input variables distributions for the signal point $m_N= 100$ GeV and $\alpha=0.2$, and backgrounds.} }
 \end{figure}

\begin{figure}[tbp]
\centering
{\includegraphics[width=0.83\textwidth]{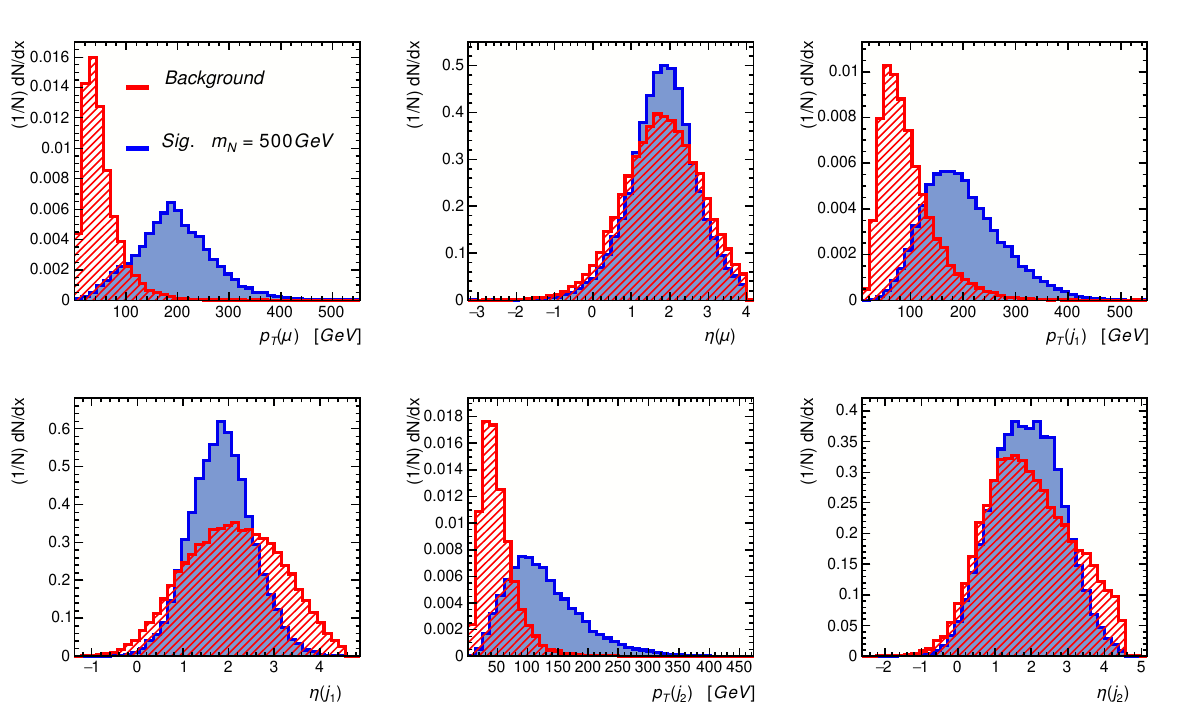}}

{\includegraphics[width=0.83\textwidth]{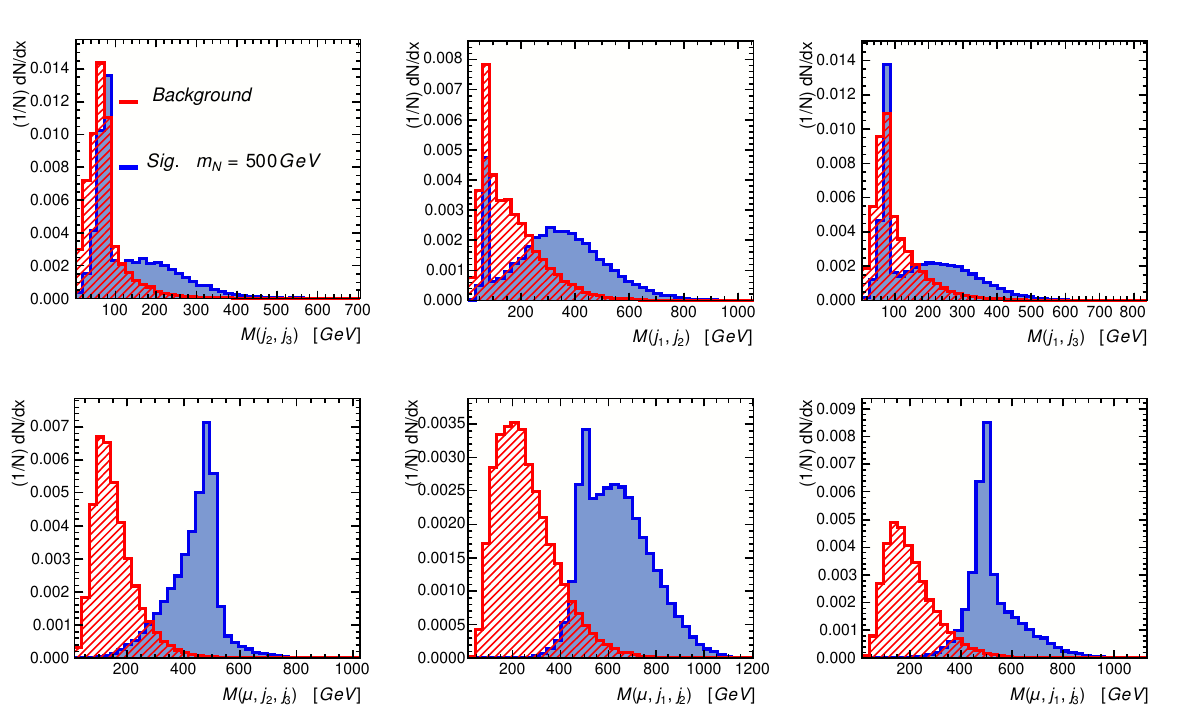}}
\caption{\label{fig:TMVAinput_500_02} {TMVA input variables distributions for the signal point $m_N= 500$ GeV and $\alpha=0.2$, and backgrounds.} }
\end{figure}

The discriminating power of the BDT relies on the fact that the signal and the background are characterized by different features that can be entangled. The most relevant distributions are shown in figure \ref{fig:TMVAinput_100_02} for $m_N=100$ GeV and figure \ref{fig:TMVAinput_500_02} for $m_N=500$ GeV, for the LNV channel. These are the transverse momenta and pseudo-rapidities of the final muon (or anti-muon), jets and the missing energy, and also the invariant mass of each jet, of jet pairs, and the invariant masses of pairs of jets and the muons $M(\mathrm{j}_{a}, \mathrm{j}_{b}, \mu^{\pm})$.

We expect the jet beam to be the one with higher $p_T$, identified as $\mathrm{j}_1$ in our analysis, and thus, the invariant mass $M(\mathrm{j}_{2}, \mathrm{j}_{3}, \mu^{\pm})$ is expected to reconstruct the value of $m_N$ for each signal benchmark point. This is the case for the lower $m_N$ values, but as can be seen comparing figures \ref{fig:TMVAinput_100_02} and \ref{fig:TMVAinput_500_02}, this criterion fails when the $N$ is heavy (and energetic) enough to produce the hardest jet on its decay. As expected, the invariant masses of jet pairs in the background events peak mostly around the vector boson masses, but this also happens in the signal events when the bosonic charged current operator $\mathcal{O}_{Nl\phi}$ in \eqref{eq: app BosonCC} drives the $N$ decay.

It can be seen from the plots in figures \ref{fig:TMVAinput_100_02} and \ref{fig:TMVAinput_500_02} that the distributions do not favor a cut-and-count strategy to discriminate the signals. Although for the higher mass benchmark sample ($m_N=500$ GeV) the invariant mass distributions in the last row of figure \ref{fig:TMVAinput_500_02} seem to allow for a significant separation, it must be kept in mind the short number of events obtained for this signal benchmark, given the low cross section value (see also figure \ref{fig:crossect}). 

\begin{figure}[tbp]
\center{
{\includegraphics[width=0.48\textwidth]{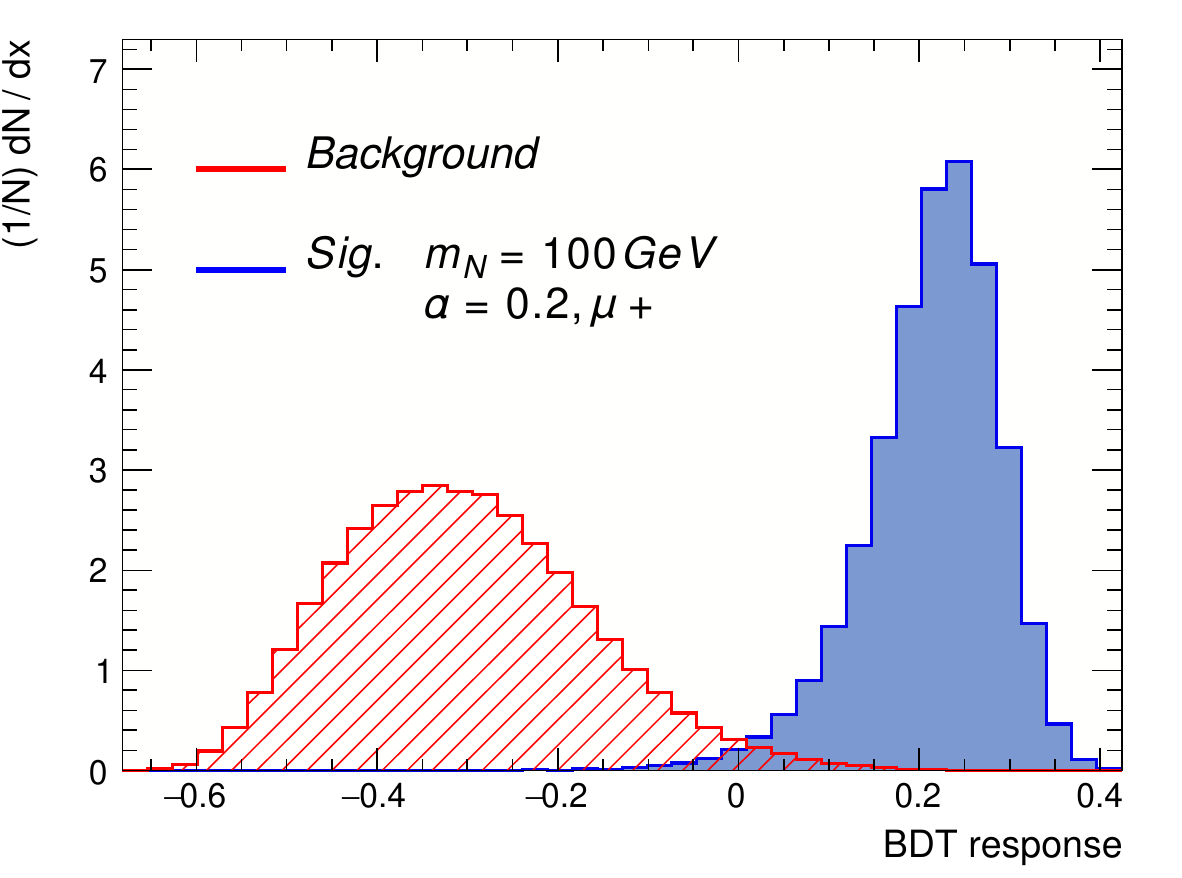}}~
{\includegraphics[width=0.48\textwidth]{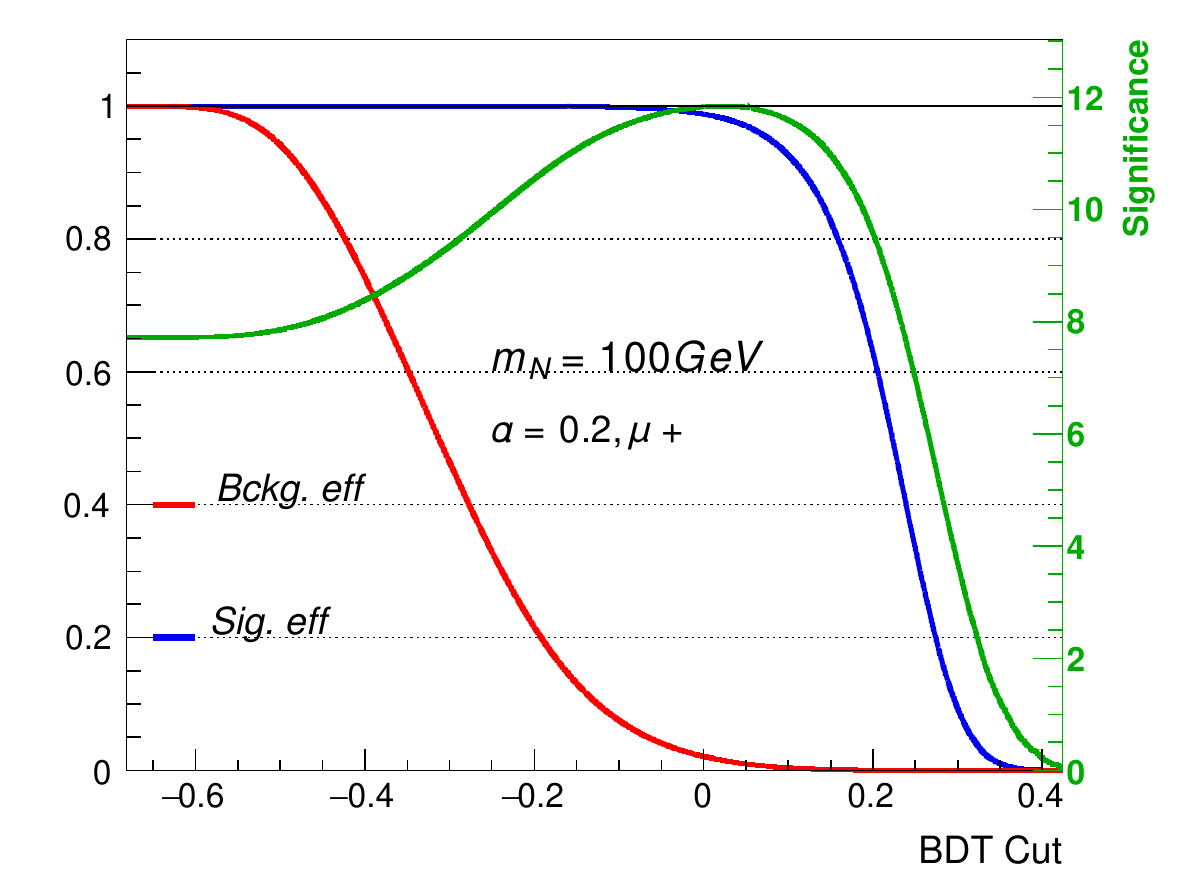}}

{\includegraphics[width=0.48\textwidth]{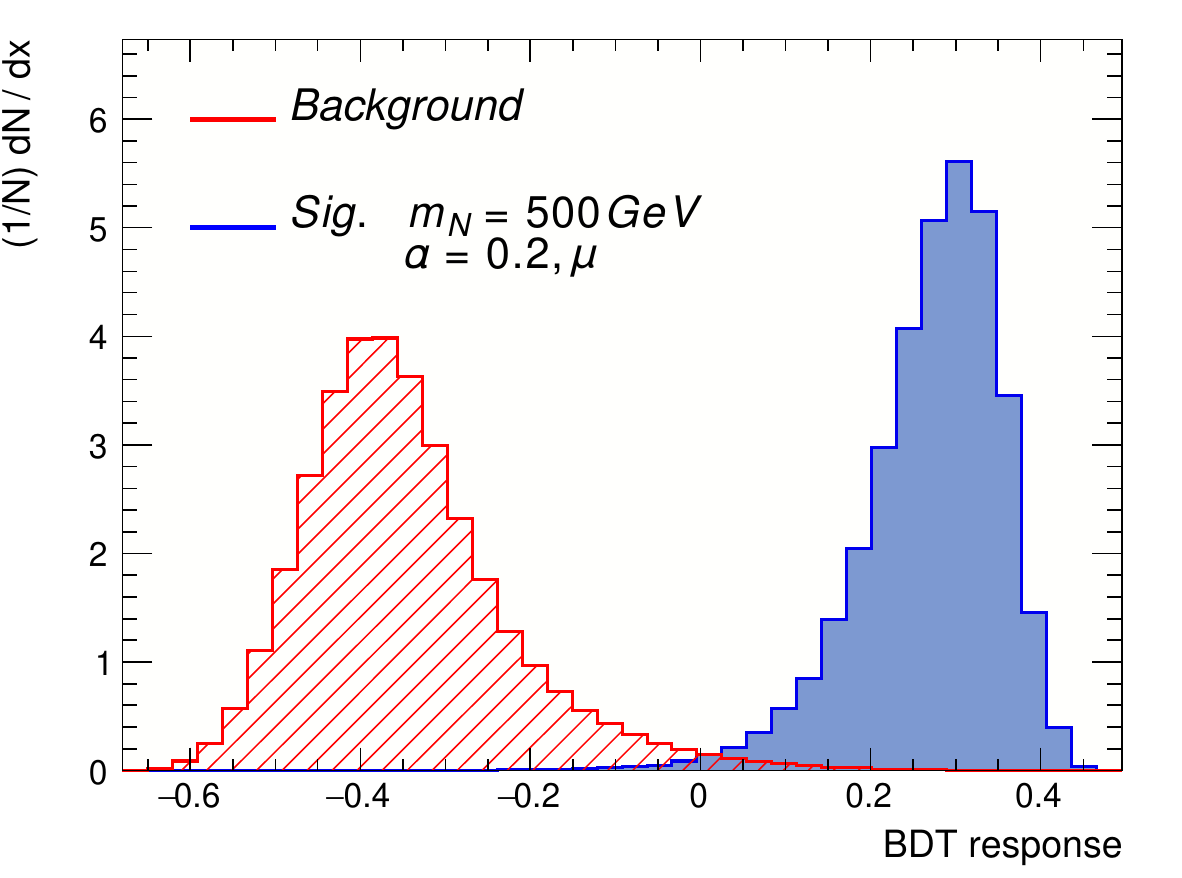}}~
{\includegraphics[width=0.48\textwidth]{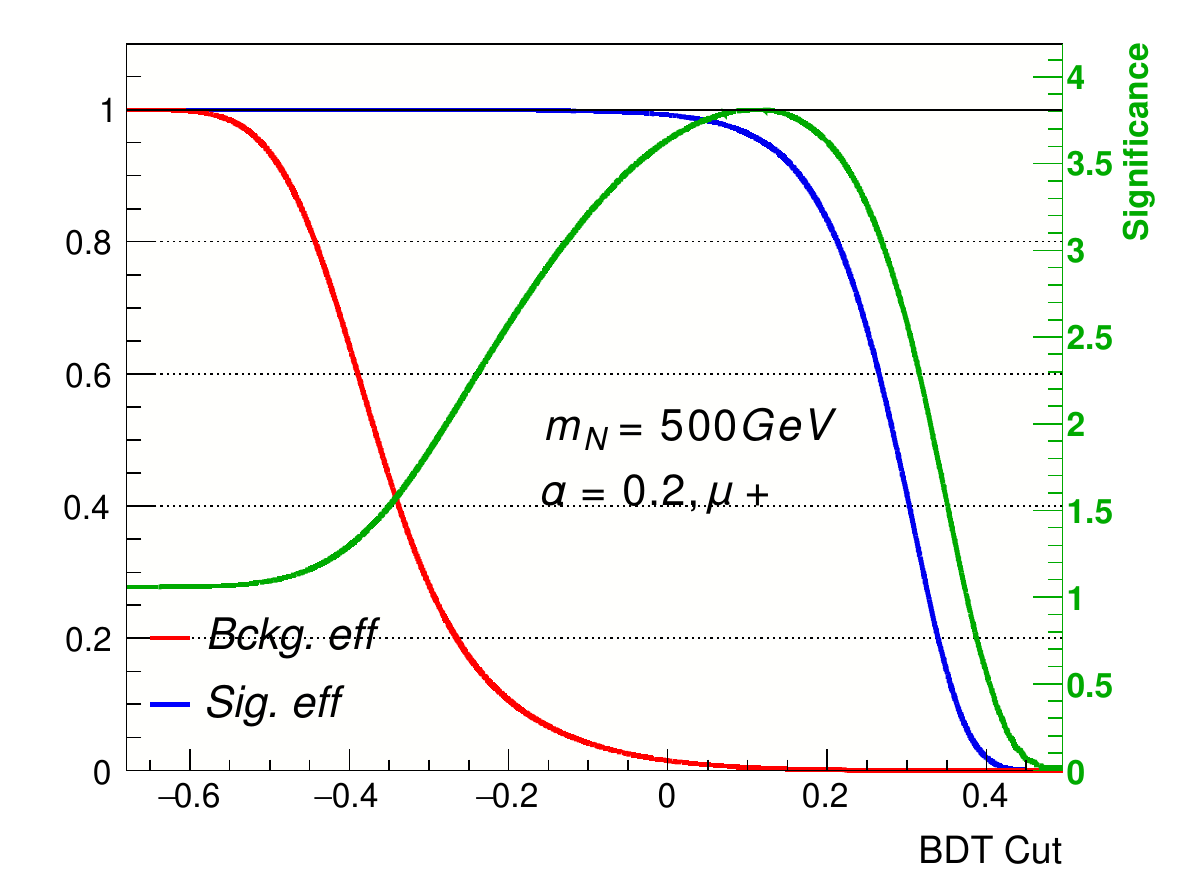}}}
\caption{\label{fig:BDTdist}{BDT normalized distributions and Cut efficiencies for signal and background samples for the LNV process $p e^{-} \to \mu^{+} +3 \mathrm{j}$ at LHeC. Up: $m_N=100$ GeV, $\alpha= 0.2$. For $N_s=146$ and $N_b=212$, BDT cut $\geq 0.0247$ gives significance ${N_s}/{\sqrt{N_s+N_b}}=11.8419$. Down: $m_N=500$ GeV, $\alpha= 0.2$. For $N_{s}=16$ and $N_b =212$, BDT cut $\geq 0.1186$ gives significance ${N_s}/{\sqrt{N_s +N_b}}=3.8117$.} }
 \end{figure}

In order to obtain the optimized values for the BDT cut for each benchmark sample, we input the expected number of physical background and signal events after applying the pre-selection cuts. As an example, the BDT variable normalized distributions for the LNV ($pe^{-}\to \mu^{+} + 3 \mathrm{j}$) signal datasets with $m_N=100$ GeV, $m_N=500$ GeV and $\alpha=0.2$ and the combined backgrounds are shown in figure \ref{fig:BDTdist}, together with the optimal BDT cut efficiency curves. The plots in figures \ref{fig:TMVAinput_100_02}, \ref{fig:TMVAinput_500_02} and \ref{fig:BDTdist} show the results for the LNV channel $p e^{-} \to \mu^{+} +3 \mathrm{j}$. The efficiencies and  distributions for the LFV case are similar and we do not show them here. The final number of physical events classified as signal and background for both the LFV and LNV processes are given in table \ref{tab:benchmarksAMU}.

%%%

We use the multivariate analysis to classify events as signal or background for the benchmark signal generated datasets. The numbers of physical events at the LHeC after applying both the pre-selection and BDT classification cuts are shown in table \ref{tab:benchmarksAMU}, for an integrated luminosity of $\mathcal{L}=100 ~ \rm fb^{-1}$. 
\begin{table}[tbp]
% \centering
 \resizebox{0.49\textwidth}{!}{
 \begin{tabular}{l | c c c c c c}
 \hline
 \multicolumn{7}{c}{\bf{LFV $(p e^- \to \mu^{-} + 3 \mathrm{j} )$}} \\
 \firsthline 
\diagbox{$m_N$}{$\alpha$}   &   & $0.10$ & $0.15$ & $0.20$ & $0.25$ & $0.30$ \\   %%{\bf{LFV $(\mu^{-})$}}
\hline
\multirow{2}{*}{$100$ GeV}    & $N_s  ~$ &  $45.25$ 	&  $80.16$  & $131.62$ & $195.14$ & $279.24$  \\ 
  & $N_b  ~$	&  $246.45$ 	&  $253.47$  & $260.02$ & $270.76$ & $278.39$  \\ 
\hline  
\multirow{2}{*}{$125$ GeV}  &	&  $49.62$ 	&  $82.16$   & $126.01$ & $187.22$ & $257.29$ \\  
   & 	&  $249.96$ 	&  $256.43 $  & $265.78$ & $273.86$ & $287.10$  \\
\hline
\multirow{2}{*}{$150$ GeV}  &	&  $22.47 $ 	&  $33.27$  & $48.91$ & $69.69$ & $95.78$  \\  
 & 	&  $242.67$ 	&  $245.84$  & $249.51 $ & $253.18$ & $257.22$  \\ 
\hline
\multirow{2}{*}{$200$ GeV}  & 	&  $14.99$ 	&  $20.04$  & $27.49$ & $36.13$ & $47.83$  \\
   & 	&  $240.04$ 	&  $241.30$  & $242.51$ & $244.83$ & $246.65$  \\ 
\hline
\multirow{2}{*}{$300$ GeV}    & 	&  $17.66$ 	&  $21.00$  & $26.19$ & $31.25$ & $38.36$  \\
 & 	&  $238.31$ 	&  $238.85$  & $238.48$ & $239.42$ & $239.71$  \\ 
\hline
\multirow{2}{*}{$500$ GeV}   &	&  $12.63$ 	&  $14.19 $  & $15.60$ & $16.48$ & $18.02$  \\
 & 	&  $235.90$ 	&  $235.96 $  & $235.80  $ & $236.37$ & $236.45$  \\ 
\lasthline
 \end{tabular}
 }
 \resizebox{0.49\textwidth}{!}{
 \begin{tabular}{l | c c c c c c}
 \hline
 \multicolumn{7}{c}{\bf{LNV $(p e^- \to \mu^{+} + 3 \mathrm{j} )$}} \\
 \firsthline 
\diagbox{$m_N$}{$\alpha$}   &   & $0.10$ & $0.15$ & $0.20$ & $0.25$ & $0.30$ \\
\hline
\multirow{2}{*}{$100$ GeV}    & $N_s  ~$ &  $42.81$ 	&  $76.61$  & $124.76$ & $186.90$ & $264.87$  \\ 
  & $N_b  ~$	&  $221.89$ 	&  $227.43$  & $233.80$ & $241.06$ & $248.65$  \\ 
\hline  
\multirow{2}{*}{$125$ GeV}  &	&  $49.51$ 	&  $81.60$  & $127.19$ & $187.30$ & $261.75$  \\  
   & 	&  $224.83$ 	&  $230.32$  & $236.31$ & $243.02$ & $250.31$  \\
\hline
\multirow{2}{*}{$150$ GeV}  &	&  $23.00$ 	&  $33.90$  & $51.66$ & $72.06$ & $100.07$  \\ 
   & 	&  $218.59$ 	&  $221.46$  & $222.89$ & $226.87$ & $228.41$  \\ 
\hline
\multirow{2}{*}{$200$ GeV}  &	&  $15.22$ 	&  $20.72	$  & $28.59$ & $37.46$ & $49.38$  \\
   & 	&  $217.05$ 	&  $217.93$  & $218.70$ & $220.83$ & $222.38$  \\ 
\hline
\multirow{2}{*}{$300$ GeV}  &	&  $17.87$ 	&  $21.53$  & $25.61 $ & $31.22$ & $38.77$  \\
    & 	&  $215.53 $ 	&  $215.74$  & $216.52$ & $216.76$ & $216.79$  \\ 
\hline
\multirow{2}{*}{$500$ GeV}  &	&  $12.59 $ 	&  $13.93$  & $15.02$ & $16.74$ & $18.09$  \\
 & 	&  $213.35$ 	&  $213.59$  & $213.84$ & $213.53$ & $213.57$  \\ 
\lasthline 
 \end{tabular}
 }
 \caption{{Number of events of LFV signal (left) and LNV signal (right) and backgrounds after TMVA classification at the LHeC for an integrated luminosity $\mathcal{L}=100 ~\rm fb^{-1}$.}}\label{tab:benchmarksAMU} 
\end{table}

\subsection{Results}\label{sec:results}

\begin{figure}[tpb]
\centering
{\includegraphics[width=0.48\textwidth]{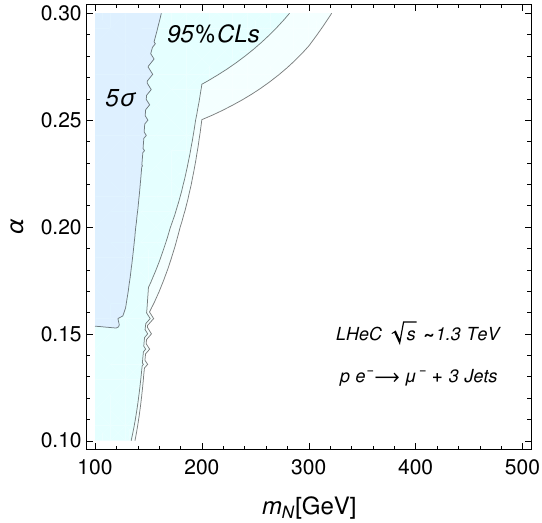}}~
{\includegraphics[width=0.48\textwidth]{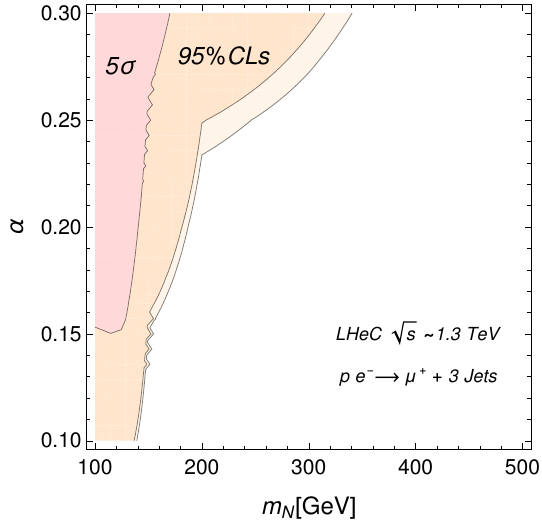}}
 \caption{ \label{fig:contmuamu}{{$5\sigma$-Discovery and 95\% CLs limits, for the muon-trijet (left) and anti-muon-trijet (right) channels at the LHeC (Taking $\Lambda=1$ TeV).}}}
 \end{figure}

The projected sensitivity limits on the effective couplings $\alpha$ for the heavy Majorana neutrino masses $m_N$ in the $100-500$ GeV range obtained from both studied processes at the LHeC are shown in figure \ref{fig:contmuamu}. 

%% % 

The 95\% CLs exclusion limits in the $m_N- \alpha$ plane are calculated following the PDG review on Statistics \cite{ParticleDataGroup:2020ssz} and Apprendix B in \cite{Magill:2018jla}. For each signal point in table \ref{tab:benchmarksAMU} we calculate the upper number of signal events $s^{up}$ consistent at 95\% CLs with the observation of the expected number of background events, by supposing that the data collected in the experiment exactly matches the integer part of the number of events for the background prediction. The shaded areas (lower mass, higher couplings) correspond to the parameter regions where the interpolated expected number of signal events exceeds the upper allowed value $s^{up}$, and thus the limits are imposed directly on the parameter space values. As the number of events classified as background after the BDT cut changes from one signal benchmark point to another, we show the curves corresponding to the greatest (and lowest) upper number of signal events $s^{up}$ for each channel. The region between this two curves is displayed in a lighter color in the plots in figure \ref{fig:contmuamu}.  

We also show the $5\sigma$- discovery contours, obtained using the well known formula for the signal statistical significance \cite{Cowan:2010js}: 
\begin{equation*}
 Z \sigma= \sqrt{2 \left[\rm(N_s+N_b) ~\rm{ln} \left(1+ \frac{N_s}{N_b}\right)-N_s\right]}=5 \sigma.
\end{equation*}
The lower mass and higher coupling ($m_N<150$ GeV, $\alpha>0.15$) regions in the parameter space could be separated with $5\sigma$ significance from the expected backgrounds for both the LFV and LNV channels, meaning the LHeC would be able to discover this lepton-trijet signals after collecting $100 ~\rm fb^{-1}$ of data.

The obtained sensitivity prospects show that the LHeC could be able to probe the scenario of a heavy $N$ with a mass near and above the electroweak scale, and constrain the effective couplings (mostly those of the muon family) to a region of the parameter space as tight as the bounds that are currently considered for the $\mathcal{O}(10)$ GeV scale masses (see refs. \cite{Alcaide:2019pnf, Fernandez-Martinez:2023phj, Mitra:2022nri} for comparable $\nu$SMEFT phenomenologic studies). Also, we find the results for discovery and exclusion regions are similar for both the muon-trijet (LFV) and the anti-muon-trijet (LNV) signals studied. 

It is important to stress that the bounds in figure \ref{fig:contmuamu} are obtained for a realistic benchmark scenario configuration where we fix all the couplings of the $d=6$ $\nu$SMEFT operators in table \ref{tab:Operators} for every flavor to the same numerical value $\alpha$, excepting those of the operators contributing to the unobserved neutrinoless double beta decay, corresponding to the first fermions family ($i=1$), which are set to the value of the corresponding bound $\alpha_{0\nu\beta\beta}(m_N) = 3.2\times 10^{-2}\left(\frac{m_N}{100 ~ ~\rm{GeV}} \right)^{1/2}$ for $\Lambda=1$ TeV. The loop factor in the couplings of the $\mathcal{O}_{NB}$ and $\mathcal{O}_{NW}$ dipole operators is included in the numerical calculations: we fix $\alpha_{NB}=\alpha_{NW}=\alpha$, but take into account the loop factor in the interaction vertices, as presented in the Lagrangian terms \eqref{eq: app ONB} and \eqref{eq: app ONW}.

Also, the $\nu$SMEFT approach considered here only includes one heavy Majorana neutrino $N$ as observable degree of freedom, and neglects the renormalizable Type-I seesaw mixing terms it could have with the active neutrinos. In this sense, the obtained sensitivity prospects for the LHeC appear to be looser than those obtained in refs. \cite{Antusch:2019eiz, Gu:2022muc}, but they are just non-comparable.

The interested reader can find sensitivity prospects for the near-future experiments concerning the dimension-6 $\nu$SMEFT interactions in \cite{Beltran:2021hpq} for a long-lived $N$ at the LHC exploiting  possible displaced vertices searches, and in \cite{Barducci:2022hll} for prompt and displaced $N$ decays at future Higgs factories, in both cases for lighter $N$ benchmark scenarios with $m_N \lesssim 60 ~\text{GeV}$. Also, a variety of testable signals in planned experiments are discussed in \cite{Mitra:2022nri}. We have explored the discovery potential for the heavy $N$ at the ILC in \cite{Zapata:2022qwo}. However, we are leaving for future work a systematic comparison of the sensitivity reaches of near-future planned experiments.

\section{Summary}\label{sec:summary}

In this paper we study the prospects of the future LHeC electron-proton collider to discover or constrain the $\nu$SMEFT interactions, performing the first dedicated and realistic analysis of the well known lepton-trijet signals, both for the lepton flavor violating $p ~ e^{-} \rightarrow \mu^{-} +  3 \mathrm{j}$ (LFV) and the lepton number violating $p ~ e^{-} \rightarrow \mu^{+} +  3 \mathrm{j}$ (LNV) channels. Despite of both processes being irreducible SM background-free, we take into account the possible backgrounds due to charge misidentification and final states with extra unobserved light neutrinos, performing a dedicated simulation and analysis at the reconstructed level. 

The effective field theory extending the standard model with sterile right-handed neutrinos $\nu$SMEFT is the adequate tool for parameterizing new high-scale weakly coupled physics in a model independent manner, and allows for a systematic study of the HNLs phenomenology in current and future experiments. We consider heavy Majorana neutrinos coupled to ordinary matter by dimension 6 effective operators, focusing on a simplified scenario with only one right-handed neutrino added, and consider every operator contributing to the $N$ production and decay channels.

A thorough signals kinematical characterization for the LHeC environment would suggest the use of the separation of the $N$ decay products from the beam jet for a search strategy, but we find the detailed simulation of the complete $N$ production and decay to the lepton-trijet final states cannot be significantly discriminated from the backgrounds on a cut-and-count approach at the reconstructed level. We thus focus on testing the performance of a multivariate analysis with a boosted decision tree (BDT) algorithm. 

The obtained 95\% CLs exclusion limits in the $m_N- \alpha$ plane are presented in figure \ref{fig:contmuamu}, showing that the LHeC could constrain the effective couplings (mostly those of the muon family) to a region of the parameter space as tight as the bounds that are currently considered for the sub-electroweak scale masses (see Refs. \cite{Alcaide:2019pnf, Fernandez-Martinez:2023phj, Mitra:2022nri}). 

Our results demonstrate that the LHeC is also an excellent facility for discovering heavy Majorana neutrinos around the electroweak scale, even in the limit of the $\nu$SMEFT where one discards their mixing with the active neutrino states. The LNV and LFV lepton-trijet signatures could be “golden channels” for HNLs searches. A discovery of heavy neutrinos would have far-reaching consequences, but also constraining the possible new physics involved in neutrino mass generation can be a path to resolve the origin of the observed neutrino masses, which is one of the most challenging open questions in particle physics.

\acknowledgments

We thank CONICET (Argentina) and PEDECIBA, CSIC (Uruguay) for their financial support.
%%%%%%%%%%%%%%%%%%%%%%%%%%

\appendix

\section{Explicit Lagrangian terms from the effective operators}\label{app:explicitLagOpe}

For completeness we list here the explicit Lagrangian terms given by each $d=6$ effective operator listed in table \ref{tab:Operators}, with $\phi= (0, \frac{v+h}{\sqrt{2}})^T$, after electroweak symmetry breaking (the hermitian conjugate must be added). 

\noindent The Higgs dressed mixing operator gives the following Lagrangian terms,
\begin{equation}  \label{eq: app LNHiggs}
\mathcal{O}^{(i)}_{LN\phi}=(\phi^{\dagger}\phi)(\bar L_i N \tilde{\phi}) \to \frac{\alpha^{(i)}_{LN\phi}}{\Lambda^2}
\left( \frac{3v^2}{2\sqrt{2}}~\bar \nu_{L,i} N_R~ h + \frac{3v}{2\sqrt{2}}~\bar \nu_{L,i} N_R~ h h+ \frac{1}{2\sqrt{2}}~\bar \nu_{L,i} N_R~ h h h \right)
\end{equation}
plus a contriubution to neutrino mass (see \cite{Mitra:2022nri}):
$ \frac{\alpha^{(i)}_{LN\phi}}{\Lambda^2} \frac{v^3}{2\sqrt{2}} \nu_{L,i} N_R $. 

\noindent The neutral bosonic current contributes to the $Z \to N N $ decays as well as interactions with the Higgs, and leads also to new Higgs vertices 
\begin{equation} \label{eq: app BosonNC}
\mathcal{O}_{NN\phi}=i(\phi^{\dagger}\overleftrightarrow{D_{\mu}}\phi)(\bar N \gamma^{\mu} N)\to - \frac{\alpha^{(i)}_{NN\phi}}{\Lambda^2} \left( (\bar N_R \gamma^{\mu} N_R) \left( \frac{m_Z}{v} Z_{\mu} \right) \left( v^2 + 2 v h + h h \right) \right).
\end{equation}

\noindent The bosonic charged current resembles the SM CC interaction, here substituting active for sterile neutrinos
\begin{equation}\label{eq: app BosonCC}
\mathcal{O}^{(i)}_{N l\phi}= i(\phi^T \epsilon D_{\mu}\phi)(\bar N \gamma^{\mu} l_i) \to \frac{\alpha^{(i)}_{N l\phi}}{\Lambda^2} \frac{ m_{W}}{\sqrt{2}v }  (\bar N_R \gamma^{\mu} l_{R,i}) W^{+}_{\mu} ( v^2 + 2 v h +  h h).
\end{equation}

\noindent The dipole operators must be generated at one-loop level in the unknown UV complete theory:
thus the Lagrangian terms they generate include a loop factor $1/16\pi^2$, which we explicitly write here and consider in our numerical calculations. 

\begin{equation}\label{eq: app ONB}
\mathcal{O}^{(i)}_{NB}=(\bar L_i \sigma^{\mu\nu} N) \tilde \phi B_{\mu\nu} \to -i \frac{\alpha^{(i)}_{NB}}{\Lambda^2 (16\pi^2)} (\bar \nu_{L,i} \sigma^{\mu\nu} N_R)  [2 p_{\mu}^{(A)}A_{\nu} c_W -2 p_{\mu}^{(Z)}Z_{\nu} s_W]\frac{(v+h)}{\sqrt{2}}
\end{equation}

\begin{eqnarray}\label{eq: app ONW}
 && \mathcal{O}^{(i)}_{NW} =(\bar L_i \sigma^{\mu\nu} \sigma^I N) \tilde \phi W_{\mu\nu}^I \to  
\nonumber 
\\ &&  -i \frac{\alpha^{(i)}_{NW}}{\Lambda^2 (16\pi^2)} 
\left\lbrace 
(\bar \nu_{L,i} \sigma^{\mu\nu} N_R) \left[2 p^{(Z)}_{\mu} Z_{\nu} c_W + 2 p^{(A)}_{\mu}A_{\nu} s_W 
+2 g W^+_{\mu} W^-_{\nu} \right] \frac{(v+h)}{\sqrt{2}}
\right.
\nonumber 
\\ && \left.
+  (\bar \ell_{L,i} \sigma^{\mu\nu} N_R) 
\left[2 g W^-_{\mu} (Z_{\nu} c_W+ A_{\nu} s_W) + 2 p^{(W)}_{\mu} W^-_{\nu} \right] 
(v+ h) 
\right\rbrace .
\end{eqnarray}
Here the action of derivatives on the fields is substituted for the corresponding (in-going to vertex) momenta, and $g$ is the $SU(2)_L$ coupling.

\noindent The four-fermion contact terms include neutral currents mediated by vectors:
\begin{equation} \label{eq: app QNN}
\mathcal{O}^{(i)}_{QNN}=(\bar{Q_i} \gamma^\mu Q_i) (\bar{N} \gamma_\mu N)\to \frac{\alpha^{(i)}_{QNN}}{\Lambda^2} (\bar u_{L,i} \gamma^{\mu} u_{L,i} + \bar d_{L,i} \gamma^{\mu} d_{L,i}) (\bar N_R \gamma_{\mu} N_R)
\end{equation}
\begin{equation} \label{eq: app LNN}
\mathcal{O}^{(i)}_{LNN}=(\bar{L_i} \gamma^\mu L_i) (\bar{N} \gamma_\mu N)\to \frac{\alpha^{(i)}_{LNN}}{\Lambda^2} (\bar \nu_{L,i} \gamma^{\mu} \nu_{L,i} + \bar \ell_{L,i}\gamma^{\mu} \ell_{L,i}) (\bar N_R \gamma_{\mu} N_R)
\end{equation}
\begin{equation}\label{eq: app fNN}
\mathcal{O}^{(i)}_{fNN}= (\bar{f_i} \gamma^\mu f_i) (\bar{N} \gamma_\mu N)\to \frac{\alpha^{(i)}_{fNN}}{\Lambda^2} (\bar f_{i} \gamma^{\mu} f_{i}) (\bar N_R \gamma_{\mu} N_R), \;\;\;  f= u_R, d_R, \ell_R
\end{equation}
A charged current mediated by a vector:
\begin{equation} \label{eq: app duNl}
\mathcal{O}^{(i, j)}_{duNl}= (\bar d _j \gamma^{\mu} u _j) (\bar N \gamma_{\mu} l_i)\to  \frac{\alpha^{(i, j)}_{duNl}}{\Lambda^2} (\bar d_{R,j} \gamma^{\mu} u_{R,j}) (\bar N_R \gamma_{\mu} \ell_{R,i}) 
\end{equation}
And interactions that can be mediated by both charged and neutral scalars:
\begin{equation}\label{eq: app QuNL}
\mathcal{O}^{(i,j)}_{QuNL}= (\bar Q _j u _j)(\bar N L_i)\to \frac{\alpha^{(i,j)}_{QuNL}}{\Lambda^2}(\bar
u_{L,j}u_{R,j}\bar N_R \nu_{L,i}+\bar d_{L,j}u_{R,j} \bar N_R \ell_{L,i})
\end{equation}
\begin{equation}\label{eq: app LNQd}
\mathcal{O}^{(i, j)}_{LNQd}=(\bar L_i N) \epsilon (\bar Q _j d _j) \to \frac{\alpha^{(i,j)}_{LNQd}}{\Lambda^2}(\bar \nu_{L,i}N_R \bar d_{L,j}d_{R,j}-\bar \ell_{L,i}N_R \bar u_{L,j}d_{R,j}) 
\end{equation}
\begin{equation}\label{eq: app QNLd}
\mathcal{O}^{(i, j)}_{QNLd}=(\bar Q _i N)\epsilon (\bar L_j, d_j) \to \frac{\alpha^{(i,j)}_{QNLd}}{\Lambda^2}(\bar u_{L,i}N_R \bar \ell_{L,j} d_{R,j}-\bar d_{L,i}N_R \bar \nu_{L,j}d_{R,j}) 
\end{equation}
\begin{equation}\label{eq: app LNLl}
\mathcal{O}^{(i, j)}_{LNLl}=(\bar L_i N)\epsilon (\bar L_j l_j) \to \frac{\alpha^{(i, j)}_{LNLl}}{\Lambda^2}(\bar \nu_{L,i}N_R \bar \ell_{L,j} \ell_{R,j}-\bar \ell_{L,i}N_R \bar \nu_{L,j} \ell_{R,j}).
\end{equation}

\bibliographystyle{bibstyle}
\bibliography{Bib_N_12_23}

%Merlin.mbs v4.21 2009-07-09.
\begin{thebibliography}{10}%
\makeatletter
\providecommand \@ifxundefined [1]{%
 \ifx #1\undefined \expandafter \@firstoftwo
 \else \expandafter \@secondoftwo
\fi
}%
\providecommand \@ifnum [1]{%
 \ifnum #1\expandafter \@firstoftwo
 \else \expandafter \@secondoftwo
\fi
}%
\providecommand \enquote [1]{``#1''}%
\providecommand \bibnamefont  [1]{#1}%
\providecommand \bibfnamefont [1]{#1}%
\providecommand \citenamefont [1]{#1}%
\providecommand\href[0]{\@sanitize\@href}%
\providecommand\@href[1]{\endgroup\@@startlink{#1}\endgroup\@@href}%
\providecommand\@@href[1]{#1\@@endlink}%
\providecommand \@sanitize [0]{\begingroup\catcode`\&12\catcode`\#12\relax}%
\@ifxundefined \pdfoutput {\@firstoftwo}{%
 \@ifnum{\z@=\pdfoutput}{\@firstoftwo}{\@secondoftwo}%
}{%
 \providecommand\@@startlink[1]{\leavevmode\special{html:<a href="#1">}}%
 \providecommand\@@endlink[0]{\special{html:</a>}}%
}{%
 \providecommand\@@startlink[1]{%
  \leavevmode
  \pdfstartlink
   attr{/Border[0 0 1 ]/H/I/C[0 1 1]}%
   user{/Subtype/Link/A<</Type/Action/S/URI/URI(#1)>>}%
  \relax
 }%
 \providecommand\@@endlink[0]{\pdfendlink}%
}%
\providecommand \url  [0]{\begingroup\@sanitize \@url }%
\providecommand \@url [1]{\endgroup\@href {#1}{\urlprefix}}%
\providecommand \urlprefix [0]{URL }%
\providecommand \Eprint[0]{\href }%
\@ifxundefined \urlstyle {%
  \providecommand \doi [1]{doi:\discretionary{}{}{}#1}%
}{%
  \providecommand \doi [0]{doi:\discretionary{}{}{}\begingroup
  \urlstyle{rm}\Url }%
}%
\providecommand \doibase [0]{http://dx.doi.org/}%
\providecommand \Doi[1]{\href{\doibase#1}}%
\providecommand \bibAnnote [3]{%
  \BibitemShut{#1}%
  \begin{quotation}\noindent
    \textsc{Key:}\ #2\\\textsc{Annotation:}\ #3%
  \end{quotation}%
}%
\providecommand \bibAnnoteFile [2]{%
  \IfFileExists{#2}{\bibAnnote {#1} {#2} {\input{#2}}}{}%
}%
\providecommand \typeout [0]{\immediate \write \m@ne }%
\providecommand \selectlanguage [0]{\@gobble}%
\providecommand \bibinfo [0]{\@secondoftwo}%
\providecommand \bibfield [0]{\@secondoftwo}%
\providecommand \translation [1]{[#1]}%
\providecommand \BibitemOpen[0]{}%
\providecommand \bibitemStop [0]{}%
\providecommand \bibitemNoStop [0]{.\EOS\space}%
\providecommand \EOS [0]{\spacefactor3000\relax}%
\providecommand \BibitemShut [1]{\csname bibitem#1\endcsname}%
%</preamble>
\bibitem{Minkowski:1977sc}%
  \BibitemOpen
  \bibfield{author}{%
  \bibinfo {author} {\bibfnamefont{P.}~\bibnamefont{Minkowski}},\ }%
  \emph{\bibinfo {title} {{$\mu \rightarrow e \gamma$ at a rate of one out of
  1-billion muon decays?}}},\ \bibfield{journal}{%
  \Doi{10.1016/0370-2693(77)90435-X}{\bibinfo {journal} {Phys.Lett.}}\ }%
  \textbf{\bibinfo {volume} {B67}},\ \bibinfo {pages} {421} (\bibinfo {year}
  {1977}).%
  \bibAnnoteFile{Stop}{Minkowski:1977sc}%
%%CITATION = PHLTA,B67,421;%%
\bibitem{Mohapatra:1979ia}%
  \BibitemOpen
  \bibfield{author}{%
  \bibinfo {author} {\bibfnamefont{R.~N.}\ \bibnamefont{Mohapatra}}\ and\
  \bibinfo {author} {\bibfnamefont{G.}~\bibnamefont{Senjanovic}},\ }%
  \emph{\bibinfo {title} {{Neutrino Mass and Spontaneous Parity Violation}}},\
  \bibfield{journal}{%
  \Doi{10.1103/PhysRevLett.44.912}{\bibinfo {journal} {Phys.Rev.Lett.}}\ }%
  \textbf{\bibinfo {volume} {44}},\ \bibinfo {pages} {912} (\bibinfo {year}
  {1980}).%
  \bibAnnoteFile{Stop}{Mohapatra:1979ia}%
%%CITATION = PRLTA,44,912;%%
\bibitem{Yanagida:1980xy}%
  \BibitemOpen
  \bibfield{author}{%
  \bibinfo {author} {\bibfnamefont{T.}~\bibnamefont{Yanagida}},\ }%
  \emph{\bibinfo {title} {{Horizontal Symmetry and Masses of Neutrinos}}},\
  \bibfield{journal}{%
  \Doi{10.1143/PTP.64.1103}{\bibinfo {journal} {Prog.Theor.Phys.}}\ }%
  \textbf{\bibinfo {volume} {64}},\ \bibinfo {pages} {1103} (\bibinfo {year}
  {1980}).%
  \bibAnnoteFile{Stop}{Yanagida:1980xy}%
%%CITATION = PTPKA,64,1103;%%
\bibitem{GellMann:1980vs}%
  \BibitemOpen
  \bibfield{author}{%
  \bibinfo {author} {\bibfnamefont{M.}~\bibnamefont{Gell-Mann}}, \bibinfo
  {author} {\bibfnamefont{P.}~\bibnamefont{Ramond}}\ and\ \bibinfo {author}
  {\bibfnamefont{R.}~\bibnamefont{Slansky}},\ }%
  \emph{\bibinfo {title} {{Complex Spinors and Unified Theories}}},\
  \bibfield{journal}{%
  \bibinfo {journal} {Conf.Proc.}\ }%
  \textbf{\bibinfo {volume} {C790927}},\ \bibinfo {pages} {315} (\bibinfo
  {year} {1979}),\ \Eprint{http://arxiv.org/abs/1306.4669}{arXiv:1306.4669
  [hep-th]}.%
  \bibAnnoteFile{Stop}{GellMann:1980vs}%
%%CITATION = ARXIV:1306.4669;%%
\bibitem{Schechter:1980gr}%
  \BibitemOpen
  \bibfield{author}{%
  \bibinfo {author} {\bibfnamefont{J.}~\bibnamefont{Schechter}}\ and\ \bibinfo
  {author} {\bibfnamefont{J.~W.~F.}\ \bibnamefont{Valle}},\ }%
  \emph{\bibinfo {title} {{Neutrino Masses in SU(2) x U(1) Theories}}},\
  \bibfield{journal}{%
  \Doi{10.1103/PhysRevD.22.2227}{\bibinfo {journal} {Phys. Rev.}}\ }%
  \textbf{\bibinfo {volume} {D22}},\ \bibinfo {pages} {2227} (\bibinfo {year}
  {1980}).%
  \bibAnnoteFile{Stop}{Schechter:1980gr}%
%%CITATION = PHRVA,D22,2227;%%
\bibitem{Malinsky:2005bi}%
  \BibitemOpen
  \bibfield{author}{%
  \bibinfo {author} {\bibfnamefont{M.}~\bibnamefont{Malinsky}}, \bibinfo
  {author} {\bibfnamefont{J.}~\bibnamefont{Romao}}\ and\ \bibinfo {author}
  {\bibfnamefont{J.}~\bibnamefont{Valle}},\ }%
  \emph{\bibinfo {title} {{Novel supersymmetric SO(10) seesaw mechanism}}},\
  \bibfield{journal}{%
  \Doi{10.1103/PhysRevLett.95.161801}{\bibinfo {journal} {Phys. Rev. Lett.}}\
  }%
  \textbf{\bibinfo {volume} {95}},\ \bibinfo {pages} {161801} (\bibinfo {year}
  {2005}),\
  \Eprint{http://arxiv.org/abs/hep-ph/0506296}{arXiv:hep-ph/0506296}.%
  \bibAnnoteFile{Stop}{Malinsky:2005bi}%
\bibitem{Mohapatra:1986bd}%
  \BibitemOpen
  \bibfield{author}{%
  \bibinfo {author} {\bibfnamefont{R.}~\bibnamefont{Mohapatra}}\ and\ \bibinfo
  {author} {\bibfnamefont{J.}~\bibnamefont{Valle}},\ }%
  \emph{\bibinfo {title} {{Neutrino Mass and Baryon Number Nonconservation in
  Superstring Models}}},\ \bibfield{journal}{%
  \Doi{10.1103/PhysRevD.34.1642}{\bibinfo {journal} {Phys. Rev. D}}\ }%
  \textbf{\bibinfo {volume} {34}},\ \bibinfo {pages} {1642} (\bibinfo {year}
  {1986}).%
  \bibAnnoteFile{Stop}{Mohapatra:1986bd}%
\bibitem{Abdullahi:2022jlv}%
  \BibitemOpen
  \bibfield{author}{%
  \bibinfo {author} {\bibfnamefont{A.~M.}\ \bibnamefont{Abdullahi}}
  \emph{et~al.},\ }%
  \emph{\bibinfo {title} {{The present and future status of heavy neutral
  leptons}}},\ \bibfield{journal}{%
  \Doi{10.1088/1361-6471/ac98f9}{\bibinfo {journal} {J. Phys. G}}\ }%
  \textbf{\bibinfo {volume} {50}},\ \bibinfo {pages} {020501} (\bibinfo {year}
  {2023}),\ \Eprint{http://arxiv.org/abs/2203.08039}{arXiv:2203.08039
  [hep-ph]}.%
  \bibAnnoteFile{Stop}{Abdullahi:2022jlv}%
\bibitem{Anisimov:2006hv}%
  \BibitemOpen
  \bibfield{author}{%
  \bibinfo {author} {\bibfnamefont{A.}~\bibnamefont{Anisimov}},\ }%
  in\ \Doi{10.1142/9789812770288_0058}{\emph{\bibinfo {booktitle} {{6th
  International Workshop on the Identification of Dark Matter}}}}\ (\bibinfo
  {year} {2006})\ pp.\ \bibinfo {pages} {439--449},\
  \Eprint{http://arxiv.org/abs/hep-ph/0612024}{arXiv:hep-ph/0612024}%
  \bibAnnoteFile{NoStop}{Anisimov:2006hv}%
\bibitem{Graesser:2007yj}%
  \BibitemOpen
  \bibfield{author}{%
  \bibinfo {author} {\bibfnamefont{M.~L.}\ \bibnamefont{Graesser}},\ }%
  \emph{\bibinfo {title} {{Broadening the Higgs boson with right-handed
  neutrinos and a higher dimension operator at the electroweak scale}}},\
  \bibfield{journal}{%
  \Doi{10.1103/PhysRevD.76.075006}{\bibinfo {journal} {Phys. Rev.}}\ }%
  \textbf{\bibinfo {volume} {D76}},\ \bibinfo {pages} {075006} (\bibinfo {year}
  {2007}),\ \Eprint{http://arxiv.org/abs/0704.0438}{arXiv:0704.0438 [hep-ph]}.%
  \bibAnnoteFile{Stop}{Graesser:2007yj}%
%%CITATION = ARXIV:0704.0438;%%
\bibitem{Graesser:2007pc}%
  \BibitemOpen
  \bibfield{author}{%
  \bibinfo {author} {\bibfnamefont{M.~L.}\ \bibnamefont{Graesser}},\ }%
  \emph{\bibinfo {title} {{Experimental Constraints on Higgs Boson Decays to
  TeV-scale Right-Handed Neutrinos}}} (\bibinfo {month} {5}\ \bibinfo {year}
  {2007}),\ \Eprint{http://arxiv.org/abs/0705.2190}{arXiv:0705.2190 [hep-ph]}.%
  \bibAnnoteFile{Stop}{Graesser:2007pc}%
\bibitem{delAguila:2008ir}%
  \BibitemOpen
  \bibfield{author}{%
  \bibinfo {author} {\bibfnamefont{F.}~\bibnamefont{del Aguila}}, \bibinfo
  {author} {\bibfnamefont{S.}~\bibnamefont{Bar-Shalom}}, \bibinfo {author}
  {\bibfnamefont{A.}~\bibnamefont{Soni}}\ and\ \bibinfo {author}
  {\bibfnamefont{J.}~\bibnamefont{Wudka}},\ }%
  \emph{\bibinfo {title} {{Heavy Majorana Neutrinos in the Effective Lagrangian
  Description: Application to Hadron Colliders}}},\ \bibfield{journal}{%
  \Doi{10.1016/j.physletb.2008.11.031}{\bibinfo {journal} {Phys.Lett.}}\ }%
  \textbf{\bibinfo {volume} {B670}},\ \bibinfo {pages} {399} (\bibinfo {year}
  {2009}),\ \Eprint{http://arxiv.org/abs/0806.0876}{arXiv:0806.0876 [hep-ph]}.%
  \bibAnnoteFile{Stop}{delAguila:2008ir}%
%%CITATION = ARXIV:0806.0876;%%
\bibitem{Aparici:2009fh}%
  \BibitemOpen
  \bibfield{author}{%
  \bibinfo {author} {\bibfnamefont{A.}~\bibnamefont{Aparici}}, \bibinfo
  {author} {\bibfnamefont{K.}~\bibnamefont{Kim}}, \bibinfo {author}
  {\bibfnamefont{A.}~\bibnamefont{Santamaria}}\ and\ \bibinfo {author}
  {\bibfnamefont{J.}~\bibnamefont{Wudka}},\ }%
  \emph{\bibinfo {title} {{Right-handed neutrino magnetic moments}}},\
  \bibfield{journal}{%
  \Doi{10.1103/PhysRevD.80.013010}{\bibinfo {journal} {Phys. Rev.}}\ }%
  \textbf{\bibinfo {volume} {D80}},\ \bibinfo {pages} {013010} (\bibinfo {year}
  {2009}),\ \Eprint{http://arxiv.org/abs/0904.3244}{arXiv:0904.3244 [hep-ph]}.%
  \bibAnnoteFile{Stop}{Aparici:2009fh}%
%%CITATION = ARXIV:0904.3244;%%
\bibitem{Liao:2016qyd}%
  \BibitemOpen
  \bibfield{author}{%
  \bibinfo {author} {\bibfnamefont{Y.}~\bibnamefont{Liao}}\ and\ \bibinfo
  {author} {\bibfnamefont{X.-D.}\ \bibnamefont{Ma}},\ }%
  \emph{\bibinfo {title} {{Operators up to Dimension Seven in Standard Model
  Effective Field Theory Extended with Sterile Neutrinos}}},\
  \bibfield{journal}{%
  \Doi{10.1103/PhysRevD.96.015012}{\bibinfo {journal} {Phys. Rev.}}\ }%
  \textbf{\bibinfo {volume} {D96}},\ \bibinfo {pages} {015012} (\bibinfo {year}
  {2017}),\ \Eprint{http://arxiv.org/abs/1612.04527}{arXiv:1612.04527
  [hep-ph]}.%
  \bibAnnoteFile{Stop}{Liao:2016qyd}%
%%CITATION = ARXIV:1612.04527;%%
\bibitem{Bhattacharya:2015vja}%
  \BibitemOpen
  \bibfield{author}{%
  \bibinfo {author} {\bibfnamefont{S.}~\bibnamefont{Bhattacharya}}\ and\
  \bibinfo {author} {\bibfnamefont{J.}~\bibnamefont{Wudka}},\ }%
  \emph{\bibinfo {title} {{Dimension-seven operators in the standard model with
  right handed neutrinos}}},\ \bibfield{journal}{%
  \Doi{10.1103/PhysRevD.94.055022, 10.1103/PhysRevD.95.039904}{\bibinfo
  {journal} {Phys. Rev.}}\ }%
  \textbf{\bibinfo {volume} {D94}},\ \bibinfo {pages} {055022} (\bibinfo {year}
  {2016}),\ \bibinfo {note} {[Erratum: Phys. Rev.D95,no.3,039904(2017)]},\
  \Eprint{http://arxiv.org/abs/1505.05264}{arXiv:1505.05264 [hep-ph]}.%
  \bibAnnoteFile{Stop}{Bhattacharya:2015vja}%
%%CITATION = ARXIV:1505.05264;%%
\bibitem{Li:2021tsq}%
  \BibitemOpen
  \bibfield{author}{%
  \bibinfo {author} {\bibfnamefont{H.-L.}\ \bibnamefont{Li}}, \bibinfo {author}
  {\bibfnamefont{Z.}~\bibnamefont{Ren}}, \bibinfo {author}
  {\bibfnamefont{M.-L.}\ \bibnamefont{Xiao}}, \bibinfo {author}
  {\bibfnamefont{J.-H.}\ \bibnamefont{Yu}}\ and\ \bibinfo {author}
  {\bibfnamefont{Y.-H.}\ \bibnamefont{Zheng}},\ }%
  \emph{\bibinfo {title} {{Operator bases in effective field theories with
  sterile neutrinos: d \ensuremath{\leq} 9}}},\ \bibfield{journal}{%
  \Doi{10.1007/JHEP11(2021)003}{\bibinfo {journal} {JHEP}}\ }%
  \textbf{\bibinfo {volume} {11}},\ \bibinfo {pages} {003} (\bibinfo {year}
  {2021}),\ \Eprint{http://arxiv.org/abs/2105.09329}{arXiv:2105.09329
  [hep-ph]}.%
  \bibAnnoteFile{Stop}{Li:2021tsq}%
\bibitem{Peressutti:2014lka}%
  \BibitemOpen
  \bibfield{author}{%
  \bibinfo {author} {\bibfnamefont{J.}~\bibnamefont{Peressutti}}\ and\ \bibinfo
  {author} {\bibfnamefont{O.~A.}\ \bibnamefont{Sampayo}},\ }%
  \emph{\bibinfo {title} {{Majorana neutrinos in $e$ $\gamma$ colliders from an
  effective Lagrangian approach}}},\ \bibfield{journal}{%
  \Doi{10.1103/PhysRevD.90.013003}{\bibinfo {journal} {Phys. Rev.}}\ }%
  \textbf{\bibinfo {volume} {D90}},\ \bibinfo {pages} {013003} (\bibinfo {year}
  {2014}).%
  \bibAnnoteFile{Stop}{Peressutti:2014lka}%
%%CITATION = PHRVA,D90,013003;%%
\bibitem{Duarte:2014zea}%
  \BibitemOpen
  \bibfield{author}{%
  \bibinfo {author} {\bibfnamefont{L.}~\bibnamefont{Duarte}}, \bibinfo {author}
  {\bibfnamefont{G.~A.}\ \bibnamefont{González-Sprinberg}}\ and\ \bibinfo
  {author} {\bibfnamefont{O.~A.}\ \bibnamefont{Sampayo}},\ }%
  \emph{\bibinfo {title} {{Majorana neutrinos production at LHeC in an
  effective approach}}},\ \bibfield{journal}{%
  \Doi{10.1103/PhysRevD.91.053007}{\bibinfo {journal} {Phys. Rev.}}\ }%
  \textbf{\bibinfo {volume} {D91}},\ \bibinfo {pages} {053007} (\bibinfo {year}
  {2015}),\ \Eprint{http://arxiv.org/abs/1412.1433}{arXiv:1412.1433 [hep-ph]}.%
  \bibAnnoteFile{Stop}{Duarte:2014zea}%
%%CITATION = ARXIV:1412.1433;%%
\bibitem{Duarte:2015iba}%
  \BibitemOpen
  \bibfield{author}{%
  \bibinfo {author} {\bibfnamefont{L.}~\bibnamefont{Duarte}}, \bibinfo {author}
  {\bibfnamefont{J.}~\bibnamefont{Peressutti}}\ and\ \bibinfo {author}
  {\bibfnamefont{O.~A.}\ \bibnamefont{Sampayo}},\ }%
  \emph{\bibinfo {title} {{Majorana neutrino decay in an Effective
  Approach}}},\ \bibfield{journal}{%
  \Doi{10.1103/PhysRevD.92.093002}{\bibinfo {journal} {Phys. Rev.}}\ }%
  \textbf{\bibinfo {volume} {D92}},\ \bibinfo {pages} {093002} (\bibinfo {year}
  {2015}),\ \Eprint{http://arxiv.org/abs/1508.01588}{arXiv:1508.01588
  [hep-ph]}.%
  \bibAnnoteFile{Stop}{Duarte:2015iba}%
%%CITATION = ARXIV:1508.01588;%%
\bibitem{Duarte:2016miz}%
  \BibitemOpen
  \bibfield{author}{%
  \bibinfo {author} {\bibfnamefont{L.}~\bibnamefont{Duarte}}, \bibinfo {author}
  {\bibfnamefont{I.}~\bibnamefont{Romero}}, \bibinfo {author}
  {\bibfnamefont{J.}~\bibnamefont{Peressutti}}\ and\ \bibinfo {author}
  {\bibfnamefont{O.~A.}\ \bibnamefont{Sampayo}},\ }%
  \emph{\bibinfo {title} {{Effective Majorana neutrino decay}}},\
  \bibfield{journal}{%
  \Doi{10.1140/epjc/s10052-016-4301-8}{\bibinfo {journal} {Eur. Phys. J. C}}\
  }%
  \textbf{\bibinfo {volume} {76}},\ \bibinfo {pages} {453} (\bibinfo {year}
  {2016}),\ \Eprint{http://arxiv.org/abs/1603.08052}{arXiv:1603.08052
  [hep-ph]}.%
  \bibAnnoteFile{Stop}{Duarte:2016miz}%
\bibitem{Duarte:2016smd}%
  \BibitemOpen
  \bibfield{author}{%
  \bibinfo {author} {\bibfnamefont{L.}~\bibnamefont{Duarte}}, \bibinfo {author}
  {\bibfnamefont{I.}~\bibnamefont{Romero}}, \bibinfo {author}
  {\bibfnamefont{G.}~\bibnamefont{Zapata}}\ and\ \bibinfo {author}
  {\bibfnamefont{O.~A.}\ \bibnamefont{Sampayo}},\ }%
  \emph{\bibinfo {title} {{Effects of Majorana physics on the UHE $\nu _{\tau
  }$ flux traversing the Earth}}},\ \bibfield{journal}{%
  \Doi{10.1140/epjc/s10052-017-4638-7}{\bibinfo {journal} {Eur. Phys. J. C}}\
  }%
  \textbf{\bibinfo {volume} {77}},\ \bibinfo {pages} {68} (\bibinfo {year}
  {2017}),\ \Eprint{http://arxiv.org/abs/1609.07661}{arXiv:1609.07661
  [hep-ph]}.%
  \bibAnnoteFile{Stop}{Duarte:2016smd}%
\bibitem{Duarte:2016caz}%
  \BibitemOpen
  \bibfield{author}{%
  \bibinfo {author} {\bibfnamefont{L.}~\bibnamefont{Duarte}}, \bibinfo {author}
  {\bibfnamefont{J.}~\bibnamefont{Peressutti}}\ and\ \bibinfo {author}
  {\bibfnamefont{O.~A.}\ \bibnamefont{Sampayo}},\ }%
  \emph{\bibinfo {title} {{Not-that-heavy Majorana neutrino signals at the
  LHC}}},\ \bibfield{journal}{%
  \Doi{10.1088/1361-6471/aa99f5}{\bibinfo {journal} {J. Phys. G}}\ }%
  \textbf{\bibinfo {volume} {45}},\ \bibinfo {pages} {025001} (\bibinfo {year}
  {2018}),\ \Eprint{http://arxiv.org/abs/1610.03894}{arXiv:1610.03894
  [hep-ph]}.%
  \bibAnnoteFile{Stop}{Duarte:2016caz}%
\bibitem{Caputo:2017pit}%
  \BibitemOpen
  \bibfield{author}{%
  \bibinfo {author} {\bibfnamefont{A.}~\bibnamefont{Caputo}}, \bibinfo {author}
  {\bibfnamefont{P.}~\bibnamefont{Hernandez}}, \bibinfo {author}
  {\bibfnamefont{J.}~\bibnamefont{Lopez-Pavon}}\ and\ \bibinfo {author}
  {\bibfnamefont{J.}~\bibnamefont{Salvado}},\ }%
  \emph{\bibinfo {title} {{The seesaw portal in testable models of neutrino
  masses}}},\ \bibfield{journal}{%
  \Doi{10.1007/JHEP06(2017)112}{\bibinfo {journal} {JHEP}}\ }%
  \textbf{\bibinfo {volume} {06}},\ \bibinfo {pages} {112} (\bibinfo {year}
  {2017}),\ \Eprint{http://arxiv.org/abs/1704.08721}{arXiv:1704.08721
  [hep-ph]}.%
  \bibAnnoteFile{Stop}{Caputo:2017pit}%
%%CITATION = ARXIV:1704.08721;%%
\bibitem{Duarte:2018xst}%
  \BibitemOpen
  \bibfield{author}{%
  \bibinfo {author} {\bibfnamefont{L.}~\bibnamefont{Duarte}}, \bibinfo {author}
  {\bibfnamefont{G.}~\bibnamefont{Zapata}}\ and\ \bibinfo {author}
  {\bibfnamefont{O.~A.}\ \bibnamefont{Sampayo}},\ }%
  \emph{\bibinfo {title} {{Angular and polarization trails from effective
  interactions of Majorana neutrinos at the LHeC}}},\ \bibfield{journal}{%
  \Doi{10.1140/epjc/s10052-018-5833-x}{\bibinfo {journal} {Eur. Phys. J.}}\ }%
  \textbf{\bibinfo {volume} {C78}},\ \bibinfo {pages} {352} (\bibinfo {year}
  {2018}),\ \Eprint{http://arxiv.org/abs/1802.07620}{arXiv:1802.07620
  [hep-ph]}.%
  \bibAnnoteFile{Stop}{Duarte:2018xst}%
%%CITATION = ARXIV:1802.07620;%%
\bibitem{Yue:2018hci}%
  \BibitemOpen
  \bibfield{author}{%
  \bibinfo {author} {\bibfnamefont{C.-X.}\ \bibnamefont{Yue}}\ and\ \bibinfo
  {author} {\bibfnamefont{J.-P.}\ \bibnamefont{Chu}},\ }%
  \emph{\bibinfo {title} {{Sterile neutrino and leptonic decays of the
  pseudoscalar mesons}}},\ \bibfield{journal}{%
  \Doi{10.1103/PhysRevD.98.055012}{\bibinfo {journal} {Phys. Rev.}}\ }%
  \textbf{\bibinfo {volume} {D98}},\ \bibinfo {pages} {055012} (\bibinfo {year}
  {2018}),\ \Eprint{http://arxiv.org/abs/1808.09139}{arXiv:1808.09139
  [hep-ph]}.%
  \bibAnnoteFile{Stop}{Yue:2018hci}%
%%CITATION = ARXIV:1808.09139;%%
\bibitem{Duarte:2018kiv}%
  \BibitemOpen
  \bibfield{author}{%
  \bibinfo {author} {\bibfnamefont{L.}~\bibnamefont{Duarte}}, \bibinfo {author}
  {\bibfnamefont{G.}~\bibnamefont{Zapata}}\ and\ \bibinfo {author}
  {\bibfnamefont{O.~A.}\ \bibnamefont{Sampayo}},\ }%
  \emph{\bibinfo {title} {{Final taus and initial state polarization signatures
  from effective interactions of Majorana neutrinos at future $e^{+}e^{-}$
  colliders}}},\ \bibfield{journal}{%
  \Doi{10.1140/epjc/s10052-019-6734-3}{\bibinfo {journal} {Eur. Phys. J.}}\ }%
  \textbf{\bibinfo {volume} {C79}},\ \bibinfo {pages} {240} (\bibinfo {year}
  {2019}),\ \Eprint{http://arxiv.org/abs/1812.01154}{arXiv:1812.01154
  [hep-ph]}.%
  \bibAnnoteFile{Stop}{Duarte:2018kiv}%
%%CITATION = ARXIV:1812.01154;%%
\bibitem{Duarte:2019rzs}%
  \BibitemOpen
  \bibfield{author}{%
  \bibinfo {author} {\bibfnamefont{L.}~\bibnamefont{Duarte}}, \bibinfo {author}
  {\bibfnamefont{J.}~\bibnamefont{Peressutti}}, \bibinfo {author}
  {\bibfnamefont{I.}~\bibnamefont{Romero}}\ and\ \bibinfo {author}
  {\bibfnamefont{O.~A.}\ \bibnamefont{Sampayo}},\ }%
  \emph{\bibinfo {title} {{Majorana neutrinos with effective interactions in B
  decays}}},\ \bibfield{journal}{%
  \Doi{10.1140/epjc/s10052-019-7104-x}{\bibinfo {journal} {Eur. Phys. J.}}\ }%
  \textbf{\bibinfo {volume} {C79}},\ \bibinfo {pages} {593} (\bibinfo {year}
  {2019}),\ \Eprint{http://arxiv.org/abs/1904.07175}{arXiv:1904.07175
  [hep-ph]}.%
  \bibAnnoteFile{Stop}{Duarte:2019rzs}%
%%CITATION = ARXIV:1904.07175;%%
\bibitem{Bischer:2019ttk}%
  \BibitemOpen
  \bibfield{author}{%
  \bibinfo {author} {\bibfnamefont{I.}~\bibnamefont{Bischer}}\ and\ \bibinfo
  {author} {\bibfnamefont{W.}~\bibnamefont{Rodejohann}},\ }%
  \emph{\bibinfo {title} {{General neutrino interactions from an effective
  field theory perspective}}},\ \bibfield{journal}{%
  \Doi{10.1016/j.nuclphysb.2019.114746}{\bibinfo {journal} {Nucl. Phys. B}}\ }%
  \textbf{\bibinfo {volume} {947}},\ \bibinfo {pages} {114746} (\bibinfo {year}
  {2019}),\ \Eprint{http://arxiv.org/abs/1905.08699}{arXiv:1905.08699
  [hep-ph]}.%
  \bibAnnoteFile{Stop}{Bischer:2019ttk}%
\bibitem{Alcaide:2019pnf}%
  \BibitemOpen
  \bibfield{author}{%
  \bibinfo {author} {\bibfnamefont{J.}~\bibnamefont{Alcaide}}, \bibinfo
  {author} {\bibfnamefont{S.}~\bibnamefont{Banerjee}}, \bibinfo {author}
  {\bibfnamefont{M.}~\bibnamefont{Chala}}\ and\ \bibinfo {author}
  {\bibfnamefont{A.}~\bibnamefont{Titov}},\ }%
  \emph{\bibinfo {title} {{Probes of the Standard Model effective field theory
  extended with a right-handed neutrino}}},\ \bibfield{journal}{%
  \Doi{10.1007/JHEP08(2019)031}{\bibinfo {journal} {JHEP}}\ }%
  \textbf{\bibinfo {volume} {08}},\ \bibinfo {pages} {031} (\bibinfo {year}
  {2019}),\ \Eprint{http://arxiv.org/abs/1905.11375}{arXiv:1905.11375
  [hep-ph]}.%
  \bibAnnoteFile{Stop}{Alcaide:2019pnf}%
\bibitem{Butterworth:2019iff}%
  \BibitemOpen
  \bibfield{author}{%
  \bibinfo {author} {\bibfnamefont{J.~M.}\ \bibnamefont{Butterworth}}, \bibinfo
  {author} {\bibfnamefont{M.}~\bibnamefont{Chala}}, \bibinfo {author}
  {\bibfnamefont{C.}~\bibnamefont{Englert}}, \bibinfo {author}
  {\bibfnamefont{M.}~\bibnamefont{Spannowsky}}\ and\ \bibinfo {author}
  {\bibfnamefont{A.}~\bibnamefont{Titov}},\ }%
  \emph{\bibinfo {title} {{Higgs phenomenology as a probe of sterile
  neutrinos}}},\ \bibfield{journal}{%
  \Doi{10.1103/PhysRevD.100.115019}{\bibinfo {journal} {Phys. Rev. D}}\ }%
  \textbf{\bibinfo {volume} {100}},\ \bibinfo {pages} {115019} (\bibinfo {year}
  {2019}),\ \Eprint{http://arxiv.org/abs/1909.04665}{arXiv:1909.04665
  [hep-ph]}.%
  \bibAnnoteFile{Stop}{Butterworth:2019iff}%
\bibitem{Jones-Perez:2019plk}%
  \BibitemOpen
  \bibfield{author}{%
  \bibinfo {author} {\bibfnamefont{J.}~\bibnamefont{Jones-P\'erez}}, \bibinfo
  {author} {\bibfnamefont{J.}~\bibnamefont{Masias}}\ and\ \bibinfo {author}
  {\bibfnamefont{J.~D.}\ \bibnamefont{Ruiz-\'Alvarez}},\ }%
  \emph{\bibinfo {title} {{Search for Long-Lived Heavy Neutrinos at the LHC
  with a VBF Trigger}}},\ \bibfield{journal}{%
  \Doi{10.1140/epjc/s10052-020-8188-z}{\bibinfo {journal} {Eur. Phys. J. C}}\
  }%
  \textbf{\bibinfo {volume} {80}},\ \bibinfo {pages} {642} (\bibinfo {year}
  {2020}),\ \Eprint{http://arxiv.org/abs/1912.08206}{arXiv:1912.08206
  [hep-ph]}.%
  \bibAnnoteFile{Stop}{Jones-Perez:2019plk}%
\bibitem{Chala:2020vqp}%
  \BibitemOpen
  \bibfield{author}{%
  \bibinfo {author} {\bibfnamefont{M.}~\bibnamefont{Chala}}\ and\ \bibinfo
  {author} {\bibfnamefont{A.}~\bibnamefont{Titov}},\ }%
  \emph{\bibinfo {title} {{One-loop matching in the SMEFT extended with a
  sterile neutrino}}},\ \bibfield{journal}{%
  \Doi{10.1007/JHEP05(2020)139}{\bibinfo {journal} {JHEP}}\ }%
  \textbf{\bibinfo {volume} {05}},\ \bibinfo {pages} {139} (\bibinfo {year}
  {2020}),\ \Eprint{http://arxiv.org/abs/2001.07732}{arXiv:2001.07732
  [hep-ph]}.%
  \bibAnnoteFile{Stop}{Chala:2020vqp}%
\bibitem{Dekens:2020ttz}%
  \BibitemOpen
  \bibfield{author}{%
  \bibinfo {author} {\bibfnamefont{W.}~\bibnamefont{Dekens}}, \bibinfo {author}
  {\bibfnamefont{J.}~\bibnamefont{de~Vries}}, \bibinfo {author}
  {\bibfnamefont{K.}~\bibnamefont{Fuyuto}}, \bibinfo {author}
  {\bibfnamefont{E.}~\bibnamefont{Mereghetti}}\ and\ \bibinfo {author}
  {\bibfnamefont{G.}~\bibnamefont{Zhou}},\ }%
  \emph{\bibinfo {title} {{Sterile neutrinos and neutrinoless double beta decay
  in effective field theory}}},\ \bibfield{journal}{%
  \Doi{10.1007/JHEP06(2020)097}{\bibinfo {journal} {JHEP}}\ }%
  \textbf{\bibinfo {volume} {06}},\ \bibinfo {pages} {097} (\bibinfo {year}
  {2020}),\ \Eprint{http://arxiv.org/abs/2002.07182}{arXiv:2002.07182
  [hep-ph]}.%
  \bibAnnoteFile{Stop}{Dekens:2020ttz}%
\bibitem{Barducci:2020ncz}%
  \BibitemOpen
  \bibfield{author}{%
  \bibinfo {author} {\bibfnamefont{D.}~\bibnamefont{Barducci}}, \bibinfo
  {author} {\bibfnamefont{E.}~\bibnamefont{Bertuzzo}}, \bibinfo {author}
  {\bibfnamefont{A.}~\bibnamefont{Caputo}}\ and\ \bibinfo {author}
  {\bibfnamefont{P.}~\bibnamefont{Hernandez}},\ }%
  \emph{\bibinfo {title} {{Minimal flavor violation in the see-saw portal}}},\
  \bibfield{journal}{%
  \Doi{10.1007/JHEP06(2020)185}{\bibinfo {journal} {JHEP}}\ }%
  \textbf{\bibinfo {volume} {06}},\ \bibinfo {pages} {185} (\bibinfo {year}
  {2020}),\ \Eprint{http://arxiv.org/abs/2003.08391}{arXiv:2003.08391
  [hep-ph]}.%
  \bibAnnoteFile{Stop}{Barducci:2020ncz}%
\bibitem{Duarte:2020vgj}%
  \BibitemOpen
  \bibfield{author}{%
  \bibinfo {author} {\bibfnamefont{L.}~\bibnamefont{Duarte}}, \bibinfo {author}
  {\bibfnamefont{G.}~\bibnamefont{Zapata}}\ and\ \bibinfo {author}
  {\bibfnamefont{O.}~\bibnamefont{Sampayo}},\ }%
  \emph{\bibinfo {title} {{Angular and polarization observables for
  Majorana-mediated B decays with effective interactions}}},\
  \bibfield{journal}{%
  \Doi{10.1140/epjc/s10052-020-08471-0}{\bibinfo {journal} {Eur. Phys. J. C}}\
  }%
  \textbf{\bibinfo {volume} {80}},\ \bibinfo {pages} {896} (\bibinfo {year}
  {2020}),\ \Eprint{http://arxiv.org/abs/2006.11216}{arXiv:2006.11216
  [hep-ph]}.%
  \bibAnnoteFile{Stop}{Duarte:2020vgj}%
\bibitem{Biekotter:2020tbd}%
  \BibitemOpen
  \bibfield{author}{%
  \bibinfo {author} {\bibfnamefont{A.}~\bibnamefont{Biek\"otter}}, \bibinfo
  {author} {\bibfnamefont{M.}~\bibnamefont{Chala}}\ and\ \bibinfo {author}
  {\bibfnamefont{M.}~\bibnamefont{Spannowsky}},\ }%
  \emph{\bibinfo {title} {{The effective field theory of low scale see-saw at
  colliders}}},\ \bibfield{journal}{%
  \Doi{10.1140/s10052-020-8339-2}{\bibinfo {journal} {Eur. Phys. J. C}}\ }%
  \textbf{\bibinfo {volume} {80}},\ \bibinfo {pages} {743} (\bibinfo {year}
  {2020}),\ \Eprint{http://arxiv.org/abs/2007.00673}{arXiv:2007.00673
  [hep-ph]}.%
  \bibAnnoteFile{Stop}{Biekotter:2020tbd}%
\bibitem{DeVries:2020jbs}%
  \BibitemOpen
  \bibfield{author}{%
  \bibinfo {author} {\bibfnamefont{J.}~\bibnamefont{De~Vries}}, \bibinfo
  {author} {\bibfnamefont{H.~K.}\ \bibnamefont{Dreiner}}, \bibinfo {author}
  {\bibfnamefont{J.~Y.}\ \bibnamefont{G\"unther}}, \bibinfo {author}
  {\bibfnamefont{Z.~S.}\ \bibnamefont{Wang}}\ and\ \bibinfo {author}
  {\bibfnamefont{G.}~\bibnamefont{Zhou}},\ }%
  \emph{\bibinfo {title} {{Long-lived Sterile Neutrinos at the LHC in Effective
  Field Theory}}},\ \bibfield{journal}{%
  \Doi{10.1007/JHEP03(2021)148}{\bibinfo {journal} {JHEP}}\ }%
  \textbf{\bibinfo {volume} {03}},\ \bibinfo {pages} {148} (\bibinfo {year}
  {2021}),\ \Eprint{http://arxiv.org/abs/2010.07305}{arXiv:2010.07305
  [hep-ph]}.%
  \bibAnnoteFile{Stop}{DeVries:2020jbs}%
\bibitem{Barducci:2020icf}%
  \BibitemOpen
  \bibfield{author}{%
  \bibinfo {author} {\bibfnamefont{D.}~\bibnamefont{Barducci}}, \bibinfo
  {author} {\bibfnamefont{E.}~\bibnamefont{Bertuzzo}}, \bibinfo {author}
  {\bibfnamefont{A.}~\bibnamefont{Caputo}}, \bibinfo {author}
  {\bibfnamefont{P.}~\bibnamefont{Hernandez}}\ and\ \bibinfo {author}
  {\bibfnamefont{B.}~\bibnamefont{Mele}},\ }%
  \emph{\bibinfo {title} {{The see-saw portal at future Higgs Factories}}},\
  \bibfield{journal}{%
  \Doi{10.1007/JHEP03(2021)117}{\bibinfo {journal} {JHEP}}\ }%
  \textbf{\bibinfo {volume} {03}},\ \bibinfo {pages} {117} (\bibinfo {year}
  {2021}),\ \Eprint{http://arxiv.org/abs/2011.04725}{arXiv:2011.04725
  [hep-ph]}.%
  \bibAnnoteFile{Stop}{Barducci:2020icf}%
\bibitem{Dekens:2021qch}%
  \BibitemOpen
  \bibfield{author}{%
  \bibinfo {author} {\bibfnamefont{W.}~\bibnamefont{Dekens}}, \bibinfo {author}
  {\bibfnamefont{J.}~\bibnamefont{de~Vries}}\ and\ \bibinfo {author}
  {\bibfnamefont{T.}~\bibnamefont{Tong}},\ }%
  \emph{\bibinfo {title} {{Sterile neutrinos with non-standard interactions in
  \ensuremath{\beta}- and
  0\ensuremath{\nu}\ensuremath{\beta}\ensuremath{\beta}-decay experiments}}},\
  \bibfield{journal}{%
  \Doi{10.1007/JHEP08(2021)128}{\bibinfo {journal} {JHEP}}\ }%
  \textbf{\bibinfo {volume} {08}},\ \bibinfo {pages} {128} (\bibinfo {year}
  {2021}),\ \Eprint{http://arxiv.org/abs/2104.00140}{arXiv:2104.00140
  [hep-ph]}.%
  \bibAnnoteFile{Stop}{Dekens:2021qch}%
\bibitem{Cirigliano:2021peb}%
  \BibitemOpen
  \bibfield{author}{%
  \bibinfo {author} {\bibfnamefont{V.}~\bibnamefont{Cirigliano}}, \bibinfo
  {author} {\bibfnamefont{W.}~\bibnamefont{Dekens}}, \bibinfo {author}
  {\bibfnamefont{J.}~\bibnamefont{de~Vries}}, \bibinfo {author}
  {\bibfnamefont{K.}~\bibnamefont{Fuyuto}}, \bibinfo {author}
  {\bibfnamefont{E.}~\bibnamefont{Mereghetti}}\ and\ \bibinfo {author}
  {\bibfnamefont{R.}~\bibnamefont{Ruiz}},\ }%
  \emph{\bibinfo {title} {{Leptonic anomalous magnetic moments in
  \ensuremath{\nu} SMEFT}}},\ \bibfield{journal}{%
  \Doi{10.1007/JHEP08(2021)103}{\bibinfo {journal} {JHEP}}\ }%
  \textbf{\bibinfo {volume} {08}},\ \bibinfo {pages} {103} (\bibinfo {year}
  {2021}),\ \Eprint{http://arxiv.org/abs/2105.11462}{arXiv:2105.11462
  [hep-ph]}.%
  \bibAnnoteFile{Stop}{Cirigliano:2021peb}%
\bibitem{Cottin:2021lzz}%
  \BibitemOpen
  \bibfield{author}{%
  \bibinfo {author} {\bibfnamefont{G.}~\bibnamefont{Cottin}}, \bibinfo {author}
  {\bibfnamefont{J.~C.}\ \bibnamefont{Helo}}, \bibinfo {author}
  {\bibfnamefont{M.}~\bibnamefont{Hirsch}}, \bibinfo {author}
  {\bibfnamefont{A.}~\bibnamefont{Titov}}\ and\ \bibinfo {author}
  {\bibfnamefont{Z.~S.}\ \bibnamefont{Wang}},\ }%
  \emph{\bibinfo {title} {{Heavy neutral leptons in effective field theory and
  the high-luminosity LHC}}},\ \bibfield{journal}{%
  \Doi{10.1007/JHEP09(2021)039}{\bibinfo {journal} {JHEP}}\ }%
  \textbf{\bibinfo {volume} {09}},\ \bibinfo {pages} {039} (\bibinfo {year}
  {2021}),\ \Eprint{http://arxiv.org/abs/2105.13851}{arXiv:2105.13851
  [hep-ph]}.%
  \bibAnnoteFile{Stop}{Cottin:2021lzz}%
\bibitem{Beltran:2021hpq}%
  \BibitemOpen
  \bibfield{author}{%
  \bibinfo {author} {\bibfnamefont{R.}~\bibnamefont{Beltr\'an}}, \bibinfo
  {author} {\bibfnamefont{G.}~\bibnamefont{Cottin}}, \bibinfo {author}
  {\bibfnamefont{J.~C.}\ \bibnamefont{Helo}}, \bibinfo {author}
  {\bibfnamefont{M.}~\bibnamefont{Hirsch}}, \bibinfo {author}
  {\bibfnamefont{A.}~\bibnamefont{Titov}}\ and\ \bibinfo {author}
  {\bibfnamefont{Z.~S.}\ \bibnamefont{Wang}},\ }%
  \emph{\bibinfo {title} {{Long-lived heavy neutral leptons at the LHC:
  four-fermion single-N$_{R}$ operators}}},\ \bibfield{journal}{%
  \Doi{10.1007/JHEP01(2022)044}{\bibinfo {journal} {JHEP}}\ }%
  \textbf{\bibinfo {volume} {01}},\ \bibinfo {pages} {044} (\bibinfo {year}
  {2022}),\ \Eprint{http://arxiv.org/abs/2110.15096}{arXiv:2110.15096
  [hep-ph]}.%
  \bibAnnoteFile{Stop}{Beltran:2021hpq}%
\bibitem{Zhou:2021ylt}%
  \BibitemOpen
  \bibfield{author}{%
  \bibinfo {author} {\bibfnamefont{G.}~\bibnamefont{Zhou}}, \bibinfo {author}
  {\bibfnamefont{J.~Y.}\ \bibnamefont{G\"unther}}, \bibinfo {author}
  {\bibfnamefont{Z.~S.}\ \bibnamefont{Wang}}, \bibinfo {author}
  {\bibfnamefont{J.}~\bibnamefont{de~Vries}}\ and\ \bibinfo {author}
  {\bibfnamefont{H.~K.}\ \bibnamefont{Dreiner}},\ }%
  \emph{\bibinfo {title} {{Long-lived sterile neutrinos at Belle II in
  effective field theory}}},\ \bibfield{journal}{%
  \Doi{10.1007/JHEP04(2022)057}{\bibinfo {journal} {JHEP}}\ }%
  \textbf{\bibinfo {volume} {04}},\ \bibinfo {pages} {057} (\bibinfo {year}
  {2022}),\ \Eprint{http://arxiv.org/abs/2111.04403}{arXiv:2111.04403
  [hep-ph]}.%
  \bibAnnoteFile{Stop}{Zhou:2021ylt}%
\bibitem{Zhou:2021lnl}%
  \BibitemOpen
  \bibfield{author}{%
  \bibinfo {author} {\bibfnamefont{G.}~\bibnamefont{Zhou}},\ }%
  \emph{\bibinfo {title} {{Light sterile neutrinos and lepton-number-violating
  kaon decays in effective field theory}}},\ \bibfield{journal}{%
  \Doi{10.1007/JHEP06(2022)127}{\bibinfo {journal} {JHEP}}\ }%
  \textbf{\bibinfo {volume} {06}},\ \bibinfo {pages} {127} (\bibinfo {year}
  {2022}),\ \Eprint{http://arxiv.org/abs/2112.00767}{arXiv:2112.00767
  [hep-ph]}.%
  \bibAnnoteFile{Stop}{Zhou:2021lnl}%
\bibitem{Beltran:2022ast}%
  \BibitemOpen
  \bibfield{author}{%
  \bibinfo {author} {\bibfnamefont{R.}~\bibnamefont{Beltr\'an}}, \bibinfo
  {author} {\bibfnamefont{G.}~\bibnamefont{Cottin}}, \bibinfo {author}
  {\bibfnamefont{J.~C.}\ \bibnamefont{Helo}}, \bibinfo {author}
  {\bibfnamefont{M.}~\bibnamefont{Hirsch}}, \bibinfo {author}
  {\bibfnamefont{A.}~\bibnamefont{Titov}}\ and\ \bibinfo {author}
  {\bibfnamefont{Z.~S.}\ \bibnamefont{Wang}},\ }%
  \emph{\bibinfo {title} {{Long-lived heavy neutral leptons from mesons in
  effective field theory}}},\ \bibfield{journal}{%
  \Doi{10.1007/JHEP01(2023)015}{\bibinfo {journal} {JHEP}}\ }%
  \textbf{\bibinfo {volume} {01}},\ \bibinfo {pages} {015} (\bibinfo {year}
  {2023}),\ \Eprint{http://arxiv.org/abs/2210.02461}{arXiv:2210.02461
  [hep-ph]}.%
  \bibAnnoteFile{Stop}{Beltran:2022ast}%
\bibitem{Delgado:2022fea}%
  \BibitemOpen
  \bibfield{author}{%
  \bibinfo {author} {\bibfnamefont{F.}~\bibnamefont{Delgado}}, \bibinfo
  {author} {\bibfnamefont{L.}~\bibnamefont{Duarte}}, \bibinfo {author}
  {\bibfnamefont{J.}~\bibnamefont{Jones-Perez}}, \bibinfo {author}
  {\bibfnamefont{C.}~\bibnamefont{Manrique-Chavil}}\ and\ \bibinfo {author}
  {\bibfnamefont{S.}~\bibnamefont{Pe\~na}},\ }%
  \emph{\bibinfo {title} {{Assessment of the dimension-5 seesaw portal and
  impact of exotic Higgs decays on non-pointing photon searches}}},\
  \bibfield{journal}{%
  \Doi{10.1007/JHEP09(2022)079}{\bibinfo {journal} {JHEP}}\ }%
  \textbf{\bibinfo {volume} {09}},\ \bibinfo {pages} {079} (\bibinfo {year}
  {2022}),\ \Eprint{http://arxiv.org/abs/2205.13550}{arXiv:2205.13550
  [hep-ph]}.%
  \bibAnnoteFile{Stop}{Delgado:2022fea}%
\bibitem{Barducci:2022gdv}%
  \BibitemOpen
  \bibfield{author}{%
  \bibinfo {author} {\bibfnamefont{D.}~\bibnamefont{Barducci}}, \bibinfo
  {author} {\bibfnamefont{E.}~\bibnamefont{Bertuzzo}}, \bibinfo {author}
  {\bibfnamefont{M.}~\bibnamefont{Taoso}}\ and\ \bibinfo {author}
  {\bibfnamefont{C.}~\bibnamefont{Toni}},\ }%
  \emph{\bibinfo {title} {{Probing right-handed neutrinos dipole operators}}},\
  \bibfield{journal}{%
  \Doi{10.1007/JHEP03(2023)239}{\bibinfo {journal} {JHEP}}\ }%
  \textbf{\bibinfo {volume} {03}},\ \bibinfo {pages} {239} (\bibinfo {year}
  {2023}),\ \Eprint{http://arxiv.org/abs/2209.13469}{arXiv:2209.13469
  [hep-ph]}.%
  \bibAnnoteFile{Stop}{Barducci:2022gdv}%
\bibitem{Zapata:2022qwo}%
  \BibitemOpen
  \bibfield{author}{%
  \bibinfo {author} {\bibfnamefont{G.}~\bibnamefont{Zapata}}, \bibinfo {author}
  {\bibfnamefont{T.}~\bibnamefont{Urruzola}}, \bibinfo {author}
  {\bibfnamefont{O.~A.}\ \bibnamefont{Sampayo}}\ and\ \bibinfo {author}
  {\bibfnamefont{L.}~\bibnamefont{Duarte}},\ }%
  \emph{\bibinfo {title} {{Lepton collider probes for Majorana neutrino
  effective interactions}}},\ \bibfield{journal}{%
  \Doi{10.1140/epjc/s10052-022-10448-0}{\bibinfo {journal} {Eur. Phys. J. C}}\
  }%
  \textbf{\bibinfo {volume} {82}},\ \bibinfo {pages} {544} (\bibinfo {year}
  {2022}),\ \Eprint{http://arxiv.org/abs/2201.02480}{arXiv:2201.02480
  [hep-ph]}.%
  \bibAnnoteFile{Stop}{Zapata:2022qwo}%
\bibitem{Barducci:2022hll}%
  \BibitemOpen
  \bibfield{author}{%
  \bibinfo {author} {\bibfnamefont{D.}~\bibnamefont{Barducci}}\ and\ \bibinfo
  {author} {\bibfnamefont{E.}~\bibnamefont{Bertuzzo}},\ }%
  \emph{\bibinfo {title} {{The see-saw portal at future Higgs factories: the
  role of dimension six operators}}},\ \bibfield{journal}{%
  \Doi{10.1007/JHEP06(2022)077}{\bibinfo {journal} {JHEP}}\ }%
  \textbf{\bibinfo {volume} {06}},\ \bibinfo {pages} {077} (\bibinfo {year}
  {2022}),\ \Eprint{http://arxiv.org/abs/2201.11754}{arXiv:2201.11754
  [hep-ph]}.%
  \bibAnnoteFile{Stop}{Barducci:2022hll}%
\bibitem{Talbert:2022unj}%
  \BibitemOpen
  \bibfield{author}{%
  \bibinfo {author} {\bibfnamefont{J.}~\bibnamefont{Talbert}},\ }%
  \emph{\bibinfo {title} {{The geometric \ensuremath{\nu}SMEFT: operators and
  connections}}},\ \bibfield{journal}{%
  \Doi{10.1007/JHEP01(2023)069}{\bibinfo {journal} {JHEP}}\ }%
  \textbf{\bibinfo {volume} {01}},\ \bibinfo {pages} {069} (\bibinfo {year}
  {2023}),\ \Eprint{http://arxiv.org/abs/2208.11139}{arXiv:2208.11139
  [hep-ph]}.%
  \bibAnnoteFile{Stop}{Talbert:2022unj}%
\bibitem{Mitra:2022nri}%
  \BibitemOpen
  \bibfield{author}{%
  \bibinfo {author} {\bibfnamefont{M.}~\bibnamefont{Mitra}}, \bibinfo {author}
  {\bibfnamefont{S.}~\bibnamefont{Mandal}}, \bibinfo {author}
  {\bibfnamefont{R.}~\bibnamefont{Padhan}}, \bibinfo {author}
  {\bibfnamefont{A.}~\bibnamefont{Sarkar}}\ and\ \bibinfo {author}
  {\bibfnamefont{M.}~\bibnamefont{Spannowsky}},\ }%
  \emph{\bibinfo {title} {{Reexamining right-handed neutrino EFTs up to
  dimension six}}},\ \bibfield{journal}{%
  \Doi{10.1103/PhysRevD.106.113008}{\bibinfo {journal} {Phys. Rev. D}}\ }%
  \textbf{\bibinfo {volume} {106}},\ \bibinfo {pages} {113008} (\bibinfo {year}
  {2022}),\ \Eprint{http://arxiv.org/abs/2210.12404}{arXiv:2210.12404
  [hep-ph]}.%
  \bibAnnoteFile{Stop}{Mitra:2022nri}%
\bibitem{Beltran:2023nli}%
  \BibitemOpen
  \bibfield{author}{%
  \bibinfo {author} {\bibfnamefont{R.}~\bibnamefont{Beltr\'an}}, \bibinfo
  {author} {\bibfnamefont{G.}~\bibnamefont{Cottin}}, \bibinfo {author}
  {\bibfnamefont{M.}~\bibnamefont{Hirsch}}, \bibinfo {author}
  {\bibfnamefont{A.}~\bibnamefont{Titov}}\ and\ \bibinfo {author}
  {\bibfnamefont{Z.~S.}\ \bibnamefont{Wang}},\ }%
  \emph{\bibinfo {title} {{Reinterpretation of searches for long-lived
  particles from meson decays}}},\ \bibfield{journal}{%
  \Doi{10.1007/JHEP05(2023)031}{\bibinfo {journal} {JHEP}}\ }%
  \textbf{\bibinfo {volume} {05}},\ \bibinfo {pages} {031} (\bibinfo {year}
  {2023}),\ \Eprint{http://arxiv.org/abs/2302.03216}{arXiv:2302.03216
  [hep-ph]}.%
  \bibAnnoteFile{Stop}{Beltran:2023nli}%
\bibitem{Fernandez-Martinez:2023phj}%
  \BibitemOpen
  \bibfield{author}{%
  \bibinfo {author}
  {\bibfnamefont{E.}~\bibnamefont{Fern\'andez-Mart\'\i{}nez}}, \bibinfo
  {author} {\bibfnamefont{M.}~\bibnamefont{Gonz\'alez-L\'opez}}, \bibinfo
  {author} {\bibfnamefont{J.}~\bibnamefont{Hern\'andez-Garc\'\i{}a}}, \bibinfo
  {author} {\bibfnamefont{M.}~\bibnamefont{Hostert}}\ and\ \bibinfo {author}
  {\bibfnamefont{J.}~\bibnamefont{L\'opez-Pav\'on}},\ }%
  \emph{\bibinfo {title} {{Effective portals to heavy neutral leptons}}},\
  \bibfield{journal}{%
  \Doi{10.1007/JHEP09(2023)001}{\bibinfo {journal} {JHEP}}\ }%
  \textbf{\bibinfo {volume} {09}},\ \bibinfo {pages} {001} (\bibinfo {year}
  {2023}),\ \Eprint{http://arxiv.org/abs/2304.06772}{arXiv:2304.06772
  [hep-ph]}.%
  \bibAnnoteFile{Stop}{Fernandez-Martinez:2023phj}%
\bibitem{Beltran:2023ymm}%
  \BibitemOpen
  \bibfield{author}{%
  \bibinfo {author} {\bibfnamefont{R.}~\bibnamefont{Beltr\'an}}, \bibinfo
  {author} {\bibfnamefont{R.}~\bibnamefont{Cepedello}}\ and\ \bibinfo {author}
  {\bibfnamefont{M.}~\bibnamefont{Hirsch}},\ }%
  \emph{\bibinfo {title} {{Tree-level UV completions for N$_{R}$SMEFT d = 6 and
  d = 7 operators}}},\ \bibfield{journal}{%
  \Doi{10.1007/JHEP08(2023)166}{\bibinfo {journal} {JHEP}}\ }%
  \textbf{\bibinfo {volume} {08}},\ \bibinfo {pages} {166} (\bibinfo {year}
  {2023}),\ \Eprint{http://arxiv.org/abs/2306.12578}{arXiv:2306.12578
  [hep-ph]}.%
  \bibAnnoteFile{Stop}{Beltran:2023ymm}%
\bibitem{Beltran:2023ksw}%
  \BibitemOpen
  \bibfield{author}{%
  \bibinfo {author} {\bibfnamefont{R.}~\bibnamefont{Beltr\'an}}, \bibinfo
  {author} {\bibfnamefont{J.}~\bibnamefont{G\"unther}}, \bibinfo {author}
  {\bibfnamefont{M.}~\bibnamefont{Hirsch}}, \bibinfo {author}
  {\bibfnamefont{A.}~\bibnamefont{Titov}}\ and\ \bibinfo {author}
  {\bibfnamefont{Z.~S.}\ \bibnamefont{Wang}},\ }%
  \emph{\bibinfo {title} {{Heavy neutral leptons from kaons in effective field
  theory}}} (\bibinfo {month} {9}\ \bibinfo {year} {2023}),\
  \Eprint{http://arxiv.org/abs/2309.11546}{arXiv:2309.11546 [hep-ph]}.%
  \bibAnnoteFile{Stop}{Beltran:2023ksw}%
\bibitem{Duarte:2023tdw}%
  \BibitemOpen
  \bibfield{author}{%
  \bibinfo {author} {\bibfnamefont{L.}~\bibnamefont{Duarte}}, \bibinfo {author}
  {\bibfnamefont{J.}~\bibnamefont{Jones-P\'erez}}\ and\ \bibinfo {author}
  {\bibfnamefont{C.}~\bibnamefont{Manrique-Chavil}},\ }%
  \emph{\bibinfo {title} {{Bounding the Dimension-5 Seesaw Portal with
  Non-Pointing Photon Searches}}} (\bibinfo {month} {11}\ \bibinfo {year}
  {2023}),\ \Eprint{http://arxiv.org/abs/2311.17989}{arXiv:2311.17989
  [hep-ph]}.%
  \bibAnnoteFile{Stop}{Duarte:2023tdw}%
\bibitem{LHeC:2020van}%
  \BibitemOpen
  \bibfield{author}{%
  \bibinfo {author} {\bibfnamefont{P.}~\bibnamefont{Agostini}} \emph{et~al.}
  (\bibinfo {collaboration} {LHeC, FCC-he Study Group}),\ }%
  \emph{\bibinfo {title} {{The Large Hadron\textendash{}Electron Collider at
  the HL-LHC}}},\ \bibfield{journal}{%
  \Doi{10.1088/1361-6471/abf3ba}{\bibinfo {journal} {J. Phys. G}}\ }%
  \textbf{\bibinfo {volume} {48}},\ \bibinfo {pages} {110501} (\bibinfo {year}
  {2021}),\ \Eprint{http://arxiv.org/abs/2007.14491}{arXiv:2007.14491
  [hep-ex]}.%
  \bibAnnoteFile{Stop}{LHeC:2020van}%
\bibitem{AbelleiraFernandez:2012cc}%
  \BibitemOpen
  \bibfield{author}{%
  \bibinfo {author} {\bibfnamefont{J.}~\bibnamefont{Abelleira~Fernandez}}
  \emph{et~al.} (\bibinfo {collaboration} {LHeC Study Group}),\ }%
  \emph{\bibinfo {title} {{A Large Hadron Electron Collider at CERN: Report on
  the Physics and Design Concepts for Machine and Detector}}},\
  \bibfield{journal}{%
  \Doi{10.1088/0954-3899/39/7/075001}{\bibinfo {journal} {J.Phys.}}\ }%
  \textbf{\bibinfo {volume} {G39}},\ \bibinfo {pages} {075001} (\bibinfo {year}
  {2012}),\ \Eprint{http://arxiv.org/abs/1206.2913}{arXiv:1206.2913
  [physics.acc-ph]}.%
  \bibAnnoteFile{Stop}{AbelleiraFernandez:2012cc}%
%%CITATION = ARXIV:1206.2913;%%
\bibitem{Bruening:2013bga}%
  \BibitemOpen
  \bibfield{author}{%
  \bibinfo {author} {\bibfnamefont{O.}~\bibnamefont{Bruening}}\ and\ \bibinfo
  {author} {\bibfnamefont{M.}~\bibnamefont{Klein}},\ }%
  \emph{\bibinfo {title} {{The Large Hadron Electron Collider}}},\
  \bibfield{journal}{%
  \Doi{10.1142/S0217732313300115}{\bibinfo {journal} {Mod.Phys.Lett.}}\ }%
  \textbf{\bibinfo {volume} {A28}},\ \bibinfo {pages} {1330011} (\bibinfo
  {year} {2013}),\ \Eprint{http://arxiv.org/abs/1305.2090}{arXiv:1305.2090
  [physics.acc-ph]}.%
  \bibAnnoteFile{Stop}{Bruening:2013bga}%
%%CITATION = ARXIV:1305.2090;%%
\bibitem{Li:2018wut}%
  \BibitemOpen
  \bibfield{author}{%
  \bibinfo {author} {\bibfnamefont{S.-Y.}\ \bibnamefont{Li}}, \bibinfo {author}
  {\bibfnamefont{Z.-G.}\ \bibnamefont{Si}}\ and\ \bibinfo {author}
  {\bibfnamefont{X.-H.}\ \bibnamefont{Yang}},\ }%
  \emph{\bibinfo {title} {{Heavy Majorana Neutrino Production at Future $ep$
  Colliders}}},\ \bibfield{journal}{%
  \Doi{10.1016/j.physletb.2019.06.001}{\bibinfo {journal} {Phys. Lett. B}}\ }%
  \textbf{\bibinfo {volume} {795}},\ \bibinfo {pages} {49} (\bibinfo {year}
  {2019}),\ \Eprint{http://arxiv.org/abs/1811.10313}{arXiv:1811.10313
  [hep-ph]}.%
  \bibAnnoteFile{Stop}{Li:2018wut}%
\bibitem{Antusch:2019eiz}%
  \BibitemOpen
  \bibfield{author}{%
  \bibinfo {author} {\bibfnamefont{S.}~\bibnamefont{Antusch}}, \bibinfo
  {author} {\bibfnamefont{O.}~\bibnamefont{Fischer}}\ and\ \bibinfo {author}
  {\bibfnamefont{A.}~\bibnamefont{Hammad}},\ }%
  \emph{\bibinfo {title} {{Lepton-Trijet and Displaced Vertex Searches for
  Heavy Neutrinos at Future Electron-Proton Colliders}}},\ \bibfield{journal}{%
  \Doi{10.1007/JHEP03(2020)110}{\bibinfo {journal} {JHEP}}\ }%
  \textbf{\bibinfo {volume} {03}},\ \bibinfo {pages} {110} (\bibinfo {year}
  {2020}),\ \Eprint{http://arxiv.org/abs/1908.02852}{arXiv:1908.02852
  [hep-ph]}.%
  \bibAnnoteFile{Stop}{Antusch:2019eiz}%
\bibitem{Gu:2022muc}%
  \BibitemOpen
  \bibfield{author}{%
  \bibinfo {author} {\bibfnamefont{H.}~\bibnamefont{Gu}}\ and\ \bibinfo
  {author} {\bibfnamefont{K.}~\bibnamefont{Wang}},\ }%
  \emph{\bibinfo {title} {{Search for heavy Majorana neutrinos at
  electron-proton colliders}}},\ \bibfield{journal}{%
  \Doi{10.1103/PhysRevD.106.015006}{\bibinfo {journal} {Phys. Rev. D}}\ }%
  \textbf{\bibinfo {volume} {106}},\ \bibinfo {pages} {015006} (\bibinfo {year}
  {2022}),\ \Eprint{http://arxiv.org/abs/2201.12997}{arXiv:2201.12997
  [hep-ph]}.%
  \bibAnnoteFile{Stop}{Gu:2022muc}%
\bibitem{Antusch:2016ejd}%
  \BibitemOpen
  \bibfield{author}{%
  \bibinfo {author} {\bibfnamefont{S.}~\bibnamefont{Antusch}}, \bibinfo
  {author} {\bibfnamefont{E.}~\bibnamefont{Cazzato}}\ and\ \bibinfo {author}
  {\bibfnamefont{O.}~\bibnamefont{Fischer}},\ }%
  \emph{\bibinfo {title} {{Sterile neutrino searches at future $e^-e^+$, $pp$,
  and $e^-p$ colliders}}},\ \bibfield{journal}{%
  \Doi{10.1142/S0217751X17500786}{\bibinfo {journal} {Int. J. Mod. Phys.}}\ }%
  \textbf{\bibinfo {volume} {A32}},\ \bibinfo {pages} {1750078} (\bibinfo
  {year} {2017}),\ \Eprint{http://arxiv.org/abs/1612.02728}{arXiv:1612.02728
  [hep-ph]}.%
  \bibAnnoteFile{Stop}{Antusch:2016ejd}%
%%CITATION = ARXIV:1612.02728;%%
\bibitem{Blaksley:2011ey}%
  \BibitemOpen
  \bibfield{author}{%
  \bibinfo {author} {\bibfnamefont{C.}~\bibnamefont{Blaksley}}, \bibinfo
  {author} {\bibfnamefont{M.}~\bibnamefont{Blennow}}, \bibinfo {author}
  {\bibfnamefont{F.}~\bibnamefont{Bonnet}}, \bibinfo {author}
  {\bibfnamefont{P.}~\bibnamefont{Coloma}}\ and\ \bibinfo {author}
  {\bibfnamefont{E.}~\bibnamefont{Fernandez-Martinez}},\ }%
  \emph{\bibinfo {title} {{Heavy Neutrinos and Lepton Number Violation in lp
  Colliders}}},\ \bibfield{journal}{%
  \Doi{10.1016/j.nuclphysb.2011.06.021}{\bibinfo {journal} {Nucl.Phys.}}\ }%
  \textbf{\bibinfo {volume} {B852}},\ \bibinfo {pages} {353} (\bibinfo {year}
  {2011}),\ \Eprint{http://arxiv.org/abs/1105.0308}{arXiv:1105.0308 [hep-ph]}.%
  \bibAnnoteFile{Stop}{Blaksley:2011ey}%
%%CITATION = ARXIV:1105.0308;%%
\bibitem{Liang:2010gm}%
  \BibitemOpen
  \bibfield{author}{%
  \bibinfo {author} {\bibfnamefont{H.}~\bibnamefont{Liang}}, \bibinfo {author}
  {\bibfnamefont{X.-G.}\ \bibnamefont{He}}, \bibinfo {author}
  {\bibfnamefont{W.-G.}\ \bibnamefont{Ma}}, \bibinfo {author}
  {\bibfnamefont{S.-M.}\ \bibnamefont{Wang}}\ and\ \bibinfo {author}
  {\bibfnamefont{R.-Y.}\ \bibnamefont{Zhang}},\ }%
  \emph{\bibinfo {title} {{Seesaw Type I and III at the LHeC}}},\
  \bibfield{journal}{%
  \Doi{10.1007/JHEP09(2010)023}{\bibinfo {journal} {JHEP}}\ }%
  \textbf{\bibinfo {volume} {1009}},\ \bibinfo {pages} {023} (\bibinfo {year}
  {2010}),\ \Eprint{http://arxiv.org/abs/1006.5534}{arXiv:1006.5534 [hep-ph]}.%
  \bibAnnoteFile{Stop}{Liang:2010gm}%
%%CITATION = ARXIV:1006.5534;%%
\bibitem{Ingelman:1993ve}%
  \BibitemOpen
  \bibfield{author}{%
  \bibinfo {author} {\bibfnamefont{G.}~\bibnamefont{Ingelman}}\ and\ \bibinfo
  {author} {\bibfnamefont{J.}~\bibnamefont{Rathsman}},\ }%
  \emph{\bibinfo {title} {{Heavy Majorana neutrinos at e p colliders}}},\
  \bibfield{journal}{%
  \Doi{10.1007/BF01474620}{\bibinfo {journal} {Z.Phys.}}\ }%
  \textbf{\bibinfo {volume} {C60}},\ \bibinfo {pages} {243} (\bibinfo {year}
  {1993}).%
  \bibAnnoteFile{Stop}{Ingelman:1993ve}%
%%CITATION = ZEPYA,C60,243;%%
\bibitem{Buchmuller:1991tu}%
  \BibitemOpen
  \bibfield{author}{%
  \bibinfo {author} {\bibfnamefont{W.}~\bibnamefont{Buchmuller}}\ and\ \bibinfo
  {author} {\bibfnamefont{C.}~\bibnamefont{Greub}},\ }%
  \emph{\bibinfo {title} {{Heavy Majorana neutrinos in electron - positron and
  electron - proton collisions}}},\ \bibfield{journal}{%
  \Doi{10.1016/0550-3213(91)80024-G}{\bibinfo {journal} {Nucl.Phys.}}\ }%
  \textbf{\bibinfo {volume} {B363}},\ \bibinfo {pages} {345} (\bibinfo {year}
  {1991}).%
  \bibAnnoteFile{Stop}{Buchmuller:1991tu}%
%%CITATION = NUPHA,B363,345;%%
\bibitem{Batell:2022ogj}%
  \BibitemOpen
  \bibfield{author}{%
  \bibinfo {author} {\bibfnamefont{B.}~\bibnamefont{Batell}}, \bibinfo {author}
  {\bibfnamefont{T.}~\bibnamefont{Ghosh}}, \bibinfo {author}
  {\bibfnamefont{T.}~\bibnamefont{Han}}\ and\ \bibinfo {author}
  {\bibfnamefont{K.}~\bibnamefont{Xie}},\ }%
  \emph{\bibinfo {title} {{Heavy neutral leptons at the Electron-Ion
  Collider}}},\ \bibfield{journal}{%
  \Doi{10.1007/JHEP03(2023)020}{\bibinfo {journal} {JHEP}}\ }%
  \textbf{\bibinfo {volume} {03}},\ \bibinfo {pages} {020} (\bibinfo {year}
  {2023}),\ \Eprint{http://arxiv.org/abs/2210.09287}{arXiv:2210.09287
  [hep-ph]}.%
  \bibAnnoteFile{Stop}{Batell:2022ogj}%
\bibitem{Cottin:2021tfo}%
  \BibitemOpen
  \bibfield{author}{%
  \bibinfo {author} {\bibfnamefont{G.}~\bibnamefont{Cottin}}, \bibinfo {author}
  {\bibfnamefont{O.}~\bibnamefont{Fischer}}, \bibinfo {author}
  {\bibfnamefont{S.}~\bibnamefont{Mandal}}, \bibinfo {author}
  {\bibfnamefont{M.}~\bibnamefont{Mitra}}\ and\ \bibinfo {author}
  {\bibfnamefont{R.}~\bibnamefont{Padhan}},\ }%
  \emph{\bibinfo {title} {{Displaced neutrino jets at the LHeC}}},\
  \bibfield{journal}{%
  \Doi{10.1007/JHEP06(2022)168}{\bibinfo {journal} {JHEP}}\ }%
  \textbf{\bibinfo {volume} {06}},\ \bibinfo {pages} {168} (\bibinfo {year}
  {2022}),\ \Eprint{http://arxiv.org/abs/2104.13578}{arXiv:2104.13578
  [hep-ph]}.%
  \bibAnnoteFile{Stop}{Cottin:2021tfo}%
\bibitem{Antusch:2020vul}%
  \BibitemOpen
  \bibfield{author}{%
  \bibinfo {author} {\bibfnamefont{S.}~\bibnamefont{Antusch}}, \bibinfo
  {author} {\bibfnamefont{A.}~\bibnamefont{Hammad}}\ and\ \bibinfo {author}
  {\bibfnamefont{A.}~\bibnamefont{Rashed}},\ }%
  \emph{\bibinfo {title} {{Searching for charged lepton flavor violation at
  $ep$ colliders}}},\ \bibfield{journal}{%
  \Doi{10.1007/JHEP03(2021)230}{\bibinfo {journal} {JHEP}}\ }%
  \textbf{\bibinfo {volume} {03}},\ \bibinfo {pages} {230} (\bibinfo {year}
  {2021}),\ \Eprint{http://arxiv.org/abs/2010.08907}{arXiv:2010.08907
  [hep-ph]}.%
  \bibAnnoteFile{Stop}{Antusch:2020vul}%
\bibitem{Das:2018usr}%
  \BibitemOpen
  \bibfield{author}{%
  \bibinfo {author} {\bibfnamefont{A.}~\bibnamefont{Das}}, \bibinfo {author}
  {\bibfnamefont{S.}~\bibnamefont{Jana}}, \bibinfo {author}
  {\bibfnamefont{S.}~\bibnamefont{Mandal}}\ and\ \bibinfo {author}
  {\bibfnamefont{S.}~\bibnamefont{Nandi}},\ }%
  \emph{\bibinfo {title} {{Probing right handed neutrinos at the LHeC and
  lepton colliders using fat jet signatures}}},\ \bibfield{journal}{%
  \Doi{10.1103/PhysRevD.99.055030}{\bibinfo {journal} {Phys. Rev. D}}\ }%
  \textbf{\bibinfo {volume} {99}},\ \bibinfo {pages} {055030} (\bibinfo {year}
  {2019}),\ \Eprint{http://arxiv.org/abs/1811.04291}{arXiv:1811.04291
  [hep-ph]}.%
  \bibAnnoteFile{Stop}{Das:2018usr}%
\bibitem{Wudka:1999ax}%
  \BibitemOpen
  \bibfield{author}{%
  \bibinfo {author} {\bibfnamefont{J.}~\bibnamefont{Wudka}},\ }%
  \emph{\bibinfo {title} {{A Short course in effective Lagrangians}}},\
  \bibfield{journal}{%
  \Doi{10.1063/1.1315034}{\bibinfo {journal} {AIP Conf.Proc.}}\ }%
  \textbf{\bibinfo {volume} {531}},\ \bibinfo {pages} {81} (\bibinfo {year}
  {2000}),\ \Eprint{http://arxiv.org/abs/hep-ph/0002180}{arXiv:hep-ph/0002180
  [hep-ph]}.%
  \bibAnnoteFile{Stop}{Wudka:1999ax}%
%%CITATION = HEP-PH/0002180;%%
\bibitem{Weinberg:1979sa}%
  \BibitemOpen
  \bibfield{author}{%
  \bibinfo {author} {\bibfnamefont{S.}~\bibnamefont{Weinberg}},\ }%
  \emph{\bibinfo {title} {{Baryon and Lepton Nonconserving Processes}}},\
  \bibfield{journal}{%
  \Doi{10.1103/PhysRevLett.43.1566}{\bibinfo {journal} {Phys. Rev. Lett.}}\ }%
  \textbf{\bibinfo {volume} {43}},\ \bibinfo {pages} {1566} (\bibinfo {year}
  {1979}).%
  \bibAnnoteFile{Stop}{Weinberg:1979sa}%
%%CITATION = PRLTA,43,1566;%%
\bibitem{Arzt:1994gp}%
  \BibitemOpen
  \bibfield{author}{%
  \bibinfo {author} {\bibfnamefont{C.}~\bibnamefont{Arzt}}, \bibinfo {author}
  {\bibfnamefont{M.}~\bibnamefont{Einhorn}}\ and\ \bibinfo {author}
  {\bibfnamefont{J.}~\bibnamefont{Wudka}},\ }%
  \emph{\bibinfo {title} {{Patterns of deviation from the standard model}}},\
  \bibfield{journal}{%
  \Doi{10.1016/0550-3213(94)00336-D}{\bibinfo {journal} {Nucl.Phys.}}\ }%
  \textbf{\bibinfo {volume} {B433}},\ \bibinfo {pages} {41} (\bibinfo {year}
  {1995}),\ \Eprint{http://arxiv.org/abs/hep-ph/9405214}{arXiv:hep-ph/9405214
  [hep-ph]}.%
  \bibAnnoteFile{Stop}{Arzt:1994gp}%
%%CITATION = HEP-PH/9405214;%%
\bibitem{Magill:2018jla}%
  \BibitemOpen
  \bibfield{author}{%
  \bibinfo {author} {\bibfnamefont{G.}~\bibnamefont{Magill}}, \bibinfo {author}
  {\bibfnamefont{R.}~\bibnamefont{Plestid}}, \bibinfo {author}
  {\bibfnamefont{M.}~\bibnamefont{Pospelov}}\ and\ \bibinfo {author}
  {\bibfnamefont{Y.-D.}\ \bibnamefont{Tsai}},\ }%
  \emph{\bibinfo {title} {{Dipole Portal to Heavy Neutral Leptons}}},\
  \bibfield{journal}{%
  \Doi{10.1103/PhysRevD.98.115015}{\bibinfo {journal} {Phys. Rev. D}}\ }%
  \textbf{\bibinfo {volume} {98}},\ \bibinfo {pages} {115015} (\bibinfo {year}
  {2018}),\ \Eprint{http://arxiv.org/abs/1803.03262}{arXiv:1803.03262
  [hep-ph]}.%
  \bibAnnoteFile{Stop}{Magill:2018jla}%
\bibitem{KamLAND-Zen:2016pfg}%
  \BibitemOpen
  \bibfield{author}{%
  \bibinfo {author} {\bibfnamefont{A.}~\bibnamefont{Gando}} \emph{et~al.}
  (\bibinfo {collaboration} {KamLAND-Zen}),\ }%
  \emph{\bibinfo {title} {{Search for Majorana Neutrinos near the Inverted Mass
  Hierarchy Region with KamLAND-Zen}}},\ \bibfield{journal}{%
  \Doi{10.1103/PhysRevLett.117.109903, 10.1103/PhysRevLett.117.082503}{\bibinfo
  {journal} {Phys. Rev. Lett.}}\ }%
  \textbf{\bibinfo {volume} {117}},\ \bibinfo {pages} {082503} (\bibinfo {year}
  {2016}),\ \bibinfo {note} {[Addendum: Phys. Rev.
  Lett.117,no.10,109903(2016)]},\
  \Eprint{http://arxiv.org/abs/1605.02889}{arXiv:1605.02889 [hep-ex]}.%
  \bibAnnoteFile{Stop}{KamLAND-Zen:2016pfg}%
%%CITATION = ARXIV:1605.02889;%%
\bibitem{Dekens:2023iyc}%
  \BibitemOpen
  \bibfield{author}{%
  \bibinfo {author} {\bibfnamefont{W.}~\bibnamefont{Dekens}}, \bibinfo {author}
  {\bibfnamefont{J.}~\bibnamefont{de~Vries}}, \bibinfo {author}
  {\bibfnamefont{E.}~\bibnamefont{Mereghetti}}, \bibinfo {author}
  {\bibfnamefont{J.}~\bibnamefont{Men\'endez}}, \bibinfo {author}
  {\bibfnamefont{P.}~\bibnamefont{Soriano}}\ and\ \bibinfo {author}
  {\bibfnamefont{G.}~\bibnamefont{Zhou}},\ }%
  \emph{\bibinfo {title} {{Neutrinoless double-beta decay in the
  neutrino-extended Standard Model}}} (\bibinfo {month} {3}\ \bibinfo {year}
  {2023}),\ \Eprint{http://arxiv.org/abs/2303.04168}{arXiv:2303.04168
  [hep-ph]}.%
  \bibAnnoteFile{Stop}{Dekens:2023iyc}%
\bibitem{Antel:2023hkf}%
  \BibitemOpen
  \bibfield{author}{%
  \bibinfo {author} {\bibfnamefont{C.}~\bibnamefont{Antel}} \emph{et~al.},\ }%
  \emph{\bibinfo {title} {{Feebly Interacting Particles: FIPs 2022 workshop
  report}}} (\bibinfo {month} {5}\ \bibinfo {year} {2023}),\
  \Eprint{http://arxiv.org/abs/2305.01715}{arXiv:2305.01715 [hep-ph]}.%
  \bibAnnoteFile{Stop}{Antel:2023hkf}%
\bibitem{Bolton:2019pcu}%
  \BibitemOpen
  \bibfield{author}{%
  \bibinfo {author} {\bibfnamefont{P.~D.}\ \bibnamefont{Bolton}}, \bibinfo
  {author} {\bibfnamefont{F.~F.}\ \bibnamefont{Deppisch}}\ and\ \bibinfo
  {author} {\bibfnamefont{P.}~\bibnamefont{Bhupal~Dev}},\ }%
  \emph{\bibinfo {title} {{Neutrinoless double beta decay versus other probes
  of heavy sterile neutrinos}}},\ \bibfield{journal}{%
  \Doi{10.1007/JHEP03(2020)170}{\bibinfo {journal} {JHEP}}\ }%
  \textbf{\bibinfo {volume} {03}},\ \bibinfo {pages} {170} (\bibinfo {year}
  {2020}),\ \Eprint{http://arxiv.org/abs/1912.03058}{arXiv:1912.03058
  [hep-ph]}.%
  \bibAnnoteFile{Stop}{Bolton:2019pcu}%
\bibitem{MEG:2016leq}%
  \BibitemOpen
  \bibfield{author}{%
  \bibinfo {author} {\bibfnamefont{A.~M.}\ \bibnamefont{Baldini}} \emph{et~al.}
  (\bibinfo {collaboration} {MEG}),\ }%
  \emph{\bibinfo {title} {{Search for the lepton flavour violating decay $\mu
  ^+ \rightarrow \mathrm {e}^+ \gamma $ with the full dataset of the MEG
  experiment}}},\ \bibfield{journal}{%
  \Doi{10.1140/epjc/s10052-016-4271-x}{\bibinfo {journal} {Eur. Phys. J. C}}\
  }%
  \textbf{\bibinfo {volume} {76}},\ \bibinfo {pages} {434} (\bibinfo {year}
  {2016}),\ \Eprint{http://arxiv.org/abs/1605.05081}{arXiv:1605.05081
  [hep-ex]}.%
  \bibAnnoteFile{Stop}{MEG:2016leq}%
\bibitem{Blennow:2023mqx}%
  \BibitemOpen
  \bibfield{author}{%
  \bibinfo {author} {\bibfnamefont{M.}~\bibnamefont{Blennow}}, \bibinfo
  {author} {\bibfnamefont{E.}~\bibnamefont{Fern\'andez-Mart\'\i{}nez}},
  \bibinfo {author} {\bibfnamefont{J.}~\bibnamefont{Hern\'andez-Garc\'\i{}a}},
  \bibinfo {author} {\bibfnamefont{J.}~\bibnamefont{L\'opez-Pav\'on}}, \bibinfo
  {author} {\bibfnamefont{X.}~\bibnamefont{Marcano}}\ and\ \bibinfo {author}
  {\bibfnamefont{D.}~\bibnamefont{Naredo-Tuero}},\ }%
  \emph{\bibinfo {title} {{Bounds on lepton non-unitarity and heavy neutrino
  mixing}}},\ \bibfield{journal}{%
  \Doi{10.1007/JHEP08(2023)030}{\bibinfo {journal} {JHEP}}\ }%
  \textbf{\bibinfo {volume} {08}},\ \bibinfo {pages} {030} (\bibinfo {year}
  {2023}),\ \Eprint{http://arxiv.org/abs/2306.01040}{arXiv:2306.01040
  [hep-ph]}.%
  \bibAnnoteFile{Stop}{Blennow:2023mqx}%
\bibitem{CMS:2018jxx}%
  \BibitemOpen
  \bibfield{author}{%
  \bibinfo {author} {\bibfnamefont{A.~M.}\ \bibnamefont{Sirunyan}}
  \emph{et~al.} (\bibinfo {collaboration} {CMS}),\ }%
  \emph{\bibinfo {title} {{Search for heavy Majorana neutrinos in same-sign
  dilepton channels in proton-proton collisions at $ \sqrt{s}=13 $ TeV}}},\
  \bibfield{journal}{%
  \Doi{10.1007/JHEP01(2019)122}{\bibinfo {journal} {JHEP}}\ }%
  \textbf{\bibinfo {volume} {01}},\ \bibinfo {pages} {122} (\bibinfo {year}
  {2019}),\ \Eprint{http://arxiv.org/abs/1806.10905}{arXiv:1806.10905
  [hep-ex]}.%
  \bibAnnoteFile{Stop}{CMS:2018jxx}%
\bibitem{Alloul:2013bka}%
  \BibitemOpen
  \bibfield{author}{%
  \bibinfo {author} {\bibfnamefont{A.}~\bibnamefont{Alloul}}, \bibinfo {author}
  {\bibfnamefont{N.~D.}\ \bibnamefont{Christensen}}, \bibinfo {author}
  {\bibfnamefont{C.}~\bibnamefont{Degrande}}, \bibinfo {author}
  {\bibfnamefont{C.}~\bibnamefont{Duhr}}\ and\ \bibinfo {author}
  {\bibfnamefont{B.}~\bibnamefont{Fuks}},\ }%
  \emph{\bibinfo {title} {{FeynRules 2.0 - A complete toolbox for tree-level
  phenomenology}}},\ \bibfield{journal}{%
  \Doi{10.1016/j.cpc.2014.04.012}{\bibinfo {journal} {Comput. Phys. Commun.}}\
  }%
  \textbf{\bibinfo {volume} {185}},\ \bibinfo {pages} {2250} (\bibinfo {year}
  {2014}),\ \Eprint{http://arxiv.org/abs/1310.1921}{arXiv:1310.1921 [hep-ph]}.%
  \bibAnnoteFile{Stop}{Alloul:2013bka}%
%%CITATION = ARXIV:1310.1921;%%
\bibitem{Degrande:2011ua}%
  \BibitemOpen
  \bibfield{author}{%
  \bibinfo {author} {\bibfnamefont{C.}~\bibnamefont{Degrande}}, \bibinfo
  {author} {\bibfnamefont{C.}~\bibnamefont{Duhr}}, \bibinfo {author}
  {\bibfnamefont{B.}~\bibnamefont{Fuks}}, \bibinfo {author}
  {\bibfnamefont{D.}~\bibnamefont{Grellscheid}}, \bibinfo {author}
  {\bibfnamefont{O.}~\bibnamefont{Mattelaer}}\ and\ \bibinfo {author}
  {\bibfnamefont{T.}~\bibnamefont{Reiter}},\ }%
  \emph{\bibinfo {title} {{UFO - The Universal FeynRules Output}}},\
  \bibfield{journal}{%
  \Doi{10.1016/j.cpc.2012.01.022}{\bibinfo {journal} {Comput. Phys. Commun.}}\
  }%
  \textbf{\bibinfo {volume} {183}},\ \bibinfo {pages} {1201} (\bibinfo {year}
  {2012}),\ \Eprint{http://arxiv.org/abs/1108.2040}{arXiv:1108.2040 [hep-ph]}.%
  \bibAnnoteFile{Stop}{Degrande:2011ua}%
%%CITATION = ARXIV:1108.2040;%%
\bibitem{Alwall:2014hca}%
  \BibitemOpen
  \bibfield{author}{%
  \bibinfo {author} {\bibfnamefont{J.}~\bibnamefont{Alwall}}, \bibinfo {author}
  {\bibfnamefont{R.}~\bibnamefont{Frederix}}, \bibinfo {author}
  {\bibfnamefont{S.}~\bibnamefont{Frixione}}, \bibinfo {author}
  {\bibfnamefont{V.}~\bibnamefont{Hirschi}}, \bibinfo {author}
  {\bibfnamefont{F.}~\bibnamefont{Maltoni}}, \bibinfo {author}
  {\bibfnamefont{O.}~\bibnamefont{Mattelaer}}, \bibinfo {author}
  {\bibfnamefont{H.~S.}\ \bibnamefont{Shao}}, \bibinfo {author}
  {\bibfnamefont{T.}~\bibnamefont{Stelzer}}, \bibinfo {author}
  {\bibfnamefont{P.}~\bibnamefont{Torrielli}}\ and\ \bibinfo {author}
  {\bibfnamefont{M.}~\bibnamefont{Zaro}},\ }%
  \emph{\bibinfo {title} {{The automated computation of tree-level and
  next-to-leading order differential cross sections, and their matching to
  parton shower simulations}}},\ \bibfield{journal}{%
  \Doi{10.1007/JHEP07(2014)079}{\bibinfo {journal} {JHEP}}\ }%
  \textbf{\bibinfo {volume} {07}},\ \bibinfo {pages} {079} (\bibinfo {year}
  {2014}),\ \Eprint{http://arxiv.org/abs/1405.0301}{arXiv:1405.0301 [hep-ph]}.%
  \bibAnnoteFile{Stop}{Alwall:2014hca}%
\bibitem{Alwall:2011uj}%
  \BibitemOpen
  \bibfield{author}{%
  \bibinfo {author} {\bibfnamefont{J.}~\bibnamefont{Alwall}}, \bibinfo {author}
  {\bibfnamefont{M.}~\bibnamefont{Herquet}}, \bibinfo {author}
  {\bibfnamefont{F.}~\bibnamefont{Maltoni}}, \bibinfo {author}
  {\bibfnamefont{O.}~\bibnamefont{Mattelaer}}\ and\ \bibinfo {author}
  {\bibfnamefont{T.}~\bibnamefont{Stelzer}},\ }%
  \emph{\bibinfo {title} {{MadGraph 5 : Going Beyond}}},\ \bibfield{journal}{%
  \Doi{10.1007/JHEP06(2011)128}{\bibinfo {journal} {JHEP}}\ }%
  \textbf{\bibinfo {volume} {06}},\ \bibinfo {pages} {128} (\bibinfo {year}
  {2011}),\ \Eprint{http://arxiv.org/abs/1106.0522}{arXiv:1106.0522 [hep-ph]}.%
  \bibAnnoteFile{Stop}{Alwall:2011uj}%
\bibitem{Sjostrand:2006za}%
  \BibitemOpen
  \bibfield{author}{%
  \bibinfo {author} {\bibfnamefont{T.}~\bibnamefont{Sjostrand}}, \bibinfo
  {author} {\bibfnamefont{S.}~\bibnamefont{Mrenna}}\ and\ \bibinfo {author}
  {\bibfnamefont{P.~Z.}\ \bibnamefont{Skands}},\ }%
  \emph{\bibinfo {title} {{PYTHIA 6.4 Physics and Manual}}},\
  \bibfield{journal}{%
  \Doi{10.1088/1126-6708/2006/05/026}{\bibinfo {journal} {JHEP}}\ }%
  \textbf{\bibinfo {volume} {05}},\ \bibinfo {pages} {026} (\bibinfo {year}
  {2006}),\ \Eprint{http://arxiv.org/abs/hep-ph/0603175}{arXiv:hep-ph/0603175
  [hep-ph]}.%
  \bibAnnoteFile{Stop}{Sjostrand:2006za}%
%%CITATION = HEP-PH/0603175;%%
\bibitem{deFavereau:2013fsa}%
  \BibitemOpen
  \bibfield{author}{%
  \bibinfo {author} {\bibfnamefont{J.}~\bibnamefont{de~Favereau}}, \bibinfo
  {author} {\bibfnamefont{C.}~\bibnamefont{Delaere}}, \bibinfo {author}
  {\bibfnamefont{P.}~\bibnamefont{Demin}}, \bibinfo {author}
  {\bibfnamefont{A.}~\bibnamefont{Giammanco}}, \bibinfo {author}
  {\bibfnamefont{V.}~\bibnamefont{Lema\^\i{}tre}}, \bibinfo {author}
  {\bibfnamefont{A.}~\bibnamefont{Mertens}}\ and\ \bibinfo {author}
  {\bibfnamefont{M.}~\bibnamefont{Selvaggi}} (\bibinfo {collaboration} {DELPHES
  3}),\ }%
  \emph{\bibinfo {title} {{DELPHES 3, A modular framework for fast simulation
  of a generic collider experiment}}},\ \bibfield{journal}{%
  \Doi{10.1007/JHEP02(2014)057}{\bibinfo {journal} {JHEP}}\ }%
  \textbf{\bibinfo {volume} {02}},\ \bibinfo {pages} {057} (\bibinfo {year}
  {2014}),\ \Eprint{http://arxiv.org/abs/1307.6346}{arXiv:1307.6346 [hep-ex]}.%
  \bibAnnoteFile{Stop}{deFavereau:2013fsa}%
\bibitem{Cacciari:2011ma}%
  \BibitemOpen
  \bibfield{author}{%
  \bibinfo {author} {\bibfnamefont{M.}~\bibnamefont{Cacciari}}, \bibinfo
  {author} {\bibfnamefont{G.~P.}\ \bibnamefont{Salam}}\ and\ \bibinfo {author}
  {\bibfnamefont{G.}~\bibnamefont{Soyez}},\ }%
  \emph{\bibinfo {title} {{FastJet User Manual}}},\ \bibfield{journal}{%
  \Doi{10.1140/epjc/s10052-012-1896-2}{\bibinfo {journal} {Eur. Phys. J. C}}\
  }%
  \textbf{\bibinfo {volume} {72}},\ \bibinfo {pages} {1896} (\bibinfo {year}
  {2012}),\ \Eprint{http://arxiv.org/abs/1111.6097}{arXiv:1111.6097 [hep-ph]}.%
  \bibAnnoteFile{Stop}{Cacciari:2011ma}%
\bibitem{Conte:2012fm}%
  \BibitemOpen
  \bibfield{author}{%
  \bibinfo {author} {\bibfnamefont{E.}~\bibnamefont{Conte}}, \bibinfo {author}
  {\bibfnamefont{B.}~\bibnamefont{Fuks}}\ and\ \bibinfo {author}
  {\bibfnamefont{G.}~\bibnamefont{Serret}},\ }%
  \emph{\bibinfo {title} {{MadAnalysis 5, A User-Friendly Framework for
  Collider Phenomenology}}},\ \bibfield{journal}{%
  \Doi{10.1016/j.cpc.2012.09.009}{\bibinfo {journal} {Comput. Phys. Commun.}}\
  }%
  \textbf{\bibinfo {volume} {184}},\ \bibinfo {pages} {222} (\bibinfo {year}
  {2013}),\ \Eprint{http://arxiv.org/abs/1206.1599}{arXiv:1206.1599 [hep-ph]}.%
  \bibAnnoteFile{Stop}{Conte:2012fm}%
%%CITATION = ARXIV:1206.1599;%%
\bibitem{Hocker:2007ht}%
  \BibitemOpen
  \bibfield{author}{%
  \bibinfo {author} {\bibfnamefont{A.}~\bibnamefont{Hocker}} \emph{et~al.},\ }%
  \emph{\bibinfo {title} {{TMVA - Toolkit for Multivariate Data Analysis}}}
  (\bibinfo {month} {3}\ \bibinfo {year} {2007}),\
  \Eprint{http://arxiv.org/abs/physics/0703039}{arXiv:physics/0703039}.%
  \bibAnnoteFile{Stop}{Hocker:2007ht}%
\bibitem{ParticleDataGroup:2020ssz}%
  \BibitemOpen
  \bibfield{author}{%
  \bibinfo {author} {\bibfnamefont{P.~A.}\ \bibnamefont{Zyla}} \emph{et~al.}
  (\bibinfo {collaboration} {Particle Data Group}),\ }%
  \emph{\bibinfo {title} {{Review of Particle Physics}}},\ \bibfield{journal}{%
  \Doi{10.1093/ptep/ptaa104}{\bibinfo {journal} {PTEP}}\ }%
  \textbf{\bibinfo {volume} {2020}},\ \bibinfo {pages} {083C01} (\bibinfo
  {year} {2020}).%
  \bibAnnoteFile{Stop}{ParticleDataGroup:2020ssz}%
\bibitem{Cowan:2010js}%
  \BibitemOpen
  \bibfield{author}{%
  \bibinfo {author} {\bibfnamefont{G.}~\bibnamefont{Cowan}}, \bibinfo {author}
  {\bibfnamefont{K.}~\bibnamefont{Cranmer}}, \bibinfo {author}
  {\bibfnamefont{E.}~\bibnamefont{Gross}}\ and\ \bibinfo {author}
  {\bibfnamefont{O.}~\bibnamefont{Vitells}},\ }%
  \emph{\bibinfo {title} {{Asymptotic formulae for likelihood-based tests of
  new physics}}},\ \bibfield{journal}{%
  \Doi{10.1140/epjc/s10052-011-1554-0}{\bibinfo {journal} {Eur. Phys. J. C}}\
  }%
  \textbf{\bibinfo {volume} {71}},\ \bibinfo {pages} {1554} (\bibinfo {year}
  {2011}),\ \bibinfo {note} {[Erratum: Eur.Phys.J.C 73, 2501 (2013)]},\
  \Eprint{http://arxiv.org/abs/1007.1727}{arXiv:1007.1727 [physics.data-an]}.%
  \bibAnnoteFile{Stop}{Cowan:2010js}%
\end{thebibliography}%

\end{document}